\documentclass[10pt]{article}
\usepackage{amsmath}
\usepackage{latexsym}
\usepackage{amssymb}
\usepackage{amsfonts}
\usepackage{amsthm}
\title{DIFFERENTIAL ALGEBRA \\ AND \\  MATHEMATICAL PHYSICS    }
\author{J.-F. Pommaret \\ CERMICS, Ecole des Ponts ParisTech,\\ 6/8 Av. Blaise Pascal, 77455 Marne-la-Vall\'ee Cedex 02, France \\
E-mail: jean-francois.pommaret@wanadoo.fr\\
URL: http://cermics.enpc.fr/$\sim$pommaret/home.html }
\date{  }
\begin{document}
\maketitle

\noindent
{\bf ABSTRACT}  \\

Many equations of mathematical physics are described by {\it differential polynomials}, that is by polynomials in the derivatives of a certain number of functions. However, up to the knowledge of the author, differential algebra in a modern setting has never been applied to study the specific algebraic feature of such equations. The purpose of this short but difficult paper is to revisit a few domains like general relativity, conformal geometry and contact geometry in the light of a modern approach to nonlinear systems of partial differential equations, using new methods from {\it Differential Geometry} (D.C. Spencer, 1970), {\it Differential Algebra} (J.F. Ritt, 1950 and E. Kolchin, 1973) and {\it Algebraic Analysis} (M. Kashiwara, 1970). Identifying the {\it differential indeterminates} of Ritt and Kolchin with the {\it jet coordinates} of Spencer, the idea is to study {\it Differential Duality} by using only linear differential operators with coefficients in a differential field. In particular, the linearized second order Einstein equations are parametrizing the first order Cauchy stress equations {\it but} cannot themselves be parametrized. In the framework of {\it Homological Algebra}, this result is not coherent with the vanishing of certain {\it first and second extension modules}. As a byproduct, we shall prove that gravitation {\it and} electromagnetism must only depend on the second order jets ({\it elations}) of the system of conformal Killing equations. Finally, we shall use these new methods in order to study contact transformations in arbitrary odd dimension and apply these results to 
the study of Hamilton-Jacobi equations in mechanics.

\vspace{3cm}


\newpage

\noindent
{\bf 1) INTRODUCTION}  \\

Let us consider a manifold $X$ of dimension $n$ with local coordinates $x=(x^i)=(x^1,...,x^n)$, tangent bundle $T$, cotangent bundle $T^*$, vector bundle $S_qT^*$ of $q$-symmetric covariant tensors and vector bundle ${\wedge}^rT^*$ of $r$-skewsymmetric covariant tensors or $r$-forms. The group of isometries $ y=f(x)$ of the non-degenerate metric $\omega$ with $det(\omega)\neq 0$ on $X$ is defined by the nonlinear first order system of {\it general finite Lie equations} in {\it Lie form}:  \\
\[     {\omega}_{kl}(f(x)){\partial}_if^k(x){\partial}_jf^l(x)={\omega}_{ij}(x)    \]
Linearizing at the identity transformation $y=x$, we may introduce the corresponding {\it  Killing operator} $T \rightarrow S_2T^*:\xi \rightarrow {\cal{D}}\xi={\cal{L}}(\xi)\omega=\Omega  $, which involves the Lie derivative ${\cal{L}}$ and provides twice the so-called infinitesimal deformation tensor of continuum mechanics when $\omega$ is the Euclidean metric. We may consider the linear first order system of {\it general infinitesimal Lie equations} in {\it Medolaghi form}, also called {\it system of Killing equations}:  \\
\[{\Omega}_{ij}\equiv ({\cal{L}}(\xi){\omega})_{ij} \equiv {\omega}_{rj}(x){\partial}_i{\xi}^r +{\omega}_{ir}(x){\partial}_j{\xi}^r+{\xi}^r{\partial}_r{\omega}_{ij}(x)=0 \]
which is in fact a family of systems only depending on the {\it geometric object} $\omega$ and its derivatives. Introducing the Christoffel symbols $\gamma$, we may differentiate once and add the operator ${\cal{L}}(\xi)\gamma=\Gamma \in S_2T^*\otimes T$ with the well known Levi-Civita isomorphism $j_1(\omega)=(\omega,{\partial}_x\omega)\simeq (\omega,\gamma)$ in order to obtain the linear second order system of {\it general infinitesimal Lie equations} in {\it Medolaghi form}:  \\
\[ {\Gamma}^k_{ij}\equiv ({\cal{L}}(\xi)\gamma)^k_{ij}\equiv {\partial}_{ij}{\xi}^k+{\gamma}^k_{rj}(x){\partial}_i{\xi}^r+{\gamma}^k_{ir}(x){\partial}_j{\xi}^r-{\gamma}^r_{ij}(x){\partial}_r{\xi}^k+{\xi}^r{\partial}_r{\gamma}^k_{ij}(x)=0   \]
We have thus linearized a nonlinear differential algebraic system in order to obtain a linear system with coefficients in the differential field $\mathbb{Q}(\omega, \partial \omega, ...)$ along the idea of E. Vessiot [65]. This system is formally integrable if and only if $\omega$ has a {\it constant Riemannian curvaure}.\\
Similarly, introducing the Jacobian determinant $\Delta (x)=det({\partial}_if^k(x))$, the group of conformal transformations of the metric 
$\omega$ may be defined by the nonlinear first order system of {\it general finite Lie equations} in {\it Lie form}:  \\
\[               {\hat{\omega}}_{kl}(f(x)) {\Delta}^{-\frac{2}{n}}(x){\partial}_if^k(x){\partial}_jf^l(x)={\hat{\omega}}_{ij}(x)    \]
while introducing the {\it metric density} ${\hat{\omega}}_{ij}={\mid det(\omega)\mid }^{-\frac{1}{n}}{\omega}_{ij} \Rightarrow det(\hat{\omega})=\pm 1$ as a new {\it geometric object}, rather than by eliminating a {\it conformal factor} as usual. The {\it conformal Killing operator} $\xi \rightarrow \hat{\cal{D}}\xi={\cal{L}}(\xi)\hat{\omega}=\hat{\Omega }$ may be defined by linearization as above and we obtain the first order system of {\it general infinitesimal Lie equations} in {\it Medolaghi form}, also called {\it system of conformal Killing equations}:  \\
\[ { \hat{\Omega}}_{ij}\equiv {\hat{\omega}}_{rj}(x){\partial}_i{\xi}^r +{\hat{\omega}}_{ir}(x){\partial}_j{\xi}^r 
- \frac{2}{n}{\hat{\omega}}_{ij}(x){\partial}_r{\xi}^r + {\xi}^r{\partial}_r{\hat{\omega}}_{ij}(x)=0    \]
as a system with coefficients in the differential field $\mathbb{Q}(\bar{\omega},\partial \bar{\omega},...)$. We may introduce the {\it trace} $tr(\Omega)={\omega}^{ij}{\Omega}_{ij}$ with standard notations and obtain therefore $tr(\hat{\Omega})=0$ because ${\hat{\Omega}}_{ij}={\mid det(\omega)\mid }^{-\frac{1}{n}} ({\Omega}_{ij}- \frac{1}{n}{\omega}_{ij}tr(\Omega))$. This system becomes formally interable if and only if the corresponding {\it Weyl tensor} vanishes.  \\

The reader may look at [31,32,33,34,44] for finding examples of {\it Lie groups} or {\it Lie pseudogroups} of transformations 
along the approach initiated by E. Vessiot in 1903 [53].\\

In classical elasticity, the {\it stress tensor density} $\sigma=({\sigma}^{ij}={\sigma}^{ji})$ existing inside an elastic body is a symmetric $2$-tensor density introduced by A. Cauchy in 1822. Integrating by parts the implicit summation $- \frac{1}{2}{\sigma}^{ij}
{\Omega}_{ij}$, we obtain the {\it Cauchy operator} $\sigma \rightarrow {\partial}_r{\sigma}^{ir} + {\gamma}^i_{rs}{\sigma}^{rs}=f^i$. When $\omega$ is the euclidean metric, the corresponding {\it Cauchy stress equations} can be written as ${\partial}_r{\sigma}^{ir}=f^i$ where the right member describes the local density of forces applied to the body, for example gravitation. With zero second member, we study the possibility to "{\it parametrize} " the system of PD equations ${\partial}_r{\sigma}^{ir}=0$, namely to express its general solution by means of a certain number of arbitrary functions or {\it potentials}, called {\it stress functions}. Of course, the problem is to know about the number of such functions and the order of the parametrizing operator. For $n=1,2,3$ one may introduce the Euclidean metric $\omega=({\omega}_{ij}={\omega}_{ji})$ while, for $n=4$, one may consider the Minkowski metric. A few definitions used thereafter will be provided later on.\\

When $n=2$, the stress equations become ${\partial}_1{\sigma}^{11}+{\partial}_2{\sigma}^{12}=0, {\partial}_1{\sigma}^{21}+{\partial}_2{\sigma}^{22}=0$. Their second order parametrization ${\sigma}^{11}={\partial}_{22}\phi, {\sigma}^{12}={\sigma}^{21}=-{\partial}_{12}\phi, {\sigma}^{22}={\partial}_{11}\phi$ has been provided by George Biddell Airy (1801-1892) in 1863 [1]. It can be simply recovered in the following manner: \\
\[ \begin{array}{rcl}
{\partial}_1{\sigma}^{11}- {\partial}_2( - {\sigma}^{12})= 0 \hspace{5mm} & \Rightarrow & \hspace{5mm} \exists \varphi,\, {\sigma}^{11}={\partial}_2\varphi, {\sigma}^{12}= 
- {\partial}_1\varphi \\  
{\partial}_2{\sigma}^{22}- {\partial}_1( - {\sigma}^{21})=0  \hspace{5mm} & \Rightarrow & \hspace{5mm}  \exists \psi, \,{\sigma}^{22}={\partial}_1\psi, {\sigma}^{21}= - {\partial}_2\psi \\
   {\sigma}^{12}={\sigma}^{21} \Rightarrow {\partial}_1 \varphi - {\partial}_2\psi =0 \hspace{5mm}& \Rightarrow & \hspace{5mm}\exists \phi, \,\varphi={\partial}_2\phi, \psi={\partial}_1\phi  
   \end{array}Ê \]
We get the linear second order system:  \\
\[ \left\{  \begin{array}{rll}
{\sigma}^{11} & \equiv {\partial}_{22}\phi =0 \\
-{\sigma}^{12} & \equiv {\partial}_{12}\phi =0 \\
{\sigma}^{22} & \equiv {\partial}_{11}\phi=0
\end{array}
\right. \fbox{ $ \begin{array}{ll}
1 & 2   \\
1 & \bullet \\  
1 & \bullet  
\end{array} $ } \]
which is involutive with one equation of class $2$, $2$ equations of class $1$ and it is easy to check that the $2$ corresponding first order CC are just the stress equations.\\

When constructing a long prismatic dam with concrete as in [12,13] or in the Introduction of [36], we may transform a problem of $3$-dimensional elasticity into a problem of $2$-dimensional elasticity by supposing that the axis $x^3$ of the dam is perpendicular to the river with ${\Omega}_{ij}(x^1,x^2), \forall i,j=1,2$ and ${\Omega}_{33}=0$ because of the rocky banks of the river are supposed to be fixed. We may introduce the two {\it Lam\'{e} constants} $(\lambda,\mu)$ and the {\it Poisson coefficient} $\nu=\lambda /2(\lambda + \mu)$ in order to describe the usual constitutive relations of an homogeneous isotropic medium as follows ({\it care}: $n=3 \Rightarrow n=2$):  \\
\[ {\sigma}=\frac{1}{2}\lambda \, tr(\Omega)\, {\omega} + \mu \, {\Omega}, \, \, tr(\Omega)={\Omega}_{11}+{\Omega}_{22} \hspace{4mm}  \Rightarrow   \hspace{4mm}  
 \mu \, {\Omega}=    \sigma - \frac{\lambda}{2( \lambda + \mu)}\, tr(\sigma) \, \omega , \, \, tr(\sigma)={\sigma}^{11} + {\sigma}^{22} \]
even though ${\sigma}^{33}=\frac{1}{2}\lambda ({\Omega}_{11}+{\Omega}_{22})=\frac{1}{2}\lambda tr(\Omega) \Rightarrow {\sigma}^{33}=\nu ({\sigma}^{11} + {\sigma}^{22})\neq0$. Let us consider the {\it right square} of the diagram below with locally exact rows:  \\
\[   \begin{array}{ccccc}
   2  & \stackrel{Killing}{\longrightarrow} & 3 & \stackrel{Riemann}{\longrightarrow} & 1   \\
   \vdots &  & {\downarrow\uparrow} &  &  \vdots  \\
   2 & \stackrel{Cauchy}{\longleftarrow} & 3 & \stackrel{Airy}{\longleftarrow} & 1
\end{array}  \]
Taking into account the formula $(5.1.4)$ of [15] for the linearization of the only component of the Riemann tensor at $\omega$ when $n=2$ and substituting the Airy parametrization, we obtain:  \\
\[ tr(R)\equiv d_{11}{\Omega}_{22}+d_{22}{\Omega}_{11}-2d_{12}{\Omega}_{12}=0 \hspace{3mm} \Rightarrow 
\hspace{3mm} \mu \, tr(R)\equiv \frac{\lambda + 2 \mu}{2(\lambda +\mu)} \Delta \Delta \phi=0 \hspace{3mm} \Rightarrow 
\hspace{3mm} \Delta \Delta \phi=0  \]
where the linearized {\it scalar curvature} $tr(R)$ is allowing to define the {\it Riemann operator} in the previous diagram, namely the only {\it compatibility condition} (CC) of the Killing operator. Using now the {\it left square} of the previous diagram, we may also substitute the Airy parametrization in the Cauchy stress equations and get, when $\vec{f}=\vec{g}$ ({\it gravity})({\it care}: n=3):   \\
\[  (\lambda + \mu) \vec{\nabla}(\vec{\nabla}.\vec{\xi})+\mu \Delta \vec{\xi}=\vec{f}\hspace{2mm}  \stackrel{\vec{\nabla}}{\Rightarrow} \hspace{2mm}  (\lambda + 2\mu)\Delta tr(\Omega) =0 \Rightarrow \Delta tr(\Omega)=0\Rightarrow \Delta tr(\sigma)=0 \Rightarrow  \Delta \Delta \phi=0   \]
It remains to exhibit an arbitrary homogeneous polynomial solution of degree $3$ and to determine its $4$ coefficients by the boundary pressure conditions on the upstream and downstream walls of the dam. Of course, the Airy potential $\phi$ has {\it nothing to do} with the perturbation $\Omega$ of the metric $\omega$ and {\it the Airy parametrization is nothing else but the formal adjoint of the Riemann operator}, linearization of the Riemann tensor at $\omega$,  expressing the second order {\it compatibility conditions} (CC) of the inhomogeneous system ${\cal{D}}\xi=\Omega$. Also, as we discover at once, the origin of elastic waves is shifted by {\it one step backwards}, {\it from the right square to the left square} of the diagram. Indeed, using inertial forces $\vec{f}=\rho \,{\partial}^2\vec{\xi}/\partial t^2$ for a medium with mass $\rho$ per unit volume in the right member of Cauchy stress equations because of Newton law, we discover the existence of two types of {\it elastic waves}, namely the {\it longitudinal} and {\it transversal} waves with different speeds $v_T < v_L$ that are really existing because they are responsible for earthquakes [36]:  \\
\noindent
 \[  \left\{   \begin{array}{rcrcl}
  \vec{\nabla}.\vec{\xi}=0  &  \hspace{1cm}\Rightarrow \hspace{1cm}&  \mu \triangle \vec{\xi}= \vec{f} &  \hspace{1cm} \Rightarrow \hspace{1cm} &  v_T=\sqrt{\frac{\mu}{\rho}}  \\
  \vec{\nabla}\wedge \vec{\xi}=0 & \hspace{1cm}\Rightarrow \hspace{1cm}& (\lambda + 2 \mu )\triangle \vec{\xi}= \vec{f} & \hspace{1cm}   \Rightarrow \hspace{1cm} & v_L=\sqrt{\frac{\lambda + 2\mu}{\rho}}  
\end{array} \right. \] 

It is this comment that pushed the author to use the {\it formal adjoint} of an operator, knowing already that {\it an operator and its (formal) adjoint have the same differential rank}. In the case of the conformal Killing operator, the second order CC are generated by the {\it Weyl operator}, linearization of the Weyl tensor at $\hat{\omega}$ when $n\geq 4$. The particular situation $n=3$ will be studied and its corresponding $5$ third order CC are not known after one century [44]. Finally, the {\it Bianchi operator} describing the CC of the Riemann operator does not appear in this scheme.  \\

Summarizing what we have just said, {\it the study of elastic waves in continuum mechanics only depends on group theory} because {\it it has only to do} with one {\it differential sequence} and its {\it formal adjoint}, combined together by means of constitutive relations. We have proved in many books [33,34,] and in [41,42,44] that the situation is similar for Maxwell equations, a result leading therefore to revisit the mathematical foundations of both General Relativity (GR) and Electromagnetism (EM), thus also of Gauge Theory (GT).  \\

The first motivation for studying the methods used in this paper has been a $1000 \$ $ challenge proposed in $1970$ by J. Wheeler in the physics department of Princeton University while the author of this paper was a student of D.C. Spencer in the closeby mathematics department:  \\ 

{\it Is it possible to express the generic solutions of Einstein equations in vacuum by means of the derivatives of a certain number of arbitrary functions like the potentials for Maxwell equations ?}.\\

During the next $25$ years and though surprising it may look like, no progress has been made towards any solution and we found the {\it negative} solution of this challenge in $1995$ [35]. Then, being already in contact with M.P. Malliavin as I gave a seminar on the "{\it Deformation Theory of Algebraic and Geometric Structures} " [24,44], I presented in 1995 a seminar at IHP in Paris, proving the impossibility to parametrize Einstein equations, a result I just found. One of the participants called my attention on a recently published translation from japanese of the 1970 master thesis of M. Kashiwara that he just saw on display in the library of the Institute [21]. This has been the true starting of the story because I discovered that the duality involved in the preceding approach to physics was only a particular example of a much more sophisticated framework having to do with {\it homological algebra} [2,7,27,36,50]. \\

Let us explain this point of view by means of an elementary example. With ${\partial}_{22}\xi={\eta}^2, {\partial}_{12}\xi={\eta}^1$ for $\cal{D}$, we get  ${\partial}_1{\eta}^2-{\partial}_2{\eta}^1=\zeta$ for the CC ${\cal{D}}_1$. Then $ad({\cal{D}}_1)$ is defined by ${\mu}^2=-{\partial}_1\lambda, {\mu}^1={\partial}_2\lambda$ while $ad(\cal{D})$ is defined by $\nu={\partial}_{12}{\mu}^1+{\partial}_{22}{\mu}^2$ but the CC of $ad({\cal{D}}_1)$ are generated by ${\nu}'={\partial}_1{\mu}^1+{\partial}_2{\mu}^2$. Using operators, we have the two differential sequences:\\  
\[  \begin{array}{ccccl}
 \xi & \stackrel{\cal{D}}{\longrightarrow} & \eta & \stackrel{{\cal{D}}_1}{\longrightarrow} & \zeta   \\
  \nu& \stackrel{ad(\cal{D})}{\longleftarrow} & \mu & \stackrel{ad({\cal{D}}_1)}{\longleftarrow} & \lambda \\
       &    \swarrow &  &       &  \\
    \hspace*{2mm} {\nu}' &  &  &  &
  \end{array}  \]
where ${\cal{D}}_1$ generates the CC of ${\cal{D}}$ in the upper sequence but $ad({\cal{D}})$ does not generate the CC of $ad({\cal{D}}_1)$ in the lower sequence, even though ${\cal{D}}_1\circ {\cal{D}}=0 \Rightarrow ad({\cal{D}}) \circ ad({\cal{D}}_1)=0$, {\it contrary to what happened in the previous diagram}. We shall see that this comment brings the need to introduce the {\it first extension module} ${ext}^1(M)$ of the differential module $M$ determined by ${\cal{D}}$.    \\

In the meantime, following U. Oberst [28,29], a few persons were trying to adapt these methods to control theory and, thanks to J.L. Lions, I have been able to advertise about this new approach in a european course, held with succes during 6 years [34] and continued for 5 other years in a slightly different form [37]. By chance I met A. Quadrat, a good PhD student interested by control and computer algebra and we have been staying alone because the specialists of Algebraic Analysis were pure mathematicians, not interested at all by applications. As a byproduct, it is rather strange to discover that the impossibility to parametrize Einstein equations, that we shall prove in Section $4$, has never been acknowledged by physicists but can be found in a book on control because it is now known that a control system is controllable if and only if it is parametrizable [37,56].  \\

The following example of a double pendulum will prove that this result, still not acknowledged today by engineers, is not evident at all. For this, let us consider two pendula of respective length $l_1$ and $l_2$ attached at the ends of a rigid bar sliding horizontally with a reference position $x(t)$. If the pendula move with a respective (small) angle ${\theta}_1(t)$ and ${\theta}_2(t)$ with respect to the vertical, it is easy to prove from the Newton principle that the equations of the movements does not depend on the respective masses $m_1$ and $m_2$ of the pendula but only depend on the respective lengths and gravity $g$ along the two formulas:  \\
\[  d^2 x + l_1 d^2{\theta}_1 + g {\theta}_1=0,  \hspace{2cm} d^2x +l_2 d^2 {\theta}_2 +g {\theta}_2=0  \]
where $d=d_t$ is the standard time derivative. It is {\it experimentally} visible and any reader can check it with a few dollars, that the system is controllable, that is the angles can reach any prescribed (small) values in a finite time when starting from equilibrium, {\it if and only if} $l_1\neq l_2$ and, in this case, we have the following (injective) $4^{th}$ order parametrization:  \\
\[  -l_1l_2 d^4 \phi - g(l_1 + _2) d^2 \phi - g^2\phi=x, 
\hspace{1cm} l_2d^4\phi +g d^2 \phi= {\theta}_1, \hspace{1cm} l_1 d^4 \phi + g d^2 \phi={\theta}_2  \]
0f course, if $l_1=l_2=l$, the system cannot be controllable because, setting $\theta ={\theta}_1 -{\theta}_2$, we obtain by substraction 
$ld^2 \theta + g \theta=0$ and thus ${\theta}(0)=0, d{\theta}(0)=0 \Rightarrow {\theta}(t)=0$.  \\

We end this Introduction explaining on a simple example why the {\it second extension module} ${ext}^2(M)$ must also be considered, especially in the study of Einstein equations, though surprising it may look like. To make a comparison, let us consider the following well known {\it Poincar\'{e} sequence}:  \\ \[              {\wedge}^0T^* \stackrel{d}{\longrightarrow} {\wedge}^1T^* \stackrel{d}{\longrightarrow} {\wedge}^2T^* \stackrel{d}{\longrightarrow} ... \stackrel{d}{\longrightarrow}{\wedge}^{n-1}T^* \stackrel{d}{\longrightarrow}
{\wedge}^nT^* \rightarrow 0  \]
where $d:\omega={\omega}_Idx^I\rightarrow {\partial}_i{\omega}_Idx^i\wedge dx^I$ is the exterior derivative. When $n=3$, we have:  \\
\[  \begin{array}{ccc}\hspace{7mm} {\wedge}^0T^*\, \stackrel{d}{\longrightarrow}\, {\wedge}^1T^* \, \stackrel{d}{\longrightarrow}\, {\wedge}^2T^* \, \stackrel{d}{\longrightarrow}\, {\wedge}^3T^* \rightarrow 0 \hspace{4mm} &\Leftrightarrow & \hspace{7mm}   \phi \stackrel{grad}{\longrightarrow} \xi \stackrel{curl}{\longrightarrow} \eta \stackrel{div}{\longrightarrow} \zeta \rightarrow 0  \\
   &   &   \\
   0 \leftarrow {\wedge}^3T^* \stackrel{ad(d)}{\longleftarrow}{\wedge}^2T^* \stackrel{ad(d)}{\longleftarrow} {\wedge}^1T^* \stackrel{ad(d)}{\longleftarrow} {\wedge}^0T^*  \hspace{11mm}& \Leftrightarrow  &\hspace{4mm}   0 \leftarrow \theta \stackrel{div}{\longleftarrow} \nu \stackrel{curl}{\longleftarrow} \mu \stackrel{grad}{\longleftarrow} \lambda  \hspace{10mm}
\end{array}  \]
From their definition it follows that $div$ is parametrized by $curl$ while $curl$ is parametrized by $grad$. Also, in local coordinates, we have $ad(div)= - grad , ad(curl)=  curl, ad(grad)= - div$ and the adjoint sequence is also the Poincar\'{e} sequence {\it up to the sign}. Let us nevertheless consider the new ({\it minimal}) parametrization of $div$ obtained by setting ${\xi}^3=0$, namely [45,46]:  \\
\[    d_2{\xi}^3- d_3{\xi}^2={\eta}^1, \hspace{2mm}d_3{\xi}^1-d_1{\xi}^3={\eta}^2, \hspace{2mm} d_1{\xi}^2-d_2{\xi}^1={\eta}^3 \Rightarrow
- d_3{\xi}^2= {\eta}^1, \hspace{2mm} d_3{\xi}^1= {\eta}^2, \hspace{2mm} d_1{\xi}^2 - d_2{\xi}^1={\eta}^3  \]
If we define the {\it differential rank} of an operator by the maximum number of differentially independent second member, this is clearly an involutive differential operator with differential rank equal to $2$ because $({\xi}^1,{\xi}^2)$ can be given arbitrarily and thus $({\eta}^1, {\eta}^2)$ can be given arbitrarily or, equivalently, because the differential rank of $div$ is of course equal to $1$ as $div$ has no CC. Now, the involutive system $d_3{\xi}^2=0, d_3{\xi}^1=0, d_1{\xi}^2 - d_2{\xi}^1=0$ canot be parametrized by one arbitrary function because both ${\xi}^1$ and ${\xi}^2$ are {\it autonomous} in the sense that they both satisfy to at least one partial differential equation (PDE). Accordingly, we discover that $div$ can be parametrized by the $curl$ through $3$ arbitrary functions $({\xi}^1, {\xi}^2, {\xi}^3)$ where ${\xi}^3$ may be given arbitrarily, the $curl$ being itself parametrized by the $grad$, but $div$ can also be parametrized by another operator with less arbitrary functions or {\it potentials} which, in turn, cannot be parametrized again. Such a situation is similar to the one met in hunting rifles that may have one, two or more trigger mechanisms that can be used successively. It happens that the possibility to have one parametrization of $div$ is an intrinsic property described by the vanishing of ${ext}^1(M)$ where the differential module $M$ is determined by $grad$ while the property to have two successive parametrizations is an intrinsic property described by the vanishing of ${ext}^1(M)$ as we just said {\it plus} the vanishing of  ${ext}^2(M)$, and so on. It follows that certain parametrizations are " {\it better} " than others and no student should even imagine the minimal parametrization of $div$ that we have presented above. A similar procedure has been adopted by J.C. Maxwell [25] and G. Morera [26] when they modified the parametrization of the Cauchy stress equations obtained by E. Beltrami in 1892 (See [3, 9,10,43,47,48] for more details).   \\

We now treat the case $dim(X)=3$ as the case $dim(X)=n=2p+1\geq 5$ will need much more work (See [39] for more details). Let us introduce the so-called {\it contact} $1$-form $\alpha=dx^1-x^3dx^2$ and consider the Lie pseudogroup $\Gamma\subset aut(X)$ of (local) transformations preserving $\alpha$ up to a function factor, that is $\Gamma=\{f\in aut(X){\mid}j_1(f)^{-1}(\alpha)= \rho \alpha\}$ where again $j_q(f)$ is a symbolic way for writing out the derivatives of $f$ up to order $q$ and $\alpha$ transforms like a $1$-covariant tensor. It may be tempting to look for a kind of "{\it object} " the invariance of which should characterize $\Gamma$. Introducing the exterior derivative $d\alpha=dx^2\wedge dx^3$ as a $2$-form, we obtain the volume $3$-form $\alpha\wedge d\alpha=dx^1\wedge dx^2\wedge dx^3$. As it is well known that the exterior derivative commutes with any diffeomorphism, we obtain sucessively:\\
\[      j_1(f)^{-1}(d\alpha)=d(j_1(f)^{-1}(\alpha))=d(\rho \alpha)=\rho d\alpha +d\rho \wedge \alpha  \Rightarrow j_1(f)^{-1}(\alpha \wedge d\alpha)={\rho}^2(\alpha\wedge d\alpha)   \]
As the volume $3$-form $\alpha\wedge d\alpha$ transforms through a division by the Jacobian determinant $\Delta=\partial (f^1,f^2,f^3)/\partial (x^1,x^2,x^3)\neq 0$ of the transformation $y=f(x)$ with inverse $x=f^{-1}(y)=g(y)$, {\it the desired object is thus no longer a} $1$-{\it form but a} $1$-{\it form density} $\omega=({\omega}_1,{\omega}_2,{\omega}_3)$ transforming like a $1$-form but up to a division by the square root of the Jacobian determinant. We obtain the nonlinear differential algebraic system of {\it general finite Lie equations} in Lie form:   \\
\[    {\omega}_k(y)(\frac{\partial (y^1,...,y^n)}{\partial (x^1,...,x^n)})^{ - \frac{1}{2}}y^k_i={\omega}_i(x)  \]
It follows that the infinitesimal contact transformations are vector fields $\xi\in T=T(X)$ the tangent bundle of $X$, satisfying the $3$ so-called first order system of {\it general infinitesimal Lie equations} in {\it Medolaghi form}:ÊÊÊ\\
\[  {\Omega}_i\equiv ({\cal{L}}(\xi)\omega)_i\equiv {\omega}_r(x){\partial}_i{\xi}^r-(1/2){\omega}_i(x){\partial}_r{\xi}^r+{\xi}^r{\partial}_r{\omega}_i(x)=0   \]
When $\omega=(1,-x^3,0)$, we obtain the {\it special} involutive system (See below for details):ÊÊ\\
\[  {\partial}_3{\xi}^3+{\partial}_2{\xi}^2+2x^3{\partial}_1{\xi}^2-{\partial}_1{\xi}^1=0, {\partial}_3{\xi}^1-x^3{\partial}_3{\xi}^2=0, {\partial}_2{\xi}^1-x^3{\partial}_2{\xi}^2+x^3{\partial}_1{\xi}^1-(x^3)^2{\partial}_1{\xi}^2-{\xi}^3=0   \]
with  $2$ equations of class $3$ and $1$ equation of class $2$, a result leading thus to only $1$ {\it compatibility conditions} (CC) for the second members. Equivalently, we have the system:  \\
\[ \left \{ \begin{array}{ccl}
{\Omega}_3 & \equiv & {\xi}^1_3 - x^3{\xi}^2_3=0  \\
{\Omega}_2 & \equiv & {\xi}^1_2 - x^3{\xi}^ 2_2 + \frac{1}{2} x^3 ({\xi}^1_1+{\xi}^2_2 + {\xi}^3_3) - {\xi}^3=0  \\
{\Omega}_1 & \equiv & {\xi}^1_1 - x^3 {\xi}^2_1 - \frac{1}{2} ({\xi}^1_1 + {\xi}^2_2 + {\xi}^3_3)=0
\end{array} \right.  \]
For an arbitrary $\omega$, we may ask about the differential conditions on $\omega$ such that all the equations of order $r+1$ are only obtained by differentiating $r$ times the first order equations, exactly like in the special situation just considered where the system is involutive. We notice that, in a symbolic way, $\omega \wedge d\omega$ is now a scalar $c(x)$ providing the zero order equation ${\xi}^r{\partial}_rc(x)=0$ and the condition is $c(x)=c=cst$. The {\it integrability condition} (IC) is the {\it Vessiot structure equation}:   \\
\[  I(j_1(\omega)) \equiv {\omega}_1({\partial}_2{\omega}_3-{\partial}_3{\omega}_2)+{\omega}_2({\partial}_3{\omega}_1-{\partial}_1{\omega}_3)+{\omega}_3({\partial}_1{\omega}_2-{\partial}_2{\omega}_1)=c    \]
involving the only {\it structure constant} $c$ like the Riemannian structure.   \\
For $\omega=(1,-x^3,0)$, we get $c=1$. If we choose $\bar{\omega}=(1,0,0)$ leading to $\bar{c}=0$, we may define $\bar{\Gamma}=\{f\in aut(X){\mid} j_1(f)^{-1}(\bar{\omega})=\bar{\omega}\}$ with infinitesimal transformations satisfying the involutive system:\\
\[  {\partial}_3{\xi}^3+{\partial}_2{\xi}^2-{\partial}_1{\xi}^1=0, \hspace{3mm}{\partial}_3{\xi}^1=0, 
\hspace{3mm}{\partial}_2{\xi}^1=0  \]
with again $2$ equations of class $3$ and $1$ equation of class $2$. The {\it equivalence problem} $j_1(f)^{-1}(\omega)=\bar{\omega}$ cannot be solved even locally because this system cannot have any invertible solution. Indeed, studying the system $j_1(g)^{-1}(\bar{\omega})=\omega$, we have to solve:  \\
\[       \frac{\partial g^1}{\partial y^2}+y^3\frac{\partial g^1}{\partial y^1}=0, \frac{\partial g^1}{\partial y^3}=0 \Rightarrow \frac{\partial g^1}{\partial y^1}=0, \frac{\partial g^1}{\partial y^2}=0, \frac{\partial g^1}{\partial y^3}=0  \]
by using crossed derivatives.\\

Using now the definition of contact transformations, we have the three equations:  \\
\[  ({\cal{L}}(\xi)\alpha)_i\equiv {\alpha}_r{\partial}_i{\xi}^r + {\xi}^r{\partial}_r{\alpha}_i=\rho(x) {\alpha}_i   \]
Eliminating the arbitrary factor $\rho(x)$, we obtain the two linearly independent infinitesimal Lie equations:  \\
\[ \left \{ \begin{array}{lcl}
{\alpha}_2{\alpha}_r{\partial}_3{\xi}^r -{\alpha}_3{\alpha}_r{\partial}_2{\xi}^r  +({\alpha}_2{\partial}_r{\alpha}_3-{\alpha}_3{\partial}_r{\alpha}_2){\xi}^r & =  & 0  \\
   &  &  \\
  {\alpha}_3{\alpha}_r{\partial}_1{\xi}^r -{\alpha}_1{\alpha}_r{\partial}_3{\xi}^r+({\alpha}_3{\partial}_r{\alpha}_1-{\alpha}_1{\partial}_r{\alpha}_3){\xi}^r  & =  &  0  
\end{array} \right.  \]
which are nevertheless not in the Medolaghi form because the $1$-form $\alpha$ is {\it not} a geometric object. \\
Multiplying on the left the first equation by the test function ${\lambda}^1$ and the second by the test function ${\lambda}^2$ and integrating by part, we obtain for example, separating the terms involving {\it only} ${\lambda}_1$ from the terms involving {\it only} 
${\lambda}^2$:    \\
\[ \left \{\begin{array}{lcl}
{\xi}^1 &\rightarrow &-{\partial}_3({\alpha}_1{\alpha}_2{\lambda}^1)+{\partial}_2({\alpha}_1{\alpha}_3{\lambda}^1)+({\alpha}_2{\partial}_1{\alpha}_3-{\alpha}_3{\partial}_1{\alpha}_2){\lambda}^1\\
{\xi}^2 & \rightarrow & ...  \\
  {\xi}^3& \rightarrow& -{\partial}_3({\alpha}_2{\alpha}_3{\lambda}^1)+{\partial}_2(({\alpha}_3)^2{\lambda}^1)+({\alpha}_2{\partial}_3{\alpha}_3-{\alpha}_3{\partial}_3{\alpha}_2){\lambda}^1
\end{array} \right. \]
and:  \\
\[ \left \{ \begin{array}{lcl}
{\xi}^1 &\rightarrow &-{\partial}_1({\alpha}_1{\alpha}_3{\lambda}^2)+{\partial}_3(({\alpha}_1)^2{\lambda}^2)+({\alpha}_3{\partial}_1{\alpha}_1-{\alpha}_1{\partial}_1{\alpha}_3){\lambda}^2\\
{\xi}^2 & \rightarrow & ...  \\
  {\xi}^3& \rightarrow& -{\partial}_1(({\alpha}_3)^2{\lambda}^2)+{\partial}_3({\alpha}_1{\alpha}_3{\lambda}^2)+({\alpha}_3{\partial}_3{\alpha}_1-{\alpha}_1{\partial}_3{\alpha}_3){\lambda}^2
\end{array}  \right.\]
that we may rewrite respectively as:  \\
\[ \left \{\begin{array}{c}
-({\alpha}_1{\alpha}_2){\partial}_3{\lambda}^1+({\alpha}_1{\alpha}_3){\partial}_2{\lambda}^1 - ({\alpha}_1({\partial}_2{\alpha}_3-{\partial}_3{\alpha}_2) -{\alpha}_2({\partial}_3{\alpha}_1-{\partial}_1{\alpha}_3) - {\alpha}_3({\partial}_1{\alpha}_2-{\partial}_2{\alpha}_1)){\lambda}^1\\
 ...  \\
   -({\alpha}_2{\alpha}_3){\partial}_3{\lambda}^1+({\alpha}_3)^2{\partial}_2{\lambda}^1+2{\alpha}_3({\partial}_2{\alpha}_3-{\partial}_3{\alpha}_2){\lambda}^1
\end{array} \right. \]
and:  \\
\[ \left \{ \begin{array}{c}
-({\alpha}_1{\alpha}_3){\partial}_1{\lambda}^2+({\alpha}_1)^2{\partial}_3{\lambda}^2 + 2{\alpha}_1({\partial}_3{\alpha}_1 - {\partial}_1{\alpha}_3){\lambda}^2\\
 ...  \\
  -({\alpha}_3)^2{\partial}_1{\lambda}^2+({\alpha}_1{\alpha}_3){\partial}_3{\lambda}^2+2{\alpha}_3({\partial}_3{\alpha}_1-{\partial}_1{\alpha}_3){\lambda}^2
\end{array}  \right.\]

Multiplying each first row on the left by $ - {\alpha}_3$, then each third row on the left by ${\alpha}_1$ in order to eliminate the derivatives of $\lambda$, adding and collecting the results, we discover that ${\lambda}^2$ strikingly disappears and we only obtain for the kernel of the adjoint operator:  \\
\[   {\alpha}_3 [{\alpha}_1({\partial}_2{\alpha}_3 - {\partial}_3{\alpha}_2) + {\alpha}_2({\partial}_3{\alpha}_1 - {\partial}_1{\alpha}_3) +
{\alpha}_3({\partial}_1{\alpha}_2 - {\partial}_2{\alpha}_1) ]{\lambda}^1=0  \]
that is ${\alpha}_3 I(j_1(\alpha)){\lambda}^1$=0 and all the possible permutations. As $\alpha \neq 0$, then one at least of the three components must not vanish and may even be supposed to be equal to $1$ because $\alpha$ is defined uo to a function factor. We get therefore $I(j_1(\alpha))\lambda=0$, that is $c\lambda = 0$ whenever the system is formally integrable. We let the reader treat directly the standard 
case with ${\alpha}_3=0$ ({\it care}):    \\
\[  \alpha=dx^1-x^3dx^2 \Rightarrow {\partial}_3{\xi}^1-x^3{\partial}_3{\xi}^2=0, {\partial}_2{\xi}^1-x^3{\partial}_2{\xi}^2+x^3({\partial}_1{\xi}^1-x^3{\partial}_1{\xi}^2) - {\xi}^3=0 \Rightarrow c=1 \Rightarrow \lambda=0  \]
Though it is rather surprising at first sight, let us now explain why we shall need non trivial homological algebra in order to understand the previous results. Indeed, the last system $R_1$ is neither formally integrable nor involutive, even though it has an involutive symbol $g_1$ defined by:   \\
\[          {\xi}^1_3-x^3{\xi}^2_3=0, \hspace{1cm} {\xi}^1_2-x^3{\xi}^2_2+x^3{\xi}^1_1 -(x^3)^2{\xi}^2_1=0   \]
It is now easy to check that the system $R^{(1)}_1\subset J_1(T)$ defined by the $3$ PD equations: \\
\[  \left\{  \begin{array}{lcl}
{\Phi}^3 \equiv {\xi}^3_3+{\xi}^2_2 +2x^3{\xi}^2_1 - {\xi}^1_1 & = & 0  \\
{\Phi}^2 \equiv {\xi}^1_3 - x^3{\xi}^2_3 & = &  0  \\
{\Phi}^1 \equiv {\xi}^1_2 - x^3{\xi}^2_2+x^3{\xi}^1_1 - (x^3)^2{\xi}^2_1 - {\xi}^3 & = &  0 
\end{array}
\right. \fbox{$\begin{array}{lll}
1 & 2 & 3 \\
1 & 2 & 3  \\
1 & 2 & \bullet 
\end{array}$}  \]
is involutive with $2$ equations of class $3$ and $1$ equation of class $2$. Taking into account the relations ${\Phi}^1={\Omega}_2+x^3{\Omega}_1, {\Phi}^2={\Omega}_3, {\Phi}^3= - 2{\Omega}_1$ and substituting, we obtain the only first order CC:  \\
\[   d_3{\Phi}^1 - d_2 {\Phi}^2 - x^3 d_1{\Phi}^2 + {\Phi}^3=0 \Leftrightarrow  
                (d_2{\Omega}_3 - d_3{\Omega}_2) - x^3 (d_3{\Omega}_1 - d_1{\Omega}_3) + {\Omega}_1=0   \]
and we recognize the linearization of the Vessiot structure equation, {\it following exactly the same procedure as the one used previously for the linearization of the constant Riemannian curvature}. However, in this new framework, we shall now prove and illustrate the following Lemma and striking Theorem (See next Sections for the definitions): \\

\noindent
{\bf LEMMA 1.1}: A (formally) surjective linear differential operator ${\cal{D}}$ defined over a differential field $K$ is defining a projective and thus torsion-free differential module $M$ if and only if its (formal) adjoint is (formally) injective.  \\

 \noindent
 {\it Proof}: In this specific situation, let us consider the finite free presentation over $D=K[d]$:  \\
 \[ 0 \rightarrow F_1 \stackrel{\cal{D}}{\longrightarrow} F_0 \stackrel{p}{\longrightarrow} M \rightarrow 0  \]
 If $M$ is projective, then it is well known that that such a sequence splits (See [2,7,17,27,36,50] for more details or [23], Lemma 3.3, p 212). Then, applying $hom_D(\bullet,D)$ we get again the new splitting sequence:  \\
 \[0 \leftarrow {F_1}^* \stackrel{\,\,\,{\cal{D}}^*}{\longleftarrow} {F_0}^* \stackrel{\,\,\,p^*}{\longleftarrow} M^* \leftarrow 0  \]
 and obtain $ker(ad({\cal{D}}))=0$ in the operator sense or rather $coker({\cal{D}}^*)=0$ in the module sense. \\
Conversely,we have already exhibited the long exact dual sequence:  \\
\[0 \leftarrow N \leftarrow {F_1}^* \stackrel{\,\,\,{\cal{D}}^*}{\longleftarrow} {F_0}^* \stackrel{\,\,\,p^*}{\longleftarrow} 
M^* \leftarrow 0  \] 
Accordingly, if $N=0$, as the dual $F^*$ of a free differential module $F$ is again a free differential module, thus a projective module, this sequence splits and $M^*$ is thus a projective module. \\
Applying again $hom_D(\bullet,D)$, we have the commutative and exact diagram:  \\
\[ \begin{array}{rcccccl}
0 \rightarrow &F_1 &\stackrel{\cal{D}}{\longrightarrow} &F_0 &\stackrel{p}{\longrightarrow} &M &\rightarrow 0  \\
    &  \downarrow  & & \downarrow  &  & \downarrow &    \\
0 \rightarrow & \,\,\,\,\,{F_1}^{**}& \stackrel{\,\,\,\,\,{D}^{**}}{\longrightarrow} &\,\,\,\,\,{F_0}^{**} &\stackrel{\,\,\,\,\,{p}^{**}}{\longrightarrow}& \,\,\,\,\,{M}^{**} &\rightarrow 0  
\end{array}   \]
Using the isomorphism $F\simeq F^{**}$ when $F$ has finite rank over $D$, we obtain an isomorphism $M\simeq M^{**}$. As $M^*$ is projective because $F^*_0\simeq F_1^*\oplus M^*$, then $M^{**}$ is also projective and thus $M$ is projective.  \\ 
\hspace*{12cm}   Q.E.D.  \\ 

\noindent
{\bf EXAMPLE 1.2}: In the preceding contact situation, the system $R_1$ is defined by $2$ equations only while the system $R^{(1)}_1$ is defined by $3$ equations that we have provided. Accordingly, with $K=\mathbb{Q}(x^1,x^2,x^3)$ and $D=K[d]=K[d_1,d_2,d_3]$, we obtain the free presentation $0 \rightarrow D^2 \rightarrow D^3 \rightarrow M \rightarrow 0$ and we have seen that $c\neq 0 \Rightarrow \lambda=0$. The parametrization:  \\
\[ {\xi}^1=\phi - x^3{\partial}_3\phi, \,\,{\xi}^2= - {\partial}_3\phi, \,\, {\xi}^3={\partial}_2\phi + x^3{\partial}_1\phi \,\, \Rightarrow  \,\, \phi=\alpha(\xi)\,\, \Rightarrow \,\,  {\cal{L}}(\xi)\alpha={\partial}_1\phi \,\,  \alpha \]
by means of an arbitrary function $\phi$ is well known and proves that $M\simeq D$. On the contrary, if we choose $\omega=(1,0,0)$, then the system of special Medolaghi equations that we have exhibited shows that ${\xi}^1$ is a torsion element and this system cannot be parametrized.  \\  
 
\noindent 
{\bf THEOREM 1.3}: The possibility to parametrize the system of general Medolaghi equations only depends on the structure constant $c$.  \\

\noindent
{\it Proof}: For any geometric object $\omega$ of order $q$ and the corresponding system $R_q(\omega)$ of general Medolaghi equations, let us now define an equivalence relation ${\bar{\omega}}\sim \omega  \Leftrightarrow R_q(\omega)=R_q(\bar{\omega})$. In the contact situation, we have first to study when we have ${\bar{\omega}}_r{\xi}^r_i-\frac{1}{2}{\bar{\omega}}_i{\xi}^r_r=0\Leftrightarrow {\omega}_r{\xi}^r_i-\frac{1}{2}{\omega}_i{\xi}^r_r$. Though it looks like to be a simple algebraic problem, one needs an explicit computation or computer algebra and we prefer to use another more powerful technique ([39], p 688). Introducing the completely skewsymmetrical symbol $\epsilon=({\epsilon}^{i_1i_2i_3})$ where ${\epsilon}^{i_1i_2i_3}=1$ if $(i_1i_2i_3)$ is an even permutation of $(123)$ or $-1$ if it is an odd permutation and $0$ otherwise, let us introduce the skewsymmetrical $2$-contravariant density ${\omega}^{ij}={\epsilon}^{ijk}{\omega}_k$. Then one can rewrite the system of general infinitesimal Lie equations $R_1(\omega)$ as :  \\
\[ -{\omega}^{rj}(x){\xi}^i_r-{\omega}^{ir}(x){\xi}^j_r-\frac{1}{2}{\omega}^{ij}(x){\xi}^r_r+{\xi}^r{\partial}_r{\omega}^{ij}(x)=0 \]
and we may exhibit a section ${\xi}^i_r={\omega}^{is}A_{rs}$ with $A_{rs}=A_{sr}$ and thus ${\xi}^r_r=0$. It is important to notice that $det(\omega)=0$ when $n=2p+1$, contrary to the Riemann or symplectic case and $\omega$ cannot therefore be used in order to raise or lower indices. As we must have ${\bar{R}}^0_1=R^0_1$ where the isotropy $R^0_1$ is defined by the short exact sequence $0\rightarrow R^0_1 \rightarrow R_1 \stackrel{{\pi}^1_0}{ \rightarrow} T \rightarrow 0$, the same section must satisfy $({\bar{\omega}}^{rj}{\omega}^{is}+{\bar{\omega}}^{ir}{\omega}^{js})A_{rs}=0, \forall A_{rs}=A_{sr}$, and we must have $({\bar{\omega}}^{rj}{\omega}^{is}+{\bar{\omega}}^{ir}{\omega}^{js})+({\bar{\omega}}^{sj}{\omega}^{ir}+{\bar{\omega}}^{is}{\omega}^{jr})=0$. Setting $s=j$, we get ${\bar{\omega}}^{rj}{\omega}^{ij}={\bar{\omega}}^{ij}{\omega}^{rj}\Rightarrow {\bar{\omega}}^{ij}(x)=a(x){\omega}^{ij}(x)$. Substituting and substracting, we get ${\omega}^{ij}(x){\xi}^r{\partial}_ra(x)=0\Rightarrow a(x)=a=cst\neq 0$ because $\omega\neq 0$ and one of the components at least must be nonzero. Accordingly, the {\it normalizer} $N(\Theta)=\{ \xi\in T\mid {\cal{L}}(\xi)\omega=A\omega, Ac=0\}$ and $\Theta$ is of codimension $1$ in its normalizer if $c=0$ or $N(\Theta)=\Theta$ if $c\neq 0$. For example, in the case of a contact structure with $c=1$, we have $N(\Theta)=\Theta$ but, when $\omega=(1,0,0)\Rightarrow c=0$, we have to eliminate the constant $A$ among the equations ${\partial}_3{\xi}^3+{\partial}_2{\xi}^2-{\partial}_1{\xi}^1=-2A, {\partial}_3{\xi}^1=0, {\partial}_2{\xi}^1=0$ and we may add the infinitesimal generator $x^i{\partial}_i$ of a dilatation providing $A=-\frac{1}{2}$. As we have already seen, the parametrization is only existing for $c\neq 0$. This is an "open property " because $ \bar{\omega}=a\omega, a=cst \Rightarrow \bar{c}= a^2 c$ and thus any nonzero value of $c$ can be reached because $a\neq 0$. 
\hspace*{12cm}   Q.E.D.   \\

{\it However, no one of the previous results can be extended to an arbitrary} $n=2p+1\geq 5$.  \\

It is clear from the beginning of this Introduction that an isometry is a solution of a nonlinear system in {\it Lie form} [31,34,44] and that we have linearized this system at the identity transformation in order to study elastic waves. However, in general, no explicit solution may be known but most nonlinear systems of OD or PD equations of mathematical physics (constant riemannian curvature is a good example in [14]) are defined by differential polynomials. This is particularly clear for riemannian, conformal, complex, contact, symplectic or unimodular structures on manifolds [44]. Hence, in Section $2$ we shall provide the main results that exist in the formal theory of systems of nonlinear PD equations in order to construct a {\it formal linearization}. The proof of many results is quite difficult as it involves delicate chases in $3$-dimensional diagrams [31,34,36]. In physics, the linear system obtained may have coefficients in a certain {\it differential field} and we shall need to revisit {\it differential algebra} in Section $3$ because Spencer and Kolchin never clearly understood that their respective works could be combined. It will follow that the linear systems will have coefficients in a differential field $K$ and we shall have to introduce the ring $D=K[d]=K[d_1, ... , d_n]$ of differential operators with coefficients in $K$, which is even an integral domain. This fact will be particularly useful in order to revisit {\it differential duality} in Section $4$ before applying it to the study of {\it conformal structures} in Section $5$, caring {\it separately} about the cases $n=3,n=4$ and $n\geq 5$, then to {\it contact structures} in Section $6$, caring also {\it separately} about the cases $n=3$ and $n\geq 5$, finally concluding in the last Section $7$.\\

\newpage

\noindent
{\bf 2) DIFFERENTIAL GEOMETRY}  \\

If $X$ is a manifold with local coordinates $(x^i)$ for $i=1, ... ,n=dim(X)$, let $\cal{E}$ be a {\it fibered manifold} over $X$ with ${dim}_X({\cal{E}})=m$, that is a manifold with local coordinates $(x^i,y^k)$ for $i=1,...,n$ and $k=1,...,m$ simply denoted by $(x,y)$, {\it projection} $\pi:{\cal{E}}\rightarrow X:(x,y)\rightarrow (x)$ and changes of local coordinates $\bar{x}=\varphi(x), \bar{y}=\psi(x,y)$. If $\cal{E}$ and $\cal{F}$ are two fibered manifolds over $X$ with respective local coordinates $(x,y)$ and $(x,z)$, we denote by ${\cal{E}}{\times}_X{\cal{F}}$ the {\it fibered product} of $\cal{E}$ and $\cal{F}$ over $X$ as the new fibered manifold over $X$ with local coordinates $(x,y,z)$. We denote by $f:X\rightarrow {\cal{E}}: (x)\rightarrow (x,y=f(x))$ a global {\it section} of $\cal{E}$, that is a map such that $\pi\circ f=id_X$ but local sections over an open set $U\subset X$ may also be considered when needed. Under a change of coordinates, a section transforms like $\bar{f}(\varphi(x))=\psi(x,f(x))$ and the derivatives transform like:\\
\[   \frac{\partial{\bar{f}}^l}{\partial{\bar{x}}^r}(\varphi(x)){\partial}_i{\varphi}^r(x)=\frac{\partial{\psi}^l}{\partial x^i}(x,f(x))+\frac{\partial {\psi}^l}{\partial y^k}(x,f(x)){\partial}_if^k(x)  \]
We may introduce new coordinates $(x^i,y^k,y^k_i)$ transforming like:\\
\[ {\bar{y}}^l_r{\partial}_i{\varphi}^r(x)=\frac{\partial{\psi}^l}{\partial x^i}(x,y)+\frac{\partial {\psi}^l}{\partial y^k}(x,y)y^k_i  \]
We shall denote by $J_q({\cal{E}})$ the {\it q-jet bundle} of $\cal{E}$ with local coordinates $(x^i, y^k, y^k_i, y^k_{ij},...)=(x,y_q)$ called {\it jet coordinates} and sections $f_q:(x)\rightarrow (x,f^k(x), f^k_i(x), f^k_{ij}(x), ...)=(x,f_q(x))$ transforming like the sections $j_q(f):(x) \rightarrow (x,f^k(x), {\partial}_if^k(x), {\partial}_{ij}f^k(x), ...)=(x,j_q(f)(x))$ where both $f_q$ and $j_q(f)$ are over the section $f$ of $\cal{E}$. It will be useful to introduce a {\it multi-index} $\mu=({\mu}_1, ... ,{\mu}_n)$ with length $\mid \mu \mid={\mu}_1+ ... +{\mu}_n$ and to set ${\mu}+1_i=({\mu}_1  ... ,{\mu}_{i-1}, {\mu}_i+1, {\mu}_{i+1},...,{\mu}_n)$. Finally, a jet coordinate $y^k_{\mu}$ is said to be of {\it class} $i$ if ${\mu}_1=...={\mu}_{i-1}=0, {\mu}_i\neq 0$. As the background will always be clear enough, we shall use the same notation for a vector bundle or a fibered manifold and their sets of sections [31,36]. We finally notice 
that $J_q({\cal{E}})$ is a fibered manifold over $X$ with projection ${\pi}_q$ while $J_{q+r}({\cal{E}})$ is a fibered manifold over $J_q({\cal{E}})$ with projection ${\pi}^{q+r}_q, \forall r\geq 0$ [, , ].\\

\noindent
{\bf DEFINITION 2.1}: A ({\it nonlinear}) {\it system} of order $q$ on $\cal{E}$ is a fibered submanifold ${\cal{R}}_q\subset J_q({\cal{E}})$ and a global or local {\it solution} of ${\cal{R}}_q$ is a section $f$ of $\cal{E}$ over $X$ or $U\subset X$ such that $j_q(f)$ is a section of ${\cal{R}}_q$ over $X$ or $U\subset X$.\\

\noindent
{\bf DEFINITION 2.2}: When the changes of coordinates have the linear form $\bar{x}=\varphi(x),\bar{y}= A(x)y$, we say that $\cal{E}$ is a {\it vector bundle} over $X$. Vector bundles will be denoted by capital letters $C,E,F$ and will have sections denoted by $\xi,\eta,\zeta$. In particular, we shall denote as usual by $T=T(X)$ the {\it tangent bundle} of $X$, by $T^*=T^*(X)$ the {\it cotangent bundle}, by ${\wedge}^rT^*$ the {\it bundle of r-forms} and by $S_qT^*$ the {\it bundle of q-symmetric covariant tensors}. When the changes of coordinates have the form $\bar{x}=\varphi(x),\bar{y}=A(x)y+B(x)$ we say that $\cal{E}$ is an {\it affine bundle} over $X$ and we define the {\it associated vector bundle} $E$ over $X$ by the local coordinates $(x,v)$ changing like $\bar{x}=\varphi(x),\bar{v}=A(x)v$. \\

\noindent
{\bf DEFINITION 2.3}: If the tangent bundle $T({\cal{E}})$ has local coordinates $(x,y,u,v)$ changing like ${\bar{u}}^j={\partial}_i{\varphi}^j(x)u^i, {\bar{v}}^l=\frac{\partial {\psi}^l}{\partial x^i}(x,y)u^i+\frac{\partial {\psi}^l}{\partial y^k}(x,y)v^k$, we may introduce the {\it vertical bundle} $V({\cal{E}})\subset T({\cal{E}})$ as a vector bundle over $\cal{E}$ with local coordinates $(x,y,v)$ obtained by setting $u=0$ and changes ${\bar{v}}^l=\frac{\partial {\psi}^l}{\partial y^k}(x,y)v^k$. Of course, when $\cal{E}$ is an affine bundle over $X$ with associated vector bundle $E$ over $X$, we have $V({\cal{E}})={\cal{E}}\times_XE$. With a slight abuse of language, we shall set $E=V(\cal{E})$ as a vector bundle over $\cal{E}$.\\

For a later use, if $\cal{E}$ is a fibered manifold over $X$ and $f$ is a section of $\cal{E}$, we denote by $f^{-1}(V({\cal{E}}))$ the {\it reciprocal image} of $V({\cal{E}})$ by $f$ as the vector bundle over $X$ obtained when replacing $(x,y,v)$ by $(x,f(x),v) $ in each chart. A similar construction may also be done for any affine bundle over ${\cal{E}}$. Loking at the transition rules of $J_q(\cal{E})$, we deduce easily the following results: \\

\noindent
{\bf PROPOSITION 2.4}: $J_q(\cal{E})$ is an affine bundle over $J_{q-1}(\cal{E})$ modeled on $S_qT^*{\otimes}_{\cal{E}}E$ but we shall not specify the tensor product in general.  \\

\noindent
{\bf PROPOSITION 2.5}: There is a canonical isomorphism $V(J_q({\cal{E}})) \simeq J_q(V({\cal{E}}))=J_q(E)$ of vector bundles over $J_q(\cal{E})$ given by setting $v^k_{\mu}=v^k_{,\mu}$ at any order and a short exact sequence: \\
\[  0 \rightarrow S_qT^*\otimes E \rightarrow J_q(E) \stackrel{{\pi}^q_{q-1}}{\longrightarrow} J_{q-1}(E) \rightarrow 0  \]
 of vector bundles over $J_q(\cal{E})$ allowing to establish a link with the formal theory of linear systems.  \\

\noindent
{\bf PROPOSITION 2.6}: There is an exact sequence: \\
\[  0 \rightarrow {\cal{E}} \stackrel{j_{q+1}}{\longrightarrow} J_{q+1}({\cal{E}}) \stackrel{D}{\longrightarrow} T^*\otimes J_q(E)  \]
where $Df_{q+1}=j_1(f_q) - f_{q+1}$ is over $f_q$ with components $(Df_{q+1})^k_{\mu,i}={\partial}_if^k_{\mu}-f^k_{\mu + 1_i}$ is called the (nonlinear) {\it Spencer operator}. As $J_{q+1}({\cal{E}}) \subset J_1(J_q({\cal{E}}))$, there is an induced exact sequence:  \\
\[  0 \rightarrow {\cal{E}} \stackrel{j_q}{\longrightarrow} J_{q+1}({\cal{E}}) \stackrel{D_1}{\longrightarrow} T^*\otimes J_q(E)/S_{q+1}T^*\otimes E  \]
where $D_1$ is called the {\it first Spencer operator}.  \\

\noindent
{\bf DEFINITION 2.7}: If ${\cal{R}}_q\subset J_q({\cal{E}})$ is a system of order $q$ on ${\cal{E}}$, then ${\cal{R}}_{q+1}={\rho}_1({\cal{R}}_q)=J_1({\cal{R}}_q)\cap J_{q+1}({\cal{E}})\subset J_1(J_q({\cal{E}}))$ is called the {\it first prolongation} of 
${\cal{R}}_q$ and we may define the subsets ${\cal{R}}_{q+r}$. In actual practice, if the system is defined by PDE ${\Phi}^{\tau}(x,y_q)=0$ the first prolongation is defined by adding the PDE $d_i{\Phi}^{\tau}\equiv {\partial}_i{\Phi}^{\tau} + y^k_{\mu +1_i}{\partial}{\Phi}^{\tau}/\partial y^k_{\mu}=0$. accordingly, $f_q\in {\cal{R}}_q \Leftrightarrow {\Phi}^{\tau}(x,f_q(x))=0$ and $f_{q+1}\in {\cal{R}}_{q+1} \Leftrightarrow {\partial}_i{\Phi}^{\tau}+f^k_{\mu +1_i}(x) \partial {\Phi}^{\tau}/\partial y^k_{\mu}=0$ as identities on $X$ or at least over an open subset $U\subset X$. Differentiating the first relation with respect to $x^i$ and substracting the second, we finally obtain:  \\
\[  ({\partial}_if^k_{\mu}(x) - f^k_{\mu +1_i}(x))\partial {\Phi}^{\tau}/\partial y^k_{\mu}=0 \Rightarrow Df_{q+1}\in T^*\otimes R_q \]
and the Spencer operator restricts to $D:{\cal{R}}_{q+1} \rightarrow T^*\otimes R_q$. We set ${\cal{R}}^{(1)}_{q+r}={\pi}^{q+r+1}_{q+r}( {\cal{R}}_{q+r+1})$.  \\

\noindent
{\bf DEFINITION 2.8}: The {\it symbol} of ${\cal{R}}_q$ is the family $g_q=R_q \cap S_qT^*\otimes E$ of vector spaces over ${\cal{R}}_q$. The symbol $g_{q+r}$ of ${\cal{R}}_{q+r}$ only depends on $g_q$ by a direct prolongation procedure. We may define the vector bundle $F_0$ over ${\cal{R}}_q$ by the short exact sequence $0 \rightarrow R_q \rightarrow J_q(E) \rightarrow F_0 \rightarrow 0$ and we have the exact induced sequence $0 \rightarrow g_q \rightarrow S_ qT^*\otimes E \rightarrow F_0$ .\\

Setting $a^{\tau \mu}_k(x,y_q)=\partial {\Phi}^{\tau}/\partial y^k_{\mu}(x,y_q) $ whenever $\mid \mu \mid =q$ and $(x,y_q)\in {\cal{R}}_q$, we obtain:  \\
 \[  g_q=\{ v^k_{\mu}\in S_qT^*\otimes E \mid a^{\tau \mu}_k(x,y_q)v^k_{\mu}=0\} , \mid \mu \mid =q, (x,y_q)\in {\cal{R}}_q    \]
  \[  \Rightarrow
 g_{q+r}={\rho}_r(g_q)=\{ v^k_{\mu + \nu}\in S_{q+r} T^*\otimes E \mid a^{\tau \mu}_k(x,y_q)v^k_{\mu + \nu}=0\}, \mid \mu\mid=q, \mid \nu \mid =r, (x,y_q)\in {\cal{R}}_q  \]
In general, neither $g_q$ nor $g_{q+r}$ are vector bundles over ${\cal{R}}_q$.  \\

On ${\wedge}^sT^*$ we may introduce the usual bases $\{dx^I=dx^{i_1}\wedge ... \wedge dx^{i_s}\}$ where we have set 
$I=(i_1< ... <i_s)$. In a purely algebraic setting, one has:  \\

\noindent
{\bf PROPOSITION 2.9}: There exists a map $\delta:{\wedge}^sT^*\otimes S_{q+1}T^*\otimes E\rightarrow {\wedge}^{s+1}T^*\otimes S_qT^*\otimes E$ which restricts to $\delta:{\wedge}^sT^*\otimes g_{q+1}\rightarrow {\wedge}^{s+1}T^*\otimes g_q$ and ${\delta}^2=\delta\circ\delta=0$.\\

{\it Proof}: Let us introduce the family of s-forms $\omega=\{ {\omega}^k_{\mu}=v^k_{\mu,I}dx^I \}$ and set $(\delta\omega)^k_{\mu}=dx^i\wedge{\omega}^k_{\mu+1_i}$. We obtain at once $({\delta}^2\omega)^k_{\mu}=dx^i\wedge dx^j\wedge{\omega}^k_{\mu+1_i+1_j}=0$ and $ a^{\tau \mu}_k(\delta \omega)^k_{\mu}=dx^i \wedge(  a^{\tau \mu}_k{\omega}^k_{\mu +1_i})=0$.  \\
\hspace*{12cm} Q.E.D.  \\

The kernel of each $\delta$ in the first case is equal to the image of the preceding $\delta$ but this may no longer be true in the restricted case and we set:\\

\noindent
{\bf DEFINITION 2.10}: Let $B^s_{q+r}(g_q)\subseteq Z^s_{q+r}(g_q)$ and $H^s_{q+r}(g_q)=Z^s_{q+r}(g_q)/B^s_{q+r}(g_q)$ with $H^1(g_q)=H^1_q(g_q)$ be the coboundary space $im(\delta)$, cocycle space $ker(\delta)$ and cohomology space at ${\wedge}^sT^*\otimes g_{q+r}$ of the restricted $\delta$-sequence which only depend on $g_q$ and may not be vector bundles. The symbol $g_q$ is said to be s-{\it acyclic} if $H^1_{q+r}=...=H^s_{q+r}=0, \forall r\geq 0$, {\it involutive} if it is n-acyclic and {\it finite type} if $g_{q+r}=0$ becomes trivially involutive for r large enough. In particular, if $g_q$ is involutive {\it and} finite type, then $g_q=0$. Finally, $S_qT^*\otimes E$ is involutive for any $ q\geq 0$ if we set $S_0T^*\otimes E=E$. \\

Having in mind the example of $xy_x-y=0\Rightarrow xy_{xx}=0$ with rank changing at $x=0$, we have:  \\

\noindent
{\bf PROPOSITION 2.11}: If $g_q$ is $2$-acyclic and $g_{q+1}$ is a vector bundle over ${\cal{R}}_q $, then $g_{q+r}$ is a vector bundle over ${\cal{R}}_q, \forall r\geq1$. \\

\noindent
{\it Proof}: We may define the vector bundle $F_1$ over ${\cal{R}}_q$ by the following ker/coker exact sequence where we denote by $h_1\subseteq T^*\otimes F_0$ the image of the central map:  \\
\[   0 \rightarrow g_{q+1} \rightarrow S_{q+1}T^*\otimes E \rightarrow T^*\otimes F_0 \rightarrow F_1 \rightarrow 0  \]
and we obtain by induction on $r$ the following commutative and exact diagram of vector bundles over ${\cal{R}}_q$:  \\
{\footnotesize \[  \begin{array}{lccccccc}
 & 0\hspace{3mm}& & 0 \hspace{3mm} && 0\hspace{3mm}& & 0 \hspace{3mm}  \\
 & \downarrow \hspace{3mm}& & \downarrow \hspace{3mm}& & \downarrow \hspace{3mm} & & \downarrow  \hspace{3mm}\\
0 \rightarrow & g_{q+r+1} & \rightarrow & S_{q+r+1}T^*\otimes E & \rightarrow  & S_{r+1}T^* \otimes F_0 & \rightarrow  & S_rT^*\otimes F_1  \\
  &  \downarrow \delta  & & \downarrow \delta & & \downarrow  \delta & & \downarrow \delta  \\
0 \rightarrow & T^*\otimes g_{q+r} & \rightarrow & T^* \otimes S_{q+r}T^* \otimes E & \rightarrow & T^*\otimes S_rT^*\otimes F_0 &\rightarrow & T^*\otimes S_{r-1}T^*\otimes F_1  \\
   &  \downarrow \delta  & & \downarrow \delta & & \downarrow  \delta & & \\
0 \rightarrow & {\wedge}^2T^*\otimes g_{q+r-1} & \rightarrow &{\wedge}^2T^*\otimes S_{q+r-1}T^* \otimes E & \rightarrow & {\wedge}^2T^*\otimes S_{r-1}T^*\otimes F_0 &  &   \\
  & \downarrow \delta &  & \downarrow \delta & & & &   \\
 & {\wedge}^3T^*\otimes S_{q+r-2} T^* \otimes E & = &   {\wedge}^3T^*\otimes S_{q+r-2} T^* \otimes E & & & &\\
  & & & & & & &
\end{array}   \] }
\noindent
where all the maps have been given after Definition 2.9. The image of the central map of the top row is $h_{r+1}={\rho}_r(h_1)$ and a chase proves that $h_1$ is $(s-1)$-acyclic whenever $g_q$ is $s$-acyclic by extending the diagram. The proposition finally follows by upper-semicontinuity from the relation:   \\
\[ dim(g_{q+r+1})+dim(h_{r+1})=m \, \, dim(S_{q+r+1}T^*)   \]
\hspace*{12cm}  Q.E.D.   \\   \\

\noindent
{\bf LEMMA 2.12}: If $g_q$ is involutive and $g_{q+1}$ is a vector bundle over ${\cal{R}}_q $, then $ g_q$ is also a vector bundle over 
${\cal{R}}_q$. In this case, changing linearly the local coordinates if necessary, we may look at the maximum number $\beta$ of equations that can be solved with respect to $v^k_{n...n}$ and the intrinsic number $\alpha=m-\beta$ indicates the number of $y$ that can be given arbitrarily.  \\

Using the exactness of the top row in the preceding diagram and a delicate $3$-dimensional chase, we have (See [31] and [36],p336 for the details): \\

\noindent
{\bf THEOREM 2.13}: If ${\cal{R}}_q\subset J_q({\cal{E}})$ is a system of order $q$ on ${\cal{E}}$ such that $g_{q+1}$ is a vector bundle over ${\cal{R}}_q$ and $g_q$ is $2$-acyclic, then there is an exact sequence:  \\
\[  {\cal{R}}_{q+r+1} \stackrel{{\pi}^{q+r+1}_{q+r}}{\longrightarrow} {\cal{R}}_{q+r}  \stackrel{{\kappa}_r}{\longrightarrow} S_rT^*\otimes F_1   \]
where ${\kappa}_r$ is called the $r$-{\it curvature} and $\kappa={\kappa}_0$ is simply called the {\it curvature} of ${\cal{R}}_q$.  \\

We notice that ${\cal{R}}_{q+r+1}={\rho}_r ({\cal{R}}_{q+1})$ and ${\cal{R}}_{q+r}=  {\rho}_r({\cal{R}}_q)$ in the following commutative diagram:  \\
\[  \begin{array}{ccc}
             {\cal{R}}_{q+r+1}  &  \stackrel{{\pi}^{q+r+1}_{q+1}}{\longrightarrow }& {\cal{R}}_{q+1}   \\
          \hspace{10mm}   \downarrow   {\pi}^{q+r+1}_{q+r}  &         &  \hspace{7mm}\downarrow {\pi}^{q+1}_q \\
             {\cal{R}}^{(1)}_{q+r}    &  \stackrel{{\pi}^{q+r}_q }{\longrightarrow} & {\cal{R}}^{(1)}_q    \\
               \cap   &   &   \cap   \\
               {\cal{R}}_{q+r} &  \stackrel{{\pi}^{q+r}_q}{\longrightarrow}& {\cal{R}}_q
               \end{array}  \]
We also have ${\cal{R}}^{(1)}_{q+r} \subseteq {\rho}_r({\cal{R}}^{(1)}_q)$ because we have successively:  \\
\[  \begin{array}{ccl}
{\cal{R}}^{(1)}_{q+r}={\pi}^{q+r+1}_{q+r}({\cal{R}}_{q+r+1}) & = & {\pi}^{q+r+1}_{q+r}(J_r({\cal{R}}_{q+1})\cap 
J_{q+r+1}(\cal{E}))  \\
       & \subseteq  & J_r({\pi}^{q+1}_q) (J_r({\cal{R}}_{q+1})) \cap J_{q+r}(\cal{E})  \\
          & =  &  J_r({\cal{R}}^{(1)}_q) \cap J_{q+r}(\cal{E})  \\
              & = &  {\rho}_r({\cal{R}}^{(1)}_q)
              \end{array}   \]
while chasing in the following commutative $3$-dimensional diagram:  \\
\[ÊÊ\begin{array}{rcccc}
  &  & J_r({\cal{R}}_{q+1}) & \longrightarrow & J_r(J_{q+1}({\cal{E}}))   \\
   & \nearrow  & \downarrow &  &  \nearrow \hspace{10mm} \\
   {\cal{R}}_{q+r+1}  &   & \longrightarrow & J_{q+r+1}({\cal{E}}) &  \downarrow  \\
   \downarrow \hspace{5mm} &   &  J_r({\cal{R}}_q) & \longrightarrow  &  J_r(J_q({\cal{E}}))  \\
    & \nearrow &  & \downarrow & \nearrow  \hspace{10mm} \\
   {\cal{R}}_{q+r} & & \longrightarrow & J_{q+r}({\cal{E}})
   \end{array}  \]             
with a well defined map $J_r({\pi}^{q+1}_q):J_r(J_{q+1}({\cal{E}})) \rightarrow J_r(J_q({\cal{E}}))$. We finally obtain the following crucial Theorem and its Corollary (Compare to [31], p 72-74 or [36], p 340 to [16]):  \\

\noindent
{\bf THEOREM 2.14}: Let ${\cal{R}}_q\subset J_q({\cal{E}})$ be a system of order $q$ on ${\cal{E}}$ such that ${\cal{R}}_{q+1}$ is a fibered submanifold of $J_{q+1}({\cal{E}})$. If $g_q$ is $2$-acyclic and $g_{q+1}$ is a vector bundle over ${\cal{R}}_q$, then we have ${\cal{R}}^{(1)}_{q+r}={\rho}_r ({\cal{R}}^{(1)}_q)$ for all $r\geq 0$.  \\

\noindent
{\bf DEFINITION 2.15}: A system ${\cal{R}}_q\subset J_q({\cal{E}})$ is said to be {\it formally integrable} if ${\pi}^{q+r+1}_{q+r}: {\cal{R}}_{q+r+1}\rightarrow {\cal{R}}_{q+r}$ is an epimorphism of fibered manifolds for all $r\geq 1$ and {\it involutive} if it is formally integrable with an involutive symbol $g_q$. We have the following useful test [16,31,52]: \\ 

\noindent
{\bf COROLLARY 2.16}: Let ${\cal{R}}_q\subset J_q({\cal{E}})$ be a system of order $q$ on ${\cal{E}}$ such that ${\cal{R}}_{q+1}$ is a fibered submanifold of $J_{q+1}({\cal{E}})$. If $g_q$ is $2$-acyclic (involutive) and if the map ${\pi}^{q+1}_q:{\cal{R}}_{q+1} \rightarrow {\cal{R}}_q$ is an epimorphism of fibered manifolds, then ${\cal{R}}_q$ is formally integrable (involutive).  \\

This is all what is needed in order to study systems of algebraic ordinary differential (OD) or partial differential (PD) equations.  \\

\newpage

\noindent
{\bf 3) DIFFERENTIAL ALGEBRA}  \\

We now present in an independent manner two OD examples and two PD examples, among the best ones we know, showing the difficulties met when studying differential ideals and ask the reader to revisit them later on while reading the main Theorems. As only a few results will be proved, the interested reader may look at [32,34,36] for more details and compare to [20,22,49]. \\

\noindent
{\bf EXAMPLE 3.1}: If $k=\mathbb{Q}$, $y$ is a differential indeterminate and $d_x$ is a formal derivation, we may set $d_xy=y_x, d_xy_x=y_{xx}$ and so on in order to introduce the differential ring $A=k[y,y_x, y_{xx}, ...]=k\{y\}$. We consider the (proper) differential ideal $\mathfrak{a}\subset A$ generated by the differential polynomial $P=y^2_x-4y$. We have $d_xP=2y_x(y_{xx}-2)$ and $\mathfrak{a}$ cannot 
be a prime differential ideal. Hence, looking for the "{\it solutions} " of $P=0$, we must have either $y_x=0 \Rightarrow  y=0$ or $y_{xx}=0$ and thus $y=(x+c)^2$ where $c$ should be a "{\it constant} " with no clear meaning. However, we have successively:  \\
\[    \begin{array}{ccl}
P\in \mathfrak{a} & \Rightarrow & y_x(y_{xx}-2)\in \mathfrak{a} \\
  &   \Rightarrow & y_xy_{xxx}+y_{xx}(y_{xx}-2) \in \mathfrak{a}  \\
    & \Rightarrow  & (y_x)^2y_{xxx}  \in \mathfrak{a}  \\
     & \Rightarrow & yy_{xxx} \in \mathfrak{a}  \\
     & \Rightarrow & yy_{xxxx}+y_xy_{xxx} \in \mathfrak{a}  \\
     &  \Rightarrow & y_x(y_{xxx})^2 \in \mathfrak{a}\\
     & \Rightarrow  & 2y_x y_{xxx}y_{xxxx}+y_{xx}(y_{xxx})^2 \in \mathfrak{a}\Rightarrow 2y_xy_{xxx}y_{xxxx}= - y_{xx}(y_{xxx})^2 \,\,mod(\mathfrak{a})\\
     & \Rightarrow  & 4y_{xx}y_{xxx}y_{xxxx}+2y_x(y_{xxxx})^2+ 2y_xy_{xxx}y_{xxxxx}+(y_{xxx})^3 \in \mathfrak{a}  \\
       & \Rightarrow &  4y_{xx}(y_{xxx})^2y_{xxxx}+2y_xy_{xxx}(y_{xxxx})^2+ 2y_x(y_{xxx})^2y_{xxxxx}+(y_{xxx})^4 \in \mathfrak{a}    \\
       & \Rightarrow & 4y_{xx}(y_{xxx})^2y_{xxxx}+2y_xy_{xxx}(y_{xxxx})^2+(y_{xxx})^4 \in \mathfrak{a}   \\
       &    \Rightarrow & 3y_{xx}(y_{xxx})^2y_{xxxx}+(y_{xxx})^4 \in \mathfrak{a}\\
       & \Rightarrow  &  -6y_xy_{xxx}(y_{xxxx})^2 = (y_{xxx})^4 \in \mathfrak{a}  \\
      &  \Rightarrow & (y_{xxx})^5 \in \mathfrak{a}\Rightarrow y_{xxx}\in rad(\mathfrak{a})
\end{array}  \]
and thus $\mathfrak{a}$ is neither prime nor perfect, that is equal to its radical, but $rad(\mathfrak{a})$ is perfect as it is the intersection of the prime differential ideal generated by $y$ with the prime differential ideal generated by $y_x^2-4y$ and $y_{xx}-2$, both containing $y_{xxx}$.  \\

\noindent
{\bf EXAMPLE 3.2}: With the notations of the previous Example, let us consider the (proper) differential ideal $\mathfrak{a}\subset A$ generated by the differential polynomial $P=y^2_x-4y^3$. We have $d_xP=2y_x(y_{xx}-6y^2)$ and $\mathfrak{a}$ cannot be  prime differential ideal. Hence, looking for the "{\it solutions} " of $P=0$, we must have either $y_x=0 \Rightarrow  y=0$ or $y_x^2-4y^3=0$ and $y_{xx}-6y^2=0$. However, we have successively:  \\
\[    \begin{array}{ccl}
P\in \mathfrak{a} & \Rightarrow & y_x(y_{xx}-6y^2)\in \mathfrak(a) \Rightarrow (y_x)^2(y_{xx}-6y^2)^2\in \mathfrak{a} \Rightarrow 4y^3(y_{xx}-6y^2)^2\in \mathfrak{a}\\
 & \Rightarrow & y_{xx}(y_{xx}-6y^2) + y_x(y_{xxx}-12yy_x)\in \mathfrak{a}  \\
  & \Rightarrow &  y_{xx}(y_{xx}- 6y^2)^2\in \mathfrak{a}  \\
   & \Rightarrow & (y_{xx})^2(y_{xx}-6y^2)^2-12y^2y_{xx}(y_{xx}-6y^2)^2 +36 y^4(y_{xx}-6y^2)^2 \in \mathfrak{a}  \\
   & \Rightarrow   &(y_{xx}-6y^2)^4 \in \mathfrak{a}  \Rightarrow   y_{xx}-6y^2 \in rad(\mathfrak{a})
\end{array}  \]
and thus $\mathfrak{a}$ is neither prime or perfect as before but $rad(\mathfrak{a})$ is the prime differential ideal generated by $y_x^2-4y^3$ and $y_{xx}-6y^2$.  \\

\noindent
{\bf EXAMPLE 3.3}: If $k=\mathbb{Q}$ as before, $y$ is a differential indeterminate and $(d_1,d_2)$ are two formal derivations, let us consider the differential ideal generated by $P_1=y_{22}-\frac{1}{2}(y_{11})^2$ and $P_2=y_{12}-y_{11}$ in $k\{y\}$. Using crossed derivatives, we get successively:  \\
\[  \begin{array}{ccl}
P_1,P_2 \in \mathfrak{a} &  \Rightarrow & y_{112}-y_{111}\in \mathfrak{a}, y_{122}-y_{11}y_{111}\in \mathfrak{a}, y_{222}-y_{11}y_{111}\in \mathfrak{a}  \\
      & \Rightarrow  & Q=d_2P_2-d_1P_1+d_1P_2=(y_{11}-1)y_{111} \in \mathfrak{a}  \\
      & \Rightarrow  &   d_1Q=(y_{111})^2+(y_{11}-1)y_{1111} \in \mathfrak{a} \\
      & \Rightarrow  &  ((y_{111})^3 \in \mathfrak{a} \Rightarrow y_{111} \in rad(\mathfrak{a})
\end{array}  \]
and thus $\mathfrak{a}$ is neither prime nor perfect but $rad(\mathfrak{a})$ is a perfect differential ideal and even a prime differential ideal $\mathfrak{p}$ because we obtain easily from the last section that the resisual differential ring $k\{y\}/\mathfrak{p}\simeq k[y,y_1,y_2,y_{11}]$ is a differential integral domain. Its quotient field is thus the differential field $K=Q(k\{y\}/\mathfrak{p})\simeq k(y,y_1,y_2,y_{11})$ with the rules:  \\
\[   d_1y=y_1,d_1y_1=y_{11}, d_1y_{11}=0, d_2y=y_2, d_2y_1=y_{11}, d_2y_{11}=0\] 
as a way to {\it avoid looking for solutions}. The formal linearization is the linear system $R_2\subset J_2(E)$ obtained in the last section where it was defined over ${\cal{R}}_2$, but {\it not} over $K$, by the two linear second order PDE:  \\
\[             Y_{22}-y_{11}Y_{11}=0, \hspace{1cm} Y_{12}-Y_{11}=0       \] 
changing slightly the notations for using the letter $v$ only when looking at the symbols. It is at this point that {\it the problem starts} because 
${\cal{R}}_2$ is indeed a fibered manifold with arbitrary parametric jets $(y,y_1,y_2,y_{11})$ but ${\cal{R}}_3={\rho}_1({\cal{R}}_2)$ is no longer a fibered manifold because the dimension of its symbol changes when $y_{11}=1$. We understand therefore that {\it there should be a close link existing between formal integrability and the search for prime differential ideals or differential fields}. The solution of this problem has been provided as early as in 1983 for studying the "Differential Galois Theory " but has never been acknowledged and is thus not known today ([32,34]). The idea is to add the third order PDE $y_{111}=0$ and thus the linearized PDE $Y_{111}=0$ obtaining therefore a third order involutive system well defined over $K$ with symbol $g_3=0$. We invite the reader to treat similarly the two previous examples and to compare. \\ 

\noindent
{\bf EXAMPLE 3.4}: If $k=\mathbb{Q}$ as before, $y$ is a differential indeterminate and $(d_1,d_2)$ are two formal derivations, let us consider the differential ideal generated by $P_1=y_{22}-\frac{1}{3}(y_{11})^3$ and $P_2=y_{12}-\frac{1}{2}(y_{11})^2$ in $k\{y\}$. Using crossed derivatives, we get successively:  \\
\[ P_1,P_2 \in \mathfrak{a} \Rightarrow d_2P_2 - d_1P_1 + y_{11}d_1P_2=0 \Rightarrow {\cal{R}}_2  \hspace{3mm}involutive \]
\[  \Rightarrow  y_{222}-(y_{11})^3y_{111}=0, y_{122}-(y_{11})^2y_{111}=0, y_{112}-y_{11}y_{111}=0, ... \]
and thus $dim(g_{q})=1, \forall q\geq 1$. As the symbol $g_2$ is involutive, there is an infinite number of parametric jets $(y,y_1,y_2, y_{11}, y_{111}, ...)$ and thus $k\{y\}/\mathfrak{a}\simeq k[y,y_1,y_2, y_{11}, y_{111}, ...]$ is a differential integral domain with $d_2y_2=y_{22}=\frac{1}{3}(y_{11})^3, d_2y_{11}=y_{112}=  y_{11}y_{111}, ...$. It follows that $\mathfrak{a}=\mathfrak{p}$ is a prime differential ideal with $rad(\mathfrak{p})=\mathfrak{p}$. The second order linearized system is:  \\ 
\[  Y_{22} - (y_{11})^2Y_{11}=0, \hspace{1cm}  Y_{12} - y_{11}Y_{11}=0   \]                                                                                                                                                                                                                                                                                                                                                                  
is now well defined over the differential field $K=Q(k\{y\}/\mathfrak{p})$ and is involutive.  \\

\noindent
{\bf DEFINITION 3.5}: A {\it differential ring} is a ring $A$ with a finite number of commuting derivations $({\partial}_1, ..., {\partial}_n)$ such that ${\partial }_i(a+b)={\partial}_ia +{\partial}_ib, {\partial}_i(ab)=({\partial}_ia)b+a{\partial}_ib$ that can be extended to derivations of the ring of quotients $Q(A)$ by setting ${\partial}_i(a/s)=(s{\partial}_ia-a{\partial}_is)/s^2, \forall 0\neq  s,a\in A$. We shall suppose from now on that $A$ is even an integral domain and use the differential field $K=Q(A)$. For example, if $x^1, ... , x^n$ are indeterminates over $\mathbb{Q}$, then $\mathbb{Q}[x]=\mathbb{Q}[x^1, ...,x^n]$  is a differential ring for the standard 
$({\partial}_1, ..., {\partial}_n)$ with quotient field $\mathbb{Q}(x)$.  \\

If $K$ is a differential field as above and $(y^1,...,y^m)$ are indeterminates over $K$, we transform the polynomial ring $K\{y\}={lim}_{q\rightarrow \infty}K[y_q]$ into a differential ring by introducing as usual the {\it formal derivations} $d_i={\partial}_i+y^k_{\mu+1_i}\partial/\partial y^k_{\mu}$ and we shall set $K<y>= Q(K\{y\})$.  \\

\noindent
{\bf DEFINITION 3.6}: We say that $\mathfrak{a}\subset K\{y\}$ is a {\it differential ideal} if it is stable by the $d_i$, that is if  $d_ia\in\mathfrak{a}, \forall a \in \mathfrak{a}, \forall i=1,...,n$. We shall also introduce the {\it radical} $rad(\mathfrak{a})=\{a\in A\mid \exists r,a^r\in \mathfrak{a}\}\supseteq \mathfrak{a}$ and say that $\mathfrak{a}$ is a {\it perfect} (or {\it radical}) differential ideal if $rad(\mathfrak{a})=\mathfrak{a}$. If $S$ is any subset of $A$, we shall denote by $\{S\}$ the differential ideal generated by $S$ and introduce the (non-differential) ideal ${\rho}_r(S)=\{d_{\nu}a \mid a\in S, 0 \leq  \mid\nu\mid \leq r\}$ in $A$.   \\
\newpage
\noindent
{\bf LEMMA 3.7}: If $\mathfrak{a}\subset A$ is differential ideal, then $rad(\mathfrak{a}) $ is a differential ideal containing 
$\mathfrak{a}$.  \\

\noindent
{\it Proof}: If $d$ is one of the derivations, we have $ a^{r-1}da=\frac{1}{r}da^r \in \{a^r\}$ and thus:  \\
\[ (r-1) a^{r-2}(da)^2 + a^{r-1}d^2a \in \{a^r\}\Rightarrow a^{r-2}(da)^3\in \{a^r\},... \Rightarrow (da)^{2r-1} \in \{a^r\}   \]
\hspace*{12cm}   Q.E.D.   \\

\noindent
{\bf LEMMA 3.8}: If $\mathfrak{a} \subset K\{y\}$, we set ${\mathfrak{a}}_q= \mathfrak{a}\cap K[y_q]$ with ${\mathfrak{a}}_0=\mathfrak{a}\cap K[y]$ and  ${\mathfrak{a}}_{\infty}= \mathfrak{a}$. We have in general ${\rho}_r({\mathfrak{a}}_q) \subseteq {\mathfrak{a}}_{q+r}$ and the problem will be to know when we may have equality.  \\ 

We shall say that a differential extension $L=Q(K\{y\}/\mathfrak{p})$ is a {\it finitely generated} differential extension of $K$ and we may define the {\it evaluation epimorphism} $K\{y\} \rightarrow K\{\eta \}\subset L$ with kernel $\mathfrak{p}$ where $\eta$ or $\bar{y}$ is the residual image of $y$ modulo $\mathfrak{p}$. If we study such a differential extension $L/K$, by analogy with Section 2, we shall say that $R_q$ or $g_q$ is a vector bundle over ${\cal{R}}_q$ if one can find a certain number of maximum rank determinant $D_{\alpha}$ that cannot be all zero at a generic solution of ${\mathfrak{p}}_q$ defined by differential polynomials $P_{\tau}$, that is to say, according to the Hilbert Theorem of Zeros, we may find polynomials $A_{\alpha}, B_{\tau}\in K\{y_q\}$ such that :\\
\[   {\sum}_{\alpha} A_{\alpha}D_{\alpha} + {\sum}_{\tau} B_{\tau}P_{\tau}=1   \]
In particular the following Lemma will be used in the next important Theorem:  \\

\noindent
{\bf LEMMA 3.9}: If $\mathfrak{p}$ is a prime differential ideal of $K\{y\}$, then, for $q$ sufficiently large, there is a polynomial $D\in K[y_q]$ such that $D\notin {\mathfrak{p}}_q$ and :   \\
\[          D{\mathfrak{p}}_{q+r} \subset rad ({\rho}_r({\mathfrak{p}}_q)) \subset {\mathfrak{p}}_{q+r}, \hspace {1cm}  \forall r\geq 0  \] 

\noindent
{\bf THEOREM  3.10}: ({\it Primality test}) Let ${\mathfrak{p}}_q\subset K[y_q]$ and ${\mathfrak{p}}_{q+1}\subset K[y_{q+1}]$ be prime ideals  such that ${\mathfrak{p}}_{q+1}={\rho}_1({\mathfrak{p}}_q)$ and ${\mathfrak{p}}_{q+1}\cap K[y_q]={\mathfrak{p}}_q$. If the symbol $g_q$ of the algebraic variety ${\cal{R}}_q$ defined by ${\mathfrak{p}}_q$ is $2$-acyclic and if its first prolongation $g_{q+1}$ is a vector bundle over ${\cal{R}}_q$, then $\mathfrak{p}={\rho}_{\infty}({\mathfrak{p}}_q)$ is a prime differential ideal with $\mathfrak{p} \cap K[y_{q+r}]={\rho}_r({\mathfrak{p}}_q), \forall r\geq 0 $.  \\

\noindent
{\bf COROLLARY  3.11}: Every perfect differential ideal of $\{y\}$ can be expressed in a unique way as the non-redundant intersection of a finite number of prime differential ideals.  \\

\noindent
{\bf COROLLARY 3.12}: ({\it Differential basis}) If $\mathfrak{r}$ is a perfect differential ideal of $K\{y\}$, then we have $\mathfrak{r}=rad({\rho}_{\infty} ({\mathfrak{r}}_q))$ for $q$ sufficiently large.  \\

\noindent
{\bf EXAMPLE 3.13}: As $K\{y\}$ is a polynomial ring with an infinite number of variables it is not noetherian and an ideal may not have a finite basis. With $K=\mathbb{Q}, n=1$ and $d=d_x$, then $\mathfrak{a}=\{yy_x,y_xy_{xx},y_{xx}y_{xxx}, ... \}\Rightarrow (y_x)^2+ yy_{xx}\in \mathfrak{a} \Rightarrow rad(\mathfrak{a})=\{y_x\}$ is a prime differential ideal.  \\
 
\noindent
{\bf PROPOSITION  3.14}: If $\zeta$ is differentially algebraic over $K<\eta>$ and $\eta$ is differentially algebraic over $K$, then $\zeta$ is differentially algebraic over $K$. Setting $\xi=\zeta - \eta$, it follows that, if $L/K$ is a differential extension and $\xi,\eta \in L$ are both differentially algebraic over $K$, then $\xi + \eta$, $\xi\eta$ and $d_i\xi$ are differentially algebraic over $K$.  \\

If $L=Q(K\{y\}/\mathfrak{p})$, $M=Q(K\{z\}/\mathfrak{q})$ and $N=Q(K\{y,z\}/\mathfrak{r})$ are such that $\mathfrak{p}=\mathfrak{r}\cap K\{y\}$ and $\mathfrak{q}=\mathfrak{r}\cap K\{z\}$, we have the two towers $K\subset L\subset N$ and $K\subset M\subset N$ of differential extensions and we may therefore define the new tower $K \subseteq L\cap M \subseteq <L,M> \subseteq N$. However, if only $L/K$ and $M/K$ are known and we look for such an $N$ containing both $L$ and $M$, we may use the universal property of tensor products an deduce the existence of a differential morphism $L{\otimes}_KM\rightarrow N$ by setting $d(a\otimes b)=(d_La) \otimes b+a \otimes (d_Mb)$ whenever $d_L\mid K=d_M\mid K=\partial$. The construction of an abstract {\it composite} differential field amounts therefore to look for a prime differential ideal in $L{\otimes}_K M$ which is a direct sum of integral domains [32].  \\

\noindent
{\bf DEFINITION  3.15}: A differential extension $L$ of a differential field $K$ is said to be {\it differentially algebraic} over $K$ if every element of $L$ is differentially algebraic over $K$. The set of such elements is an intermediate differential field $K' \subseteq L$, called the {\it differential algebraic closure} of $K$ in $L$. If $L/K$ is a differential extension, one can always find a maximal subset $S$ of elements of $L$ that are differentially transcendental over $K$ and such that $L$ is differentially algebraic over $K<S>$. Such a set is called a {\it differential transcedence basis} and the number of elements of $S$ is called the {\it differential transcendence degree} of $L/K$.  \\  

\noindent
{\bf THEOREM  3.16}: The number of elements in a differential basis of $L/K$ does not depent on the generators of $L/K$ and his value 
is $difftrd(L/K)=\alpha$. Moreover, if $K\subset L \subset M$ are differential fields, then $difftrd(M/K)=difftrd(M/L) + difftrd(L/K)$.  \\

\noindent
{\bf THEOREM 3.17}: If $L/K$ is a finitely generated differential extension, then any intermediate differential field $K'$ between $K$ and $L$ is also finitely generated over $K$.  \\

\noindent
{\bf EXAMPLE  3.18}: With $k=\mathbb{Q}$, let us introduce the manifolds $X$ with local coordinate $x$ and $Y$ with local coordinates $(y^1,y^2)$. We may consider the {\it algebraic Lie pseudogroup} $\Gamma \subset aut(Y)$ of (local, invertible) transformations of $Y$ preserving the $1$-form $y^2dy^1$, that is to say made up by transformations $\bar{y}=g(y)$ solutions of the Pfaffian system ${\bar{y}}^2d{\bar{y}}^1= y^2dy^1$. Equivalently, we have to look for the invertible solutions of the algebraic first order involutive system ${\cal{R}}_1\subset J_1(Y\times Y)$ defined over $k(y^1,y^2)$ by the first order involutive system of algebraic PD equations in Lie form:\\ 
\[   {\bar{y}}^2\frac{\partial {\bar{y}}^1}{\partial y^1}= y^2, \hspace{1cm} {\bar{y}}^2\frac{\partial {\bar{y}}^1}{\partial y^2}= 0 \hspace{1cm}\Rightarrow  \hspace{1cm} \frac{\partial ({\bar{y}}^1,{\bar{y}}^2)}{\partial ((y^1,y^2)}=1  \] 
By chance one can obtain the generic solution ${\bar{y}}^1=g(y^1), \hspace{2mm} {\bar{y}}^2=y^2/(\partial g(y^1)/\partial y^1)$ where $g$ is an arbitrary function of one variable. Now, if we introduce a function $y=f(x)$ and consider the corresponding transformations of the jets $(y^1,y^2,y^1_x,y^2_x, ...)$, we obtain the only generating differential invariant $\Phi\equiv {\bar{y}}^2  {\bar{y}}^1_x=y^2y^1_x$. Hence, setting $K=k<y^2y^1_x>$ and $L=k< y^1,y^2>$, we have the tower of differential extensions $k \subset K \subset L$. As any intermediate differential field $K \subset K' \subset L$ is finitely generated, let us consider $K'=k<y^2y^1_x,y^2_x>$. Then:  \\
\[  {\bar{y}}^2_x\frac{\partial {\bar{y}}^2}{\partial y^1}y^1_x + \frac{\partial {\bar{y}}^2}{\partial y^2} \Rightarrow \frac{\partial{\bar{y}}^2}{\partial y^1}=0, \frac{\partial {\bar{y}}^2}{\partial y^2}=1 \Rightarrow  {\bar{y}}^1=y^1+cst, \hspace{4mm} 
{\bar{y}}^2=y^2\]
allows to define a Lie subpseudogroup ${\Gamma}' \subset \Gamma$ with generating differential invariants $y^1_x,y^2$ in such a way that, if we set $K"=k<y^1_x, y^2>$, we have the strict inclusions $K\subset K' \subset K"  $ and it does not seem possible to obtain a {\it differential Galois correspondence} between algebraic subpseudogroups and intermediate differential fields, similar to the classical one. We have explained in [32] how to overcome this problem but this is out of the scope of this paper. It is finally important to notice that the {\it fundamental differential isomorphism} [4,5,32]:  \\
\[         Q(L{\otimes }_KL)\simeq Q(L{\otimes }_{k(y)}k[\Gamma])  \]
is the Hopf dual of the projective limit of the {\it action graph} isomorphisms between fibered manifolds:  \\
\[    {\cal{A}}_q {\times}_X{\cal{A}}_q \simeq {\cal{A}}_q{\times}_Y {\cal{R}}_q    \]
of fibered dimension $2(q+2)$. The corresponding {\it automorphic system} $y^2y^1_x=\omega$ in Lie form where $\omega$ is a geometric object as in the Introduction and its prolongations has been introduced as early as in $1903$ by E. Vessiot [53,54] as a way to study {\it principal homogeneous spaces} (PHS) for Lie pseudogroups, namely if $y=f(x)$ is a solution and $\bar{y}={\bar{f}}(x) $ is another solution, then there exists one and only one transformation $\bar{y}=g(y)$ of $\Gamma$ such that $\bar{f}=g\circ f$.  \\

This is all what is needed in order to study systems of infinitesimal Lie equations defined, like the classical and conformal Killing systems, over $\mathbb{Q}<\omega>$ where $\omega$ is a geometric object solution of a {\it system of algebraic Vessiot structure equations} (constant riemannian curvature, zero Weyl tensor).  \\

\newpage

\noindent
{\bf 4) DIFFERENTIAL DUALITY} \\

Let $A$ be a {\it unitary ring}, that is $1,a,b\in A \Rightarrow a+b,ab \in A, 1a=a1=a$ and even an {\it integral domain} ($ab=0\Rightarrow a=0$ or $b=0$) with {\it field of fractions} $K=Q(A)$. However, we shall not always assume that $A$ is commutative, that is $ab$ may be different from $ba$ in general for $a,b\in A$. We say that $M={}_AM$ is a {\it left module} over $A$ if $x,y\in M\Rightarrow ax,x+y\in M, \forall a\in A$ or a {\it right module} $M_B$ over $B$ if the operation of $B$ on $M$ is $(x,b)\rightarrow xb, \forall b\in B$. If $M$ is a left module over $A$ and a right module over $B$ with $(ax)b=a(xb), \forall a\in A,\forall  b\in B, \forall x\in M$, then we shall say that $M={ }_AM_B$ is a {\it bimodule}. Of course, $A={ }_AA_A$ is a bimodule over itself. We define the {\it torsion submodule} $t(M)=\{x\in M\mid \exists 0\neq a\in A, ax=0\}\subseteq M$ and $M$ is a {\it torsion module} if $t(M)=M$ or a {\it torsion-free module} if $t(M)=0$. We denote by $hom_A(M,N)$ the set of morphisms $f:M\rightarrow N$ such that $f(ax)=af(x)$. We finally recall that a sequence of modules and maps is exact if the kernel of any map is equal to the image of the map preceding it. \\

When $A$ is commutative, $hom(M,N)$ is again an $A$-module for the law $(bf)(x)=f(bx)$ as we have $(bf)(ax)=f(bax)=f(abx)=af(bx)=a(bf)(x)$. In the non-commutative case, things are more complicate and, given ${}_AM$ and ${}_AN_B$, then $hom_A(M,N)$ becomes a right module over $B$ for the law $(fb)(x)=f(x)b$. \\

\noindent
{\bf DEFINITION 4.1}: A module $F$ is said to be {\it free} if it is isomorphic to a (finite) power of $A$ called the {\it rank} of $F$ over $A$ and denoted by $rk_A(F)$ while the rank $rk_A(M)$ of a module $M$ is the rank of a maximum free submodule $F\subset M$. It follows from this definition that $M/F$ is a torsion module. In the sequel we shall only consider {\it finitely presented} modules, namely {\it finitely generated} modules defined by exact sequences of the type $F_1 \stackrel{d_1}{\longrightarrow} F_0 \stackrel{p}{\longrightarrow} M\longrightarrow 0$ where $F_0$ and $F_1$ are free modules of finite ranks $m_0$ and $m_1$ often denoted by $m$ and $p$ in examples. A module $P$ is called {\it projective} if there exists a free module $F$ and another (projective) module $Q$ such that $P\oplus Q\simeq F$.\\

\noindent
{\bf PROPOSITION 4.2}: For any short exact sequence $0\rightarrow M' \stackrel{f}{\longrightarrow} M \stackrel{g}{\longrightarrow} M" \rightarrow 0$, we have the relation $rk_A(M)=rk_A(M')+rk_A(M")$, even in the non-commutative case.\\

The following proposition will be used many times in Section $5$, in particular for exhibiting the Weyl tensor from the Riemann 
tensor ([2],p 73)([50],p 33) :\\

\noindent
{\bf PROPOSITION 4.3}: We shall say that the following short exact sequence {\it splits} if one of the following equivalent three conditions holds: \\
\[0\longrightarrow
M' \stackrel{\stackrel{u}{\longleftarrow}}{\stackrel{f}{\longrightarrow}} M
     \stackrel{\stackrel{v}{\longleftarrow}}{\stackrel{g}{\longrightarrow}} M''   \longrightarrow 0  \]
$\bullet$ There exists a monomorphism $v:M''\rightarrow M$ called {\it lift} of $g$ and such that $g\circ v=id_{M''}$ .\\
$\bullet$ There exists an epimorphism $u:M\rightarrow M'$ called {\it lift} of $f$ and such that $u\circ f=id_{M'}$.\\
$\bullet$ There exist isomorphisms $\varphi=(u,g):M\rightarrow M'\oplus M''$ and $\psi=f+v:M'\oplus M''\rightarrow M$ that are inverse to each other and provide an isomorphism $M\simeq M'\oplus M''$ with $f\circ u+v\circ g=id_M$ and thus $ker(u)=im(v)$.  \\
These conditions are automatically satisfied if $M"$ is free or projective.  \\

Using the notation $M^*=hom_A(M,A)$, for any morphism $f:M\rightarrow N$, we shall denote by $f^*:N^*\rightarrow M^*$ the morphism which is defined by  $f^*(h)=h\circ f, \forall h\in hom_A(N,A)$ and satisfies $rk_A(f)=rk_A(im(f))=rk_A(f^*),\forall f\in hom_A(M,N)$(See [37], Corollary 5.3, p 179). We may take out $M$ in order to obtain the {\it deleted sequence} $... \stackrel{d_2}{\longrightarrow} F_1 \stackrel{d_1}{\longrightarrow} F_0 \longrightarrow 0$ and apply  $hom_A(\bullet,A)$ in order to get the sequence $... \stackrel{d^*_2}{\longleftarrow} F^*_1 \stackrel{d^*_1}{\longleftarrow} F^*_0 \longleftarrow 0$. \\

\noindent
{\bf PROPOSITION  4.4}: The {\it extension modules}  $ext^0_A(M)=ker(d^*_1)=hom_A(M,A)=M^*$ and $ext^i(M)=ext^i_A(M)=ker(d^*_{i+1})/im(d^*_i), \forall i\geq 1$ do not depend on the resolution chosen and are torsion modules for $i\geq 1$. \\

Let $ A$ be a {\it differential ring}, that is a commutative ring with $n$ commuting {\it derivations} $\{{\partial}_1,...,{\partial}_n\}$, that is ${\partial}_i{\partial}_j={\partial}_j{\partial}_i={\partial}_{ij}, \forall i,j=1,...,n$ while ${\partial}_i(a+b)={\partial}_ia+{\partial}_ib$ and ${\partial}_i(ab)=({\partial}_ia)b+a{\partial}_ib, \forall a,b\in A$. We shall use thereafter a differential integral domain $A$ with unit $1\in A$ whenever we shall need a {\it differential field} $\mathbb{Q}\subset K=Q(A)$ of coefficients, that is a field ($a\in K\Rightarrow 1/a\in K$) with ${\partial}_i(1/a)=-(1/a^2){\partial}_ia$, in order to exhibit solved forms for systems of partial differential equations as in the preceding section. Using an implicit summation on multi-indices, we may introduce the (noncommutative) {\it ring of differential operators} $D=A[d_1,...,d_n]=A[d]$ with elements $P=a^{\mu}d_{\mu}$ such that $\mid \mu\mid<\infty$ and $d_ia=ad_i+{\partial}_ia$. The highest value of ${\mid}\mu {\mid}$ with $a^{\mu}\neq 0$ is called the {\it order} of the {\it operator} $P$ and the ring $D$ with multiplication $(P,Q)\longrightarrow P\circ Q=PQ$ is filtred by the order $q$ of the operators with the {\it filtration} $0=D_{-1}\subset D_0\subset D_1\subset  ... \subset D_q \subset ... \subset D_{\infty}=D$. Moreover, it is clear that $D$, as an algebra, is generated by $A=D_0$ and $T=D_1/D_0$ with $D_1=A\oplus T$ if we identify an element $\xi={\xi}^id_i\in T$ with the vector field $\xi={\xi}^i(x){\partial}_i$ of differential geometry, but with ${\xi}^i\in A$ now. It follows that $D={ }_DD_D$ is a {\it bimodule} over itself, being at the same time a left $D$-module ${ }_DD$ by the composition $P \longrightarrow QP$ and a right $D$-module $D_D$ by the composition $P \longrightarrow PQ$ with $D_rD_s=D_{r+s}, \forall r,s \geq 0$ in any case. \\

If we introduce {\it differential indeterminates} $y=(y^1,...,y^m)$, we may extend $d_iy^k_{\mu}=y^k_{\mu+1_i}$ to ${\Phi}^{\tau}\equiv a^{\tau\mu}_ky^k_{\mu}\stackrel{d_i}{\longrightarrow} d_i{\Phi}^{\tau}\equiv a^{\tau\mu}_ky^k_{\mu+1_i}+{\partial}_ia^{\tau\mu}_ky^k_{\mu}$ for $\tau=1,...,p$. Therefore, setting $Dy^1+...+Dy^m=Dy\simeq D^m$ and calling $I=D\Phi\subset Dy$ the {\it differential module of equations}, we obtain by residue the {\it differential module} or $D$-{\it module} $M=Dy/D\Phi$, introducing the canonical projection $Dy \stackrel{p}{\longrightarrow} M \rightarrow 0$ and denoting the residue of $y^k_{\mu}$ by ${\bar{y}}^k_{\mu}$ when there can be a confusion. Introducing the two free differential modules $F_0\simeq D^{m_0}, F_1\simeq D^{m_1}$, we obtain equivalently the {\it free presentation} $F_1\stackrel{d_1}{\longrightarrow} F_0 \stackrel{p}{\longrightarrow} M \rightarrow 0$ of order $q$ when $d_1={\cal{D}}=\Phi \circ j_q$. It follows that $M$ can be endowed with a {\it quotient filtration} obtained from that of $D^m$ which is defined by the order of the jet coordinates $y_q$ in $D_qy$. We shall suppose that the system $R_q=ker(\Phi)$ is formally integrable. We have therefore the {\it inductive limit} $0=M_{-1} \subseteq M_0 \subseteq M_1 \subseteq ... \subseteq M_q \subseteq ... \subseteq M_{\infty}=M$ with $d_iM_q\subseteq M_{q+1}$ which is the dual of the {\it projective limit} $R=R_{\infty}\rightarrow ... \rightarrow R_q \rightarrow R_0\rightarrow 0$ if we set $R=hom_K(M,K)$ with $R_q=hom_K(M_q,K)$ and $DR_{q+1}\subseteq T^*\otimes R_q$. This is the main reason for using a differential field $K$ because $hom_K(\bullet , K)$ transform any short exact sequence into a short exact sequence. We have in general 
$D_rI_s\subseteq I_{r+s}, \forall r\geq 0, \forall s<q$ with $I_r=I\cap D_ry$. \\

More generally, introducing the successive CC as in the preceding Section while changing slightly the numbering of the respective operators, we may finally obtain the {\it free resolution} of $M$, namely the exact sequence $\hspace{5mm} ... \stackrel{d_3}{\longrightarrow} F_2  \stackrel{d_2}{\longrightarrow} F_1 \stackrel{d_1}{\longrightarrow}F_0\stackrel{p}{\longrightarrow}M \longrightarrow 0 $ where $p$ is the canonical projection. Also, with a slight abuse of language, when ${\cal{D}}=\Phi \circ j_q$ is involutive, that is to say when $R_q=ker( \Phi)$ is involutive, one should say that $M$ has an {\it involutive presentation} of order $q$ or that $M_q$ is {\it involutive}. \\

\noindent
{\bf REMARK  4.5}:  In actual practice, one must never forget that ${\cal{D}}=\Phi \circ j_q$ {\it acts on the left on column vectors in the operator case and on the right on row vectors in the module case}. For this reason, when $E$ is a (finite dimensional) vector bundle over 
$X$, we may apply the correspondence $J_{\infty}(E) \leftrightarrow D{\otimes}_KE^* : J_q(E)\leftrightarrow D_q{\otimes}_K E^*$ with ${\pi}^{q+1}_q:J_{q+1}(E) \rightarrow J_q(E) \leftrightarrow D_q \subset D_{q+1}$ and $E^*=hom_K(E,K)$ between jet bundles and left differential modules in order to be able to use the {\it double dual isomorphism} $E\simeq E^{**}$ in both cases. We shall say that $D(E)=D\otimes _KE^*=ind(E^*)$ is the the left differential module {\it induced} by $E^*$. Hence, starting from a differential operator $E \stackrel{\cal{D}}{\longrightarrow}F$, we may obtain a finite presentation $D{\otimes}_KF^* \stackrel{{\cal{D}}^*}{\longrightarrow}D{\otimes}_KE^* \rightarrow M \rightarrow 0$ and conversely, keeping the same operator matrix if we act on the right of row vectors. This comment becomes particularly useful when dealing with the Poincar\'{e} sequence in electromagnetism ($n=4$) or even as we already saw in the Introduction ($n=3$).  \\

Roughly speaking, homological algebra has been created in order to find intrinsic properties of modules not depending on their presentations or even on their resolutions and we now exhibit another approach by defining the {\it formal adjoint} of an operator $P$ and an operator matrix 
${\cal{D}}$:  \\

\noindent
{\bf DEFINITION  4.6}: Setting $P=a^{\mu}d_{\mu}\in D  \stackrel{ad}{\longleftrightarrow} ad(P)=(-1)^{\mid\mu\mid}d_{\mu}a^{\mu}   \in D $, we have $ad(ad(P))=P$ and $ad(PQ)=ad(Q)ad(P), \forall P,Q\in D$. Such a definition can be extended to any matrix of operators by using the transposed matrix of adjoint operators and we get:  
\[ <\lambda,{\cal{D}} \xi>=<ad({\cal{D}})\lambda,\xi>+\hspace{1mm} {div}\hspace{1mm} ( ... )  \]
from integration by part, where $\lambda$ is a row vector of test functions and $<  > $ the usual contraction. We quote the useful formulas 
$rk_D({\cal{D}})=rk_D(ad({\cal{D}}))$ as in ([34], p 339-341 or [35]).\\

The following technical Lemma is crucially used in the next proposition:  \\

\noindent
{\bf LEMMA  4.7}: If $f\in aut(X)$ is a local diffeomorphisms on $X$, we may set $ x=f^{-1}(y)=g(y)$ and we have the {\it identity}:
\[   \frac{\partial}{\partial y^k}(\frac{1}{\Delta (g(y))} {\partial}_if^k(g(y))\equiv 0.   \]

\noindent
{\bf PROPOSITION  4.8}: If we have an operator $E\stackrel{\cal{D}}{\longrightarrow} F$, we may obtain by duality an operator 
${\wedge}^nT^*\otimes E^*\stackrel{ad(\cal{D})}{\longleftarrow} {\wedge}^nT^*\otimes F^*$. \\

Now, with operational notations, let us consider the two differential sequences:  \\
\[   \xi  \stackrel{{\cal{D}}}{\longrightarrow} \eta \stackrel{{\cal{D}}_1}{\longrightarrow} \zeta  \]
\[   \nu  \stackrel{ad({\cal{D}})}{\longleftarrow} \mu \stackrel{ad({\cal{D}}_1)}{\longleftarrow} \lambda   \]
where ${\cal{D}}_1$ generates all the CC of ${\cal{D}}$. Then ${\cal{D}}_1\circ {\cal{D}}\equiv 0 \Longleftrightarrow ad({\cal{D}})\circ ad({\cal{D}}_1)\equiv 0 $ but $ad({\cal{D}})$ may not generate all the CC of $ad({\cal{D}}_1)$ as we already saw in the Introduction. Passing to the module framework, we just recognize the definition of $ext^1(M)$ when $M$ is determined by ${\cal{D}}$. \\ 

As $D={ }_DD_D$ is a bimodule, then $M^*=hom_D(M,D)$ is a right $D$-module according to Lemma 3.1 and we may thus define a right module $N_D$ by the ker/coker long exact sequence $0\longleftarrow N_D \longleftarrow F_1^*\stackrel{{\cal{D}}^*}{ \longleftarrow} F^*_0 \longleftarrow M^* \longleftarrow 0$ but we have [6,36,43,51]:  \\

 \noindent
{\bf THEOREM  4.9}: We have the {\it side changing} procedures $M={ }_DM \rightarrow M_D={\wedge}^nT^*{\otimes}_AM$ and $N_D \rightarrow N={}_DN=hom_A({\wedge}^nT^*,N_D)$ with ${}_D((M_D))=M$ and ${}_D(N_D)=N$. \\

Now, exactly like we defined the differential module $M$ from $\cal{D}$, we may define the differential module $N$ from $ad(\cal{D})$.
For any other presentation of $M$ with an accent, we have [23,36]:  \\

\noindent
{\bf THEOREM  4.10}: The modules $N$ and $N'$ are {\it projectively equivalent}, that is one can find two projective modules $P$ and $P'$ such that $N\oplus P\simeq N' \oplus P'$ and we obtain therefore $ext^i_D(N)\simeq ext^i_D(N'), \forall i\geq 1$.  \\

\noindent
{\bf THEOREM  4.11}: The operator ${\cal{D}}$ is {\it simply parametrizable} if $ext^1(N)=0$ and {\it doubly parametrizable} if $ext^1(N)=0$ {\it and} $ext^2(N)=0$. Moreover, we have the ker/coker long exact sequence:   \\
\[ 0 \rightarrow ext^1(N) \rightarrow M \stackrel{\epsilon}{\longrightarrow}M^{**}\rightarrow ext^2(N) \rightarrow 0  \]
where $(\epsilon (m))(f)=f(m)$ whenever $f\in M^*$ and we have $t(M)=ext^1(N)=ker(\epsilon)$. \\

\noindent
{\it Proof}: We prove first that $t(M)\subseteq ker(\epsilon)$. Indeed, if $m\in t(M)$, then one may find $0\neq P \in D$ such that $Pm=0$ and thus $f(Pm)=Pf(m)=0 \Rightarrow f(m)=0$ because $D=K[d]$ is an integral domain and thus $t(M)\subseteq ker(\epsilon)$.\\
Let us now start with a free presentation of $M=cocker(d_1)$:
\[  F_1\stackrel{d_1}{\longrightarrow} F_0 \stackrel{p}{\longrightarrow }M
\longrightarrow 0 \]
Applying $hom_D(M,D)$, we may define $N_D=coker(d^*_1)$ and exhibit the following free resolution of $N$ by right $D$-modules:
\[ 0\longleftarrow N_D \longleftarrow F^*_1
\stackrel{d^*_1}{\longleftarrow}F^*_0 \stackrel{d^*_0}{\longleftarrow}F^*_{-1} 
\stackrel{d^*_{-1}}{\longleftarrow} F^*_{-2}  \]
where $M^*=ker(d^*_1)=im(d^*_0)\simeq coker (d^*_{-1})$. The deleted
sequence is:
\[ 0 \longleftarrow F^*_1\stackrel{d^*_1}{\longleftarrow}F^*_0 
\stackrel{d^*_0}{\longleftarrow}F^*_{-1} 
\stackrel{d^*_{-1}}{\longleftarrow} F^*_{-2}  \]
Applying again $hom_D(\bullet,D)$ and using the canonical isomorphism
$F^{**}\simeq F$ for any free module $F$ of finite rank, we get the sequence of left $D$-modules:
\[   \begin{array}{rcccl}
0\longrightarrow F_1\stackrel{d_1}{\longrightarrow} &F_0 &
\stackrel{d_0}{\longrightarrow}
&F_{-1}&\stackrel{d_{-1}}{\longrightarrow} F_{-2}\\
  & \downarrow & \searrow & \uparrow &   \\
  &    M  &\stackrel{\epsilon}{\longrightarrow} &M^{**} &   \\
  & \downarrow &  & \uparrow &   \\
  &   0        &  &    0     &   
\end{array}  \]
Denoting as usual a coboundary space by $B$, a cocycle space by $Z$ and the
corresponding cohomology by $H=Z/B$, we get the commutative and exact diagram:
\[
\begin{array}{rcccccl}
0\longrightarrow & B_0 & \longrightarrow & F_0 &\longrightarrow & M &
\longrightarrow 0 \\
  & \downarrow &  &  \parallel &  & \downarrow \epsilon &  \\
0\longrightarrow & Z_0 & \longrightarrow & F_0 &\longrightarrow & M^{**} &  
\end{array}  \]
An easy chase provides at once $H_0=Z_0/B_0=ext^1_D(N)\simeq 
ker(\epsilon)$. It follows that $ker(\epsilon)$ is a torsion module and, as
we already know that $t(M)\subseteq ker(\epsilon) \subseteq M$, we finally
obtain $t(M)=ker(\epsilon)$. Also, as $B_{-1}=im(\epsilon)$ and
$Z_{-1}\simeq M^{**}$, we obtain 
$H_{-1}=Z_{-1}/B_{-1}=ext^2_A(N,A)\simeq coker(\epsilon)$. Accordingly, a {\it torsion-free} ($\epsilon$ injective)/{\it reflexive} 
($\epsilon$ bijective) module is described by an operator that admits respectively a single/double step parametrization.\\
\hspace*{12cm}  Q.E.D. \\

We know turn to the operator framework;  \\

\noindent
{\bf DEFINITION  4.12}: If a differential operator $\xi \stackrel{\cal{D}}{\longrightarrow} \eta$ is given, a {\it direct problem} is to find generating {\it compatibility conditions} (CC) as an operator $\eta \stackrel{{\cal{D}}_1}{\longrightarrow} \zeta $ such that ${\cal{D}}\xi=\eta \Rightarrow {\cal{D}}_1\eta=0$. Conversely, given $\eta \stackrel{{\cal{D}}_1}{\longrightarrow} \zeta$, the {\it inverse problem} will be to look for $\xi \stackrel{\cal{D}}{\longrightarrow} \eta$ such that ${\cal{D}}_1$ generates the CC of ${\cal{D}}$ and we shall say that ${\cal{D}}_1$ {\it is parametrized by} ${\cal{D}}$ {\it if such an operator} ${\cal{D}}$ {\it is existing}. We finally notice that any operator is the adjoint of a certain operator because $ad(ad(P))=P, \forall P \in D$ and we get:  \\

\noindent
{\bf THEOREM  4.13}: ({\it reflexivity test}) In order to check whether $M$ is reflexive or not, that is to find out a parametrization if $t(M)=0$ which can be again parametrized, the test has 5 steps which are drawn in the following diagram where $ad({\cal{D}})$ generates the CC of $ad({\cal{D}}_1)$ and ${\cal{D}}_1'$ generates the CC of ${\cal{D}}=ad(ad({\cal{D}}))$ while $ad({\cal{D}}_{-1})$ generates the CC of $ad({\cal{D}})$ and ${\cal{D}}'$ generates the CC of ${\cal{D}}_{-1}$:  \\
\[  \begin{array}{rcccccccl}
 & & & & & {\eta}'     & &  {\zeta}' &\hspace{15mm} 5  \\
 & & & &  \stackrel{{\cal{D}}'}{\nearrow}   & & \stackrel{{\cal{D}}'_1}{\nearrow} &  &  \\
4 \hspace{15mm}&\phi & \stackrel{{\cal{D}}_{-1}}{\longrightarrow}& \xi  & \stackrel{{\cal{D}}}{\longrightarrow} &  \eta & \stackrel{{\cal{D}}_1}{\longrightarrow} & \zeta &\hspace{15mm}   1  \\
 &  &  &  &  &  &  &  &  \\
 &  &  &  &  &  &  &  &  \\
 3 \hspace{15mm}& \theta &\stackrel{ad({\cal{D}}_{-1})}{\longleftarrow}& \nu & \stackrel{ad({\cal{D}})}{\longleftarrow} & \mu & \stackrel{ad({\cal{D}}_1)}{\longleftarrow} & \lambda &\hspace{15mm} 2
  \end{array}  \]
\[{\cal{D}}_1 \,\,\,parametrized \,\,\,by \,\,\,{\cal{D}} \Leftrightarrow {\cal{D}}_1={\cal{D}}'_1  \Leftrightarrow ext^1(N)=0  \Leftrightarrow  \epsilon \,\,\, injective  \Leftrightarrow t(M)=0\] 
\[{\cal{D}} \,\,\,parametrized \,\,\,by \,\,\,{\cal{D}}_{-1} \Leftrightarrow {\cal{D}}={\cal{D}}' \Leftrightarrow ext^2(N)=0 \Leftrightarrow \epsilon \,\,\, surjective  \hspace{17mm}  \]

\noindent
{\bf COROLLARY 4.14}: In the differential module framework, if $F_1 \stackrel{{\cal{D}}_1}{\longrightarrow} F_0 \stackrel{p}{\longrightarrow} M \rightarrow 0$ is a finite free presentation of $M=coker({\cal{D}}_1)$ with $t(M)=0$, then we may obtain an exact sequence $F_1 \stackrel{{\cal{D}}_1}{\longrightarrow} F_0 \stackrel{{\cal{D}}}{\longrightarrow} E $ of free differential modules where ${\cal{D}}$ is the parametrizing operator. However, there may exist other parametrizations $F_1 \stackrel{{\cal{D}}_1}{\longrightarrow} F_0 \stackrel{{\cal{D}}'}{\longrightarrow} E' $ called {\it minimal parametrizations} such that $coker({\cal{D}}')$ is a torsion module and we have thus $rk_D(M)=rk_D(E')$.  \\

\noindent
{\bf REMARK 4.15}: The following chains of inclusions and short exact sequences allow to compare the main procedures used in the respective study of differential extensions and differential modules: \\
\[    \begin{array}{ccc}
       K \subset K<S> \subset L      &  \Rightarrow &    0 \rightarrow F \rightarrow M \rightarrow T \rightarrow 0  \\
                                                &       &         \\
       K \subset K' \subset  L          & \Rightarrow  &    0 \rightarrow t(M)  \rightarrow M  \rightarrow M' \rightarrow 0
       \end{array}   \]    
where $F$ is a maximum free submodule of $M$, $T=M/F$ is a torsion-module and $M'=M/t(M)$ is a torsion-free module.  The next examples open the way towards a new domain of research.  \\

\noindent
{\bf EXAMPLE 4.16}: With $n=2,m=3, K=\mathbb{Q}$, let us consider the first order nonlinear involutive system:  \\
\[    P_1\equiv  y^1_2 - y^3y^1_1=0, \hspace{1cm}  P_2\equiv y^2_2 - y^3y^2_1=0  \]
This system defines a prime differential ideal $\mathfrak{p}\subset K\{y\}$ and the differential extension $L=Q(K\{y\}/\mathfrak{p})$ is differentially algebraic over $K<y^3>$ with parametric jets $(y^1, y^2,y^1_1,y^2_1, y^1_{11}, y^2_{11}, ...)$.\\
The linearized system ${\cal{D}}_1Y=0$ over $L$ is:  \\
\[     d_2Y^1-y^3d_1Y^1-y^1_1Y^3=0, \hspace{1cm}  d_2Y^2 - y^3d_1Y^2 - y^2_1Y^3=0   \]
Multiplying by test functions $({\lambda}^1,{\lambda}^2)$ and integrating by part, we get $ad({\cal{D}}_1)\lambda=\mu$ in the form:  \\
\[  \begin{array}{ccccc}
Y^1 & \rightarrow & -d_2{\lambda}^1 + y^3 d_1{\lambda}^1+y^3_1{\lambda}^1 & = & {\mu}^1  \\
Y^2 & \rightarrow & -d_2{\lambda}^2+y^3d_1{\lambda}^2 + y^3_1{\lambda}^2  & = & {\mu}^2  \\
Y^3 & \rightarrow & -y^1_1{\lambda}^1 - y^2_1{\lambda}^2                                & = & {\mu}^3
\end{array}   \]
Using only the parametric jets for $y$ and $\lambda$ in the PD equations provided, we get:  \\
\[  -y^1_1(y^3d_1{\lambda}^1+y^3_1{\lambda}^1)-(y^3y^1_{11}+y^1_1y^3_1){\lambda}^1 - y^2_1(y^3d_1{\lambda}_2+y^3_1{\lambda}^1)-(y^3y^2_{11}+y^2_1y^3_1){\lambda}^2=d_2{\mu}^3-y^1_1{\mu}^1 -y^2_1{\mu}^2\]
\[ -y^3y^1_1d_1{\lambda}^1-y^3y^1_{11}{\lambda}^1-y^3y^2_1d_1{\lambda}^2-y^3y^2_{11}{\lambda}^2-2y^1_1y^3_1{\lambda}^1-2y^2_1y^3_1{\lambda}^2=y^3d_1{\mu}^3 + 2y^3_1{\mu}^3  \]
and the {\it only} CC $ad({\cal{D}})\mu=0$ over $L$:  \\
\[  -d_2{\mu}^3+y^3d_1{\mu}^3+y^1_1{\mu}^1+y^2_1{\mu}^2+2y^3_1{\mu}^3=0   \]
Multiplying by a test function $\xi$ and integrating by part, we get ${\cal{D}}\xi=Y$ over $L$ in the form:  \\
\[y^1_1\xi =Y^1, \hspace{1cm} y^2_1\xi=Y^2, \hspace{1cm}  d_2\xi-y^3d_1\xi+y^3_1\xi=Y^3  \]
admitting the CC ${\cal{D}}_1Y=0$ of course but also the additional zero order CC:
\[   \omega\equiv y^1_1Y^2-y^2_1Y^1=0  \]
which provides a torsion element $\omega$ satisfying $d_2\omega - y^3 d_1\omega-y^3_1\omega=0$. Setting $Y=\delta y$ as the standard variational notation used by engineer, we obtain easily $\omega \wedge \delta \omega\neq 0$ and $\omega$ {\it cannot therefore admit an integrating factor}, a result showing that $K$ is its own differential algebraic closure in $L$.  \\

\noindent
{\bf EXAMPLE  4.17}: If $\alpha=dx^1-x^3dx^2\in T^*$, the linear system obtained over $K=\mathbb{Q}(x^1,x^2,x^3)$ by eliminating the factor $\rho(x)$ in the linear system ${\cal{L}}(\xi)\alpha=\rho(x)\alpha$ admits the injective parametrization $-x^3{\partial}_3\theta + \theta={\xi}^1, -{\partial}_3\phi={\xi}^2, {\partial}_2\phi-x^3{\partial}_1\phi={\xi}^3 \Rightarrow {\xi}^1-x^3{\xi}^2=\phi $. 
It defines therefore a free differential module $M\simeq D$ which is thus reflexive and even projective. Any resolution of this module splits, 
like the short exact sequence $0 \rightarrow D^2 \rightarrow D^3 \rightarrow D  \rightarrow 0$, and we shall prove in section $6$ that the corresponding differential sequence of operators is locally exact like the Poincar\'{e} sequence ([32], p 684-691).  \\

\newpage

\noindent
{\bf 5) CONFORMAL STRUCTURE}   \\

We start this section with a general (difficult) result on the actions of Lie groups, covering at the same time the study of the classical and conformal Killing systems. For this, we notice that the involutive {\it first Spencer operator} $D_1:C_0=R_q\stackrel{j_1}{\rightarrow}J_1(R_q)\rightarrow J_1(R_q)/R_{q+1}\simeq T^*\otimes R_q/\delta (g_{q+1})=C_1$ of order one is induced by the {\it Spencer operator} $D:R_{q+1}\rightarrow T^*\otimes R_q:{\xi}_{q+1} \rightarrow j_1({\xi}_q)-{\xi}_{q+1}=\{ {\partial}_i{\xi}^k_{\mu}-{\xi}^k_{\mu +1_i}\mid 0\leq \mid \mu \mid q \}$. Introducing the {\it Spencer bundles} $C_r={\wedge}^rT^*\otimes R_q/{\delta}({\wedge}^{r-1}T^*\otimes g_{q+1})$, the first order involutive ($r+1$)-{\it Spencer operator} $D_{r+1}:C_r\rightarrow C_{r+1}$ is induced by $D:{\wedge}^rT^*\otimes R_{q+1}\rightarrow {\wedge}^{r+1}T^*\otimes R_q:\alpha\otimes {\xi}_{q+1}\rightarrow d\alpha\otimes {\xi}_q+(-1)^r\alpha\wedge D{\xi}_{q+1}$. We obtain therefore the canonical {\it linear Spencer sequence} ([34], p 150 or [52]):  \\
\[    0 \longrightarrow \Theta \stackrel{j_q}{\longrightarrow} C_0 \stackrel{D_1}{\longrightarrow} C_1 \stackrel{D_2}{\longrightarrow} C_2 \stackrel{D_3}{\longrightarrow} ... \stackrel{D_n}{\longrightarrow} C_n\longrightarrow 0  \]

\noindent
{\bf PROPOSITION 5.1}: The Spencer sequence for the Lie operator describing the infinitesimal action of a Lie group $G$ is 
(locally) isomorphic to the tensor product of the Poincar\'e sequence by the Lie algebra ${\cal{G}}=T_e(G)$ where $e\in G$ is the identity element. It follows that $D_{r+1}$ {\it generates the CC of} $D_r$ $\Leftrightarrow$ $ad(D_r)$ {\it generates the CC of} $ad(D_{r+1})$, a result not evident at all.  \\

\noindent
{\it Proof}: We may introduce a basis $\{ {\theta}_{\tau}={\theta}^i_{\tau}(x){\partial}_i \}$ of infinitesimal generators of the action with $\tau=1,...,dim(G)$ and the commutation relations $[{\theta}_{\rho},{\theta}_{\sigma}]=c^{\tau}_{\rho \sigma}{\theta}_{\tau} $ discovered by S. Lie giving the {\it structure constants} c of ${\cal{G}}$ (See [34] and [44] for more details). Any element $\lambda \in {\cal{G}}$ can be written $\lambda=\{{\lambda}^{\tau}=cst \}$. " {\it Gauging } " such an element, that is to say {\it replacing the constants by functions} or, equivalently, introducing a map $X\rightarrow {\wedge}^0T^*\otimes {\cal{G}}:(x) \rightarrow ({\lambda}^{\tau}(x))$, we may obtain  locally a map ${\wedge}^0T^*\otimes {\cal{G}}\rightarrow T: {\lambda}^{\tau}(x) \rightarrow {\lambda}^{\tau}(x){\theta}^k_{\tau}(x)$ or, equivalently, vector fields $\xi=({\xi}^i(x){\partial}_i)\in T$ of the form ${\xi}^k(x)={\lambda}^{\tau}(x){\theta}^k_{\tau}(x)$, keeping the index $i$ for $1$-forms. More generally, we can introduce a map :  \\
\[  {\wedge}^rT^*\otimes {\cal{G}} \rightarrow {\wedge}^rT^*\otimes J_q(T)=\lambda \rightarrow \lambda \otimes  j_q(\theta)=X_q: {\lambda}^{\tau}(x) \rightarrow {\lambda}^{\tau}(x) {\partial}_{\mu}{\theta}^k_{\tau}(x) ={X}^k_{\mu,I}(x)dx^I  \]
that we can lift to the element $\lambda \otimes  j_{q+1}(\theta)=X_{q+1}\in {\wedge}^r T^*\otimes J_{q+1}(T)$. It follows from the definitions that $D_rX_q=DX_{q+1}$ by introducing any element of $C_r(T)$ through its representative $X_q\in {\wedge}^rT^*\otimes J_q(T)$. We obtain therefore the {\it crucial formula}:  \\
\[  D_rX_q=DX_{q+1}=D(\lambda \otimes j_{q+1}(\theta))=d\lambda\otimes j_q(\theta)+ (-1)^r\lambda \wedge Dj_{q+1}(\theta)=d\lambda \otimes j_q(\theta)   \]
allowing to identify locally the Spencer sequence for $j_q$ with the Poincar\'e sequence. When the action is effective, the 
map ${\wedge}^0T^*\otimes {\cal{G}}\rightarrow J_q(T)$ is injective. We obtain therefore an isomorphism 
${\wedge}^0T^*\otimes {\cal{G}} \rightarrow R_q\subset J_q(T)$ when $q$ is large enough allowing to exhibit an isomorphism between the canonical Spencer sequence and the tensor product of the Poincar\'e sequence by ${\cal{G}}$ when $q$ is large enough in such a way that $R_q$ is involutive with $dim(R_q)=dim({\cal{G}})$ and $g_q=0$. \\
\hspace*{12cm}  Q.E.D.   \\

We now study what happens when $n\geq 3$ because the case $n=2$ has already been provided, proving that conformal geometry must be entirely revisited.  \\

\noindent
$\bullet$ $n=3$: Using the euclidean metric $\omega$, we have $6$ components of $\Omega \in F_0=S_2T^*$ with $dim(F_0)=n(n+1)/2=6$ in the case of the classical Killing system/operator and obtain easily the $n^2(n^2-1)/12 =6$ components of the second order Riemann operator, linearization of the Riemann tensor at $\omega$. We have  $n^2(n^2-1)(n-2)/24=3$ first order Bianchi identities ([32], p 625). Introducing the respective adjoint operators while taking into account the last Proposition and the fact that the extension modules do not depend on the resolution used ({\it a difficult result indeed} !), we get the following diagram where we have set $ad(Riemann)=Beltrami$ for historical reasons [3] and each operator generates the CC of the next one:  \\
\[  \begin{array}{rccccccl}
  &3 & \stackrel{Killing}{\longrightarrow } & 6 & \stackrel{Riemann}{\longrightarrow}& 6 & \stackrel{Bianchi}{\longrightarrow} & 3\rightarrow 0  \\
   &  &  &  &  &  &  &  \\
0 \leftarrow & 3 & \stackrel{Cauchy}{\longleftarrow} & 6 &\stackrel{Beltrami}{\longleftarrowÊ} & 6 &\longleftarrow &  3
                   \end{array}  \] 
As in the Introduction where $Airy=ad(Riemann)$, the Beltrami operator is now parametrizing the $3$ Cauchy stress equations [3] but it is rather striking to discover that {\it the central second order operator is self-adjoint} and can be given as follows:  \\
\[   \left(  \begin{array}{cccccc}
 0 & 0 & 0 & d_{33} & - 2d_{23} & d_{22} \\
 0 & - 2d_{33} & 2d_{23} & 0 & 2d_{13} & - 2d_{12}  \\
 0 & 2d_{23} & - 2d_{22} & - 2d_{13} & 2d_{12} & 0 \\
 d_{33}& 0 & - 2 d_{13} & 0 & 0 & d_{11}  \\
 - 2d_{23} & 2d_{13} & 2d_{12}& 0 & - 2d_{11} & 0 \\
 d_{22} & - 2 d_{12} & 0 & d_{11}& 0 & 0 
 \end{array} \right)     \]
The study of the conformal case is much more delicate. As ${\hat{F}}_0$ can be described by trace-free symmetric tensors, we have $dim({\hat{F}}_0)=dim(F_0) -1=5$ and it remains to discover the operator that will replace the Riemann operator. Having in mind the diagram of Proposition 2.11 and the fact that $dim({\hat{g}}_2)=3$ while ${\hat{g}}_3=0\Rightarrow {\hat{g}}_4=0$, we have successively:  \\
$\bullet$ NO CC order $1$: $0 \rightarrow {\hat{g}}_2 \rightarrow S_2T^*\otimes T \rightarrow T^*\otimes {\hat{F}}_0 \rightarrow {\hat{F}}_1 \Rightarrow 0 \Rightarrow dim({\hat{F}}_1)= 3-18 + 15=0$.\\
$\bullet$ NO CC order $2$: $0 \rightarrow {\hat{g}}_3 \rightarrow S_3T^*\otimes T \rightarrow S_2T^*\otimes {\hat{F}}_0 \rightarrow {\hat{F}}_1 \Rightarrow 0 \Rightarrow dim({\hat{F}}_1)= 0-30+30=0$.  \\
$\bullet$ OK CC order $3$: $0 \rightarrow {\hat{g}}_4 \rightarrow S_4T^*\otimes T \rightarrow S_3T^*\otimes {\hat{F}}_0 \rightarrow {\hat{F}}_1 \Rightarrow 0 \Rightarrow dim({\hat{F}}_1)=0 - 45 + 50=5$.  \\
Once again, {\it the central third order operator is self-adjoint} as can be easily seen by proving that the last $5 \rightarrow 3$ operator, obtained in [44] by means of computer algebra, can be chosen to be the transpose of the first $3 \rightarrow 5$ conformal Killing operator, {\it just by changing columns}.\\
This result can also be obtained by using the fact that, when an operator/a system is formally integrable, the order of the generating CC is equal to the number of prolongations needed to get a $2$-acyclic symbol plus $1$ ([34], p 120, [44]). In the present case, neither  ${\hat{g}}_1$ nor ${\hat{g}}_2$ are $2$-acyclic while ${\hat{g}}_3=0$ is trivially involutive, so that $(3-1)+1=3$.  \\

\noindent
$\bullet$ $n=4$: In the classical case, we may proceed as before for exibiting the $20$ components of the second order Riemann operator and the $20$ components of the first order Bianchi operator. \\
{\it The study of the conformal case is much more delicate and still unknown}. Indeed, the symbol ${\hat{g}}_2$ is $2$-acyclic when $n\geq 4$ and $3$-acyclic when $n\geq 5$. Accordingly, the Weyl operator, namely the CC for the conformal Killing operator, is second order like the Riemann operator. However, {\it when} $n=4$ {\it only} ({\it care}), the symbol ${\hat{h}}_2$ of the Weyl system is {\it not} $2$-acyclic while its first prolongation ${\hat{h}}_3$ becomes $2$-acyclic. It follows that {\it the CC for the Weyl operator are second order}, ... and so on. For example, we have the long exact sequence:  \\
\[   0 \rightarrow {\hat{g}}_5 \rightarrow S_5T^*\otimes T \rightarrow S_4T^*\otimes {\hat{F}}_0 \rightarrow S_2T^*\otimes {\hat{F}}_1 \rightarrow {\hat{F}}_2 \rightarrow 0  \]
and deduce that $dim({\hat{F}}_2)= (-0 )+ (56\times 4)-(35\times 9)+(10\times 10)=9$, a result that can be ckecked by computer algebra in a few milliseconds but is still unknown.  \\

We shall finally prove below that the {\it Einstein parametrization} of the stress equations is neither canonical nor minimal in the following diagrams:  \\
\[   \begin{array}{rcccccccccl}
 &4 & \stackrel{Killing}{\longrightarrow} & 10 & \stackrel{Riemann}{\longrightarrow} & 20 & \stackrel{Bianchi}{\longrightarrow} & 20 & \longrightarrow & 6 & \rightarrow 0 \\
  &   &                                                            & \parallel &       & \downarrow  &  & \downarrow &  &   \\
 &    &                                                            &  10     & \stackrel{Einstein}{\longrightarrow}  & 10  & \stackrel{div}{\longrightarrow} & 4    & \rightarrow & 0                      &   \\  
  &   &  &  &  &  &  &  &  \\   
0 \leftarrow & 4 & \stackrel{Cauchy}{\longleftarrow} & 10 & \stackrel{Beltrami}{\longleftarrow} & 20 & \longleftarrow & 20 &  & & \\
                     &           &                                                         & \parallel &    & \uparrow  &  &  &    \\
  &    &      & 10 &  \stackrel{Einstein}{\longleftarrow} & 10 &  &  &  & & 
\end{array}   \]
obtained by using the fact that {\it the Einstein operator is self-adjoint}, where by Einstein operator we mean the linearization of the Einstein equations at the Minkowski metric, the $6$ terms being exchanged between themselves [35,40]. Indeed, setting $E_{ij}=R_{ij}- \frac{1}{2} {\omega}_{ij}tr(R)$ with $tr(R)={\omega}^{ij}R_{ij}$, it is essential to notice that the {\it Ricci operator is not self-adjoint} because we have for example:  \\
\[  {\lambda}^{ij} ({\omega}^{rs}d_{ij}{\Omega}_{rs}) \stackrel{ad}{\longrightarrow} ({\omega}^{rs}d_{ij}{\lambda}^{ij})
{\Omega}_{rs}  \]
and $ad$ provides a term appearing in $- {\omega}_{ij}tr(R)$ but {\it not} in $2R_{ij}$ because we have, as in $(5.1.4)$ of [15]:  \\
\[  tr(\Omega)={\omega}^{rs}{\Omega}_{rs}  \hspace{1cm} \Rightarrow \hspace{1cm}
 tr(R)={\omega}^{rs}d_{rs} tr(\Omega)- d_{rs}{\Omega}^{rs}  \]

The upper $div$ induced by $Bianchi$ has {\it nothing to do} with the lower $Cauchy$ stress equations, contrary to what is still believed today while the $10$ {\it on the right} of the lower diagram has {\it nothing to do} with the perturbation of a metric which is the $10$ {\it on the left} in the upper diagram. It also follows that the Einstein equations in vacuum cannot be parametrized as we have the following diagram of operators recapitulating the five steps of the parametrizability criterion (See [35,37] for more details or [44,56] for a computer algebra exhibition of this result): \\
\[  \begin{array}{rcccl}
  &  &  &\stackrel{Riemann}{ }  & 20   \\
  & &  & \nearrow &    \\
 4 &  \stackrel{Killing}{\longrightarrow} & 10 & \stackrel{Einstein}{\longrightarrow} & 10  \\
  & & & &  \\
 4 & \stackrel{Cauchy}{\longleftarrow} & 10 & \stackrel{Einstein}{\longleftarrow} & 10 
\end{array}  \]

As a byproduct, we are facing {\it only two} possibilities, both leading to a contradiction:  \\
1) If we use the operator $S_2T^* \stackrel{Einstein}{\longrightarrow} S_2T^*$ in the geometrical setting, the $S_2T^*$ on the left has indeed {\it someting to do} with the perturbation of the metric but the $S_2T^*$ on the right has {\it nothing to do} with the stress. \\
2) If we use the adjoint operator ${\wedge}^nT^*\otimes S_2T\stackrel{Einstein}{\longleftarrow} {\wedge}^nT^*\otimes S_2T$ in the physical setting, then ${\wedge}^nT^*\otimes S_2T$ on the left has of course {\it something to do} with the stress but the ${\wedge}^nT^*\otimes S_2T$ on the right has {\it nothing to do} with the perturbation of a metric. \\

These purely mathematical results question the origin and existence of gravitational waves.  \\

We may summarize these results, which do not seem to be known, by the following differential sequences where the order of an operator is written under its arrow:  \\

\noindent 
 $\bullet$ $n=3$: $ \hspace{2cm} 3 \underset{1}{\longrightarrow} 5 \underset{3}{\longrightarrow }5 \underset{1}{ \longrightarrow }3 \rightarrow 0  $  \\
 $\bullet$ $n=4$: $ \hspace{2cm} 4 \underset{1}{\longrightarrow} 9 \underset{2}{\longrightarrow} 10 \underset{2}{\longrightarrow} 9 \underset{1}{\longrightarrow} 4  \rightarrow 0  $ \\                                           
 $\bullet$ $n=5$: $ \hspace{2cm}  5 \underset{1}{\longrightarrow} 14 \underset{2}{\longrightarrow} 35 \underset{1}{\longrightarrow} 35 \underset{2}{\longrightarrow} 14 \underset{1}{\longrightarrow} 5 \rightarrow 0  $\\

\noindent
{\bf THEOREM  5.2}: Recalling that we have $F_1=H^2(g_1)=Z^2(g_1)$ and ${\hat{F}}_1=H^2 ({\hat{g}}_1)\neq Z^2({\hat{g}}_1)$, we have the following commutative and exact "{\it fundamental diagram II} ":  \\
 \[ \begin{array}{rcccccccl}
 & & & & & & & 0 & \\
 & & & & & & & \downarrow & \\

  & & & & & 0& & S_2T^* &  \\
  & & & & & \downarrow & & \downarrow  &  \\
   & & & 0 &\longrightarrow & Z^2(g_1) & \longrightarrow & H^2(g_1)  & \longrightarrow 0  \\
   & & & \downarrow & & \downarrow & &  \downarrow  &  \\
   & 0 &\longrightarrow & T^*\otimes {\hat{g}}_2 & \stackrel{\delta}{\longrightarrow} & Z^2({\hat{g}}_1) & \longrightarrow & H^2({\hat{g}}_1) & \longrightarrow 0  \\
    & & & \downarrow & & \downarrow & & \downarrow     &   \\
 0 \longrightarrow & S_2T^* & \stackrel{\delta}{\longrightarrow}& T^*\otimes T^* &\stackrel{\delta}{\longrightarrow} & {\wedge}^2T^* & \longrightarrow & 0 &   \\
   & & & \downarrow &  & \downarrow & & &  \\
   & & & 0 & & 0 & & &  
  
     \end{array}  \]
\[   \vspace{2mm}\]
The following theorem will provide {\it all} the classical formulas of both Riemannian and conformal geometry in one piece but in a totally unusual framework {\it not depending on any conformal factor}:    \\
 
\noindent
{\bf THEOREM 5.3}: All the short exact sequences of the preceding diagram split in a canonical way, that is in a way compatible with the underlying tensorial properties of the vector bundles involved. \\
\[  \begin{array}{ccl}
T^*\otimes T^* \simeq S_2T^* \oplus {\wedge}^2 T^* &  \Rightarrow & Z^2({\hat{g}}_1)= Z^2(g_1) + \delta (T^*\otimes  {\hat{g}}_2) \simeq Z^2(g_1) \oplus  {\wedge}^2T^*  \\
 & \Rightarrow  & H^2(g_1)\simeq H^2({\hat{g}}_1) \oplus S_2T^* 
 \end{array}  \]
 
\noindent
{\it Proof}: First of all, we recall that:   \\
\[  g_1 =\{{\xi}^k_i \in T^*\otimes T\mid {\omega}_{rj}{\xi}^r_i+{\omega}_{ir}{\xi}^r_j=0 \}\subset {\hat{g}}_1=\{{\xi}^k_i\in T^*\otimes T \mid {\omega}_{rj}{\xi}^r_i+{\omega}_{ir}{\xi}^r_j - \frac{2}{n}{\omega}_{ij}{\xi}^r_r=0\}  \]
\[  \Rightarrow   \hspace{1cm} 0=g_2 \subset {\hat{g}}_2= \{ {\xi}^k_{ij}\in S_2T^*\otimes T \mid  n{\xi}^k_{ij}={\delta}^k_i{\xi}^r_{rj} +{\delta}^k_j{\xi}^r_{ri} - {\omega}_{ij}{\omega}^{ks}{\xi}^r_{rs} \}     \]
Now, if $({\tau}^k_{li,j})\in T^*\otimes {\hat{g}}_2$, then we have:  \\
\[  n {\tau}^k_{li,j}={\delta}^k_l{\tau}^r_{ri,j} + {\delta}^k_i {\tau}^r_{rl,j}-{\omega}_{li}{\omega}^{ks}{\tau}^r_{rs,j}  \] 
 and we may set ${\tau}^r_{ri,j}={\tau}_{i,j}\neq {\tau}_{j,i}$ with $({\tau}_{i,j})\in T^*\otimes T$ and such a formula does not depend on any conformal factor [53]. We have:  \\
\[  \delta ({\tau}^k_{li,j})=({\tau}^k_{li,j} - {\tau}^k_{lj,i})=({\rho}^k_{l,ij}) \in B^2( {\hat{g}}_1)\subset Z^2({\hat{g}}_1)  \]
 with:   \\
 \[  Z^2({\hat{g}}_1)= \{ ({\rho}^k_{l,ij})\in {\wedge}^2T^*\otimes {\hat{g}}_1)\mid \delta ({\rho}^k_{l,ij})=0 \}\Rightarrow {\varphi}_{ij}={\rho}^r_{r,ij}\neq 0 \]
 \[   \delta ({\rho}^k_{l,ji}) = ({\sum}_{(l,i,j)}{\rho}^k_{l,ij}={\rho}^k_{l,ij} + {\rho}^k_{i,jl} + {\rho}^k_{j,li}) \in {\wedge}^3T^*\otimes T  \]
 \noindent
$\bullet$ The splitting of the lower row is obtained by setting $({\tau}_{i,j})\in T^*\otimes T^* \rightarrow (\frac{1}{2}({\tau}_{i,j} + {\tau}_{j,i}))\in S_2T^*$ in such a way that $({\tau}_{i,j}={\tau}_{j,i}={\tau}_{ij})\in S_2T^* \Rightarrow \frac{1}{2}({\tau}_{ij}+{\tau}_{ji})={\tau}_{ij}$. \\
Similarly, $({\varphi}_{ij}= - {\varphi}_{ji})\in {\wedge}^2T^*  \rightarrow (\frac{1}{2}{\varphi}_{ij})\in T^*\otimes T^*$ and 
$ (\frac{1}{2}{\varphi}_{ij} - \frac{1}{2}{\varphi}_{ji})=({\varphi}_{ij}) \in {\wedge}^2T^*$.  \\ 

\noindent
$\bullet$ The most important result is to split the right column. For this, we first need to describe the monomorphism $0 \rightarrow S_2T^* \rightarrow H^2(g_1)$ which is in fact produced by a diagonal north-east snake type chase. Let us choose $({\tau}_{ij}={\tau}_{i,j}={\tau}_{j,i}={\tau}_{ji})\in S_2T^* \subset T^* \otimes T^*$. Then, we may find $({\tau}^k_{li,j})\in T^* \otimes {\hat{g}}_2$ by deciding that ${\tau}^r_{ri,j}={\tau}_{i,j}={\tau}_{j,i}={\tau}^r_{rj,i}$ in $Z^2({\hat{g}}_1)$ and apply 
$\delta$ in order to get ${\rho}^k_{l,ij}={\tau}^k_{li,j} - {\tau}^k_{k,lj,i}$ such that ${\rho}^r_{r,ij}={\varphi}_{ij}=0$ and thus $({\rho}^k_{l,ij})  \in Z^2(g_1)=H^2(g_1)$. We obtain:  \\
\[ \begin{array}{rcl}
n{\rho}^k_{l,ij} & = &  {\delta}^k_l{\tau}^r_{ri,j}-{\delta}^k_l{\tau}^r_{rj,i}+{\delta}^k_i {\tau}^r_{rl,j}-{\delta}^k_j{\tau}^r_{rli}-{\omega }^{ks}({\omega}_{li}{\tau}^r_{rs,j} - {\omega}_{lj}{\tau}^r_{rs,i })   \\
  & =  & ({\delta}^k_i{\tau}_{lj} - {\delta}^k_j{\tau}_{li}) -{\omega}^{ks}({\omega}_{li}{\tau}_{sj} - {\omega}_{lj}{\tau}_{si})                          \\
\end{array}  \]
Contracting in $k$ and $i$ while setting simply $tr(\tau) ={\omega}^{ij}{\tau}_{ij}, tr(\rho)={\omega}^{ij}{\rho}_{ij}$, we get:  \\
\[  n {\rho}_{ij}=n{\tau}_{ij}-{\tau}_{ij}-{\tau}_{ij}+{\omega}_{ij} tr(\tau)=(n-2){\tau}_{ij}+{\omega}_{ij}tr(\tau)=n{\rho}_{ji} \Rightarrow ntr(\rho)=2(n-1)tr(\tau) \]              
Substituting, we finally obtain ${\tau}_{ij}=\frac{n}{(n-2)}{\rho}_{ij} - \frac{n}{2(n-1)(n-2)}{\omega}_{ij}tr(\rho)$ and thus the tricky formula:  \\
\[  {\rho}^k_{l,ij}=\frac{1}{(n-2)}({\delta}^k_i{\rho}_{lj} - {\delta}^k_j{\rho}_{li}) - {\omega}^{ks}({\omega}_{li}{\rho}_{sj}-{\omega}_{lj}{\rho}_{si})) - \frac{1}{(n-1)(n-2)}({\delta}^k_i{\omega}_{lj}-{\delta}^k_j{\omega}_{li})tr(\rho)  \]
Contracting in $k$ and $i$, we check that ${\rho}_{ij}={\rho}_{ij}$ indeed, obtaining therefore the desired canonical lift $H^2(g_1) \rightarrow S_2T^* \rightarrow 0: {\rho}^k_{i,lj} \rightarrow  {\rho}^r_{i,rj}={\rho}_{ij}$. Finally, using Proposition 4.3, the epimorphism $H^2(g_1) \rightarrow H^2({\hat{g}}_1) \rightarrow 0$ is just described by the formula:  \\
\[  {\sigma}^k_{l,ij}={\rho}^k_{l,ij}-\frac{1}{(n-2)}({\delta}^k_i{\rho}_{lj} - {\delta}^k_j{\rho}_{li}-{\omega}^{ks}({\omega}_{li}{\rho}_{sj}-{\omega}_{lj}{\rho}_{si})) + \frac{1}{(n-1)(n-2)}({\delta}^k_i{\omega}_{lj}-{\delta}^k_j{\omega}_{li})tr(\rho)  \]
which is just the way to define the Weyl tensor. We notice that ${\sigma}^r_{r,ij}={\rho}^r_{r,ij}=0$ and ${\sigma}^r_{i,rj}=0$ by using indices or a circular chase showing that $Z^2({\hat{g}}_1)=Z^2(g_1) + \delta (T^*\otimes {\hat{g}}_2)$. This purely algebraic result only depends on the metric $\omega$ and does not depend on any conformal factor. In actual practice, the lift $H^2(g_1) \rightarrow S_2T^*$ is described by ${\rho}^k_{l,ij}\rightarrow {\rho}^r_{i,rj}={\rho}_{ij}={\rho}_{ji}$ but it is not evident at all that the lift $H^2({\hat{g}}_1) \rightarrow H^2(g_1)$ is described by the strict inclusion ${\sigma}^k_{l,ij} \rightarrow {\rho}^k_{l,ij}={\sigma}^k_{l,ij}$ providing a short exact sequence as in Proposition $4.3$ because ${\rho}_{ij}={\rho}^r_{i,rj}={\sigma}^r_{i,rj}=0$ by composition.\\
 \hspace*{12cm}   Q.E.D.   \\

 \noindent
 {\bf COROLLARY 5.4}: When $n\geq 4$, each component of the Weyl tensor is a torsion element killed by the Dalembertian whenever the Einstein equations in vacuum are satisfied by the metric. Hence, there exists a second order  operator ${\cal{Q}}$ such that we have an identity:  \\
 \[   \Box \circ Weyl = {\cal{Q}} \circ Ricci    \]
 
 \noindent
 {\it Proof}: According to Proposition 4.4, each extension module $ext^i(M)$ is a torsion module, $\forall i\geq 1$. It follows that each additional CC in ${\cal{D}}'_1$ which is not already in ${\cal{D}}_1$ is a torsion element as it belongs to this module. One may also notice that:  \\
 \[  rk_D(Einstein)=\frac{n(n+1)}{2}- n=\frac{n(n-1) }{2}  \hspace{3mm}, \hspace{3mm}rk_D(Riemann)=\frac{n(n+1)}{2}- n=\frac{n(n-1)}{2}   \] 
The differential ranks of the Einstein and Riemann operators are thus equal, but {\it this is a pure coincidence} because $rk_D(Einstein)$ has only to do with the $div$ operator induced by contracting the Bianchi identities, while $rk_D(Riemann)$ has only to do with the classical Killing operator and the fact that the corresponding differential module is a torsion module because we have a Lie group of transformations having $n + \frac{n(n-1)}{2}=\frac{n(n+1)}{2}$ parameters (translations + rotations). Hence, as the Riemann operator is a direct sum of the Weyl operator and the Einstein or Ricci operator according to the previous theorem, each component of the Weyl operator must be killed by a certain operator whenever the Einstein or Ricci equations in vacuum are satisfied. A direct tricky computation can be found in ([8], p 206]) and ([18], exercise 7.7]).  \\
\hspace*{12cm}   Q.E.D.   \\

\noindent
{\bf REMARK 5.5}:  In a similar manner, the EM wave equations $\Box F=0$ are easily obtained when the second set of Maxwell equations in vacuum is satisfied, {\it avoiding therefore the Lorenz (no "t") gauge condition for the EM potential}. Indeed, let us start with the Minkowski constitutive law with electric constant ${\epsilon}_0$ and magnetic constant ${\mu}_0$ such that ${\epsilon}_0{\mu}_0 c^2=1$ in vacuum:  \\
\[ {\cal{F}}^{rs}=\frac{1}{{\mu}_0}{\hat{\omega}}^{ri}{\hat{\omega}}^{sj}F_{ij}\sim {\omega}^{ri}{\omega}^{sj} F_{ij}\]
where ${\hat{\omega}}_{ij}={\mid det(\omega)\mid}^{-1/n}{\omega}_{ij}\Rightarrow  \mid det(\hat{\omega})\mid=1$, $F\in {\wedge}^2T^*$ is the EM field and the induction ${\cal{F}}$ is thus a contravariant skewsymmetric $2$-tensor density. From the Maxwell equations we have:
\[  {\partial}_rF_{ij} + {\partial}_iF_{jr} + {\partial}_jF_{ri}=0, \hspace{3mm}{\nabla}^r{\cal{F}}_{ri}=0 \hspace{3mm} \Rightarrow \hspace{3mm}{\nabla}^rF_{ri}=0 \]
\[   \Rightarrow  \hspace{3mm}  \Box F_{ij}={\nabla}^r{\nabla}_rF_{ij}={\nabla}^r ({\nabla}_iF_{rj}- {\nabla}_jF_{ri})=0   \]

\noindent
{\bf REMARK 5.6}: Using Proposition 4.3 and the splittings of Theorem 5.3 for the second column, we obtain the following commutative and exact diagram:  \\
\[  \begin{array}{cccccccl}
  &  &  0  &  &  0  &  &  0  &     \\
  & &  \downarrow & &  \downarrow  &  &  \downarrow  &  \\
  0 &  & 10 & \longrightarrow & 16 & \rightarrow & 6 &  \rightarrow 0  \\
  \downarrow &  &  \downarrow \uparrow&  & \downarrow &  & \parallel &   \\
  10  &  \stackrel{Riemann}{\longrightarrow} & 20 & \stackrel{Bianchi}{\longrightarrow} & 20 & \rightarrow & 6 & \rightarrow 0  \\
\parallel  &  &  \downarrow\uparrow &  & \downarrow &  & \downarrow &   \\
10 & \stackrel{Einstein}{\longrightarrow}&  10 &  \stackrel{div}{\longrightarrow} & 4  & \rightarrow & 0 &  \\
\downarrow &  & \downarrow & & \downarrow & & &  \\
0 & & 0 & & 0 & & & 
\end{array}   \]
It follows that the $10$ components of the Weyl tensor must satisfy a first order linear system with $16$ equations, having $6$ generating first order CC. The differential rank of the corresponding operator is thus equal to $16-6=10$ and such an operator defines a torsion module in which we have to look {\it separately} for each component of the Weyl tensor in order to obtain Corollary 5.4. The situation is similar to that of the Cauchy-Riemann equations when $n=2$. Indeed, any complex transformation $y=f(x)$ must be solution of the (linear) first order system $y^2_2 - y^1_1=0, y^1_2 + y^2_1=0$ of finite Lie equations though we obtain $y^1_{11} +y^1_{22}=0, y^2_{11}+y^2_{22}=0$, that is $y^1$ and $y^2$ are {\it separately} killed by the second order laplacian $\Delta =d_{11}+d_{22}$.  \\

\newpage

\noindent
{\bf 6) CONTACT STRUCTURE}  \\

Changing slightly the notations while setting $\alpha=1,...,p$ and $\bar{\alpha}=\alpha +p=p+1,...,2p$, we may consider the contact $1$-form 
$ \chi=dx^n-{\sum}^p_{\alpha=1}x^{\bar{\alpha}}dx^{\alpha} \Rightarrow \chi\wedge (d\chi)^p=(_1)^{p+1}p!dx^1\wedge ... \wedge dx^n\neq 0$ where the exterior power of $d\chi$ is taken $p$ times. As before, we obtain the injective parametrization:  \\
\[  {\xi}^{\alpha}= - \frac{\partial \phi}{\partial x^{\bar{\alpha}}},\,\,\, {\xi}^{\bar{\alpha}}=\frac{\partial \phi}{\partial x^{\alpha}}+x^{\bar{\alpha}}\frac{\partial \phi}{\partial x^n}, \,\,\, {\xi}^n=\phi - x^{\bar{\beta}}\frac{\partial \phi}{\partial x^{\bar{\beta}}} \, \Rightarrow \,\, \phi=i(\xi)\chi \,\, \Rightarrow \,\, {\cal{L}}(\xi)\chi= \frac{\partial \phi}{\partial x^n}\chi\]
From now on, considering $\chi$ as a $1$-form density as we did before, we may consider $\phi$ as a density section of a vector bundle $E$ with $dim(E)=1$ and we obtain the defining system in the Medolaghi form with $n$ equations:  \\
\[       {\chi}_r(x){\partial}_i{\xi}^r - \frac{1}{p+1}{\chi}_i{\partial}_r{\xi}^r+{\xi}^r{\partial}_r {\chi}_i=0   \]
We have seen that this system is involutive when $n=3$ but let the reader check as a difficult exercise that this system is not even formally 
integrable when $n\geq 5$.  \\

We may define the linear first order operator ${\cal{C}}=A\circ j_1: E \rightarrow T$ and the linear first order operator ${\cal{D}}=B\circ j_1: T \rightarrow F_0$ by the two rows of the following commutative and exact diagram:  \\
\[  \begin{array} {rcccccccc}
0 \rightarrow & Q_2 & \rightarrow & J_2(E) & \stackrel{{\rho}_1(A)}{\rightarrow}& J_1(T) & \stackrel{B}{\rightarrow} & F_0 & 
\rightarrow 0  \\
  & \downarrow &  & \downarrow & & \downarrow & &  \downarrow  \\
0 \rightarrow & Q_1 & \rightarrow & J_1(E) & \stackrel{A}{\rightarrow} & T & \rightarrow & 0 &      
\end{array}   \]
where $Q_1=ker(A)\subset J_1(E)$ and its symbol $K_1\subset T¬*\otimes E$ is easily seen to be involutive with $dim(K_{r+1})=1$. As the parametrizing operator ${\cal{C}}$ is injective with a lift $\xi \rightarrow i(\xi)\chi=\phi$, we obtain $Q^{(1)}_1={\pi}^2_1(Q_2)=0$ and thus $Q_1$ is not formally integrable. However, using Theorem 2.14, we have $Q^{(1)}_{r+1}={\rho}_r(Q^{(1)}_1)=0$ and thus $Q_{r+1}\simeq K_{r+1}\Rightarrow dim(Q_{r+1})=1$. We obtain therefore at once:  \\
\[      dim (F_0)= 1 - (n+1)(n+2)/2 +n(n+1)=n(n-1)/2      \]
and $dim(F_0)=10$ when $n=5$ instead of the $5$ equations we obtained with the $1$-form density $\chi$. Prolonging this diagram $r$-times by induction, we obtain the following commutative diagram:  \\

\[ \begin{array}{rccccccc}
  &  0  &  &  0  &  &  0  &  &  0   \\
  & \downarrow &  & \downarrow &  & \downarrow &  & \downarrow  \\
    0 \rightarrow & K_{r+2} & \rightarrow  & S_{r+2}T^*\otimes E & \stackrel{{\sigma}_{r+1}(A)}{\longrightarrow }  & S_{r+1}T^*\otimes T &  \stackrel{{\sigma}_r(B)}{\longrightarrow}  & S_rT^*\otimes F_0  \\
  & \downarrow &  & \downarrow &  & \downarrow  & & \downarrow \\
 0 \rightarrow & Q_{r+2} & \rightarrow & J_{r+2}(E) & \stackrel{{\rho}_{r+1}(A)}{\longrightarrow} & J_{r+1}(T) & \stackrel{{\rho}_r(B)}{\longrightarrow} & J_r(F_0)  \\
   & \downarrow &  & \downarrow &  & \downarrow  &  & \downarrow   \\
 0 \rightarrow & Q_{r+1} & \rightarrow & J_{r+1}(E) & \stackrel{{\rho}_r(A)}{\longrightarrow} & J_r(T) & \stackrel{{\rho}_{r-1}(B)}{\longrightarrow} & J_{r-1}(F_0)  \\
    & &  & \downarrow &  & \downarrow &  & \downarrow    \\
  &    &  &  0  &  &  0  &  &  0  
\end{array}  \]
Chasing in this diagram while cutting it in the middle by setting:  \\
\[ R_{r+1}=im({\rho}_{r+1}(A))\subseteq ker({\rho}_r(B))={\rho}_r(R_1)\Rightarrow {\pi}^{r+1}_r(R_{r+1})=R_r\Rightarrow g_{r+1}\subseteq {\rho}_r(g_1)  \]
we obtain successively:  \\
\[  \begin{array}{rcl}
dim(g_{r+1})&=&dim(R_{r+1})-dim(R_r) \\
& = & (dim(J_{r+2}(E))-dim(Q_{r+2}))-(dim(J_{r+1}(E)-dim(Q_{r+1}))\\
           &= & dim(J_{r+2}(E))-dim(J_{r+1}(E))\\
           &=&dim(S_{r+2}T^*)=(r+n+1)!/(r+2)!(n-1)!\\
    &   &     \\
dim({\rho}_r(g_1)) & = &dim(ker({\sigma}_r(B)))\\
&=&dim(im({\sigma}_{r+1}(A)))+1\\
&=&(dim(S_{r+2}T^*)-1)+1\\
&=&dim(S_{r+2}T^*)=dim(g_{r+1}) 
\end{array}  \]
It follows that $g_{r+1}={\rho}_r(g_1)$ and thus $R_{r+1}={\rho}_r(R_1)$ by induction on $r$. Hence $R_1$ is an involutive first order system because it is already formally integrable and its symbol $g_1$ is involutuve because $K_1$ is involutive. Extending step by step the previous diagram on the right, we obtain:  \\

\noindent
{\bf THEOREM 6.1}: We have the locally exact Janet sequence where all the operators are first order and involutive but ${\cal{C}}$ which is first order but not formally integrable: \\
\[  0 \rightarrow E \stackrel{{\cal{C}}}{\longrightarrow}T \stackrel{{\cal{D}}}{\longrightarrow} F_0 \stackrel{{\cal{D}}_1}{\longrightarrow}   F_1 \stackrel{{\cal{D}}_2}{\longrightarrow} ... \stackrel{{\cal{D}}_{n-2}}{\longrightarrow} F_{n-2}\rightarrow 0  \]

\noindent
{\it Proof}: We recall that a differential sequence is locally exact, like the Poincar\'{e} sequence, if any (local) section of $F_r$ killed by ${\cal{D}}_{r+1}$ is the image of a (local) section of $F_{r-1}$ by ${\cal{D}}_r$. To prove this result we just need to apply the Spencer operator $D$ to the middle row of the preceding diagram when $r$ is large enough. As it is known that all the resulting vertical Spencer sequences are locally exact (See [31,34,36,52] for more details), then the commutative diagram thus obtained is exact but perhaps the first purely algebraic column on the left which is an induced $\delta$-sequence, exact because $K_1$ is involutive. For helping the reader we provide the upper part of this diagram:  \\
\[  \begin{array}{rcccccccl}
  &  &  &  0  &  & 0 & & 0 &  \\
     &  &  & \downarrow & & \downarrow & & \downarrow &   \\
  & 0 & \rightarrow & E & \stackrel{{\cal{C}}}{\longrightarrow}&T &\stackrel{{\cal{D}}}{\longrightarrow} & F_0 & \rightarrow ... \\
  & \downarrow & & \hspace{7mm}\downarrow j_{r+2} & & \hspace{7mm}\downarrow j_{r+1} & & \hspace{4mm}\downarrow j_r & \\
0 \rightarrow &K_{r+2}& \rightarrow &J_{r+2}(E)& \longrightarrow &J_{r+1}(T)& \longrightarrow & J_r(F_0)& \rightarrow ...\\
  &  \hspace{5mm}\downarrow -\delta& & \hspace{4mm}\downarrow D& & \hspace{4mm}\downarrow D & &  \hspace{5mm} \downarrow D &  \\
0 \rightarrow &T^*\otimes K_{r+1}& \rightarrow &T^*\otimes J_{r+1}(E) &\rightarrow &T^*\otimes J_r(T) &\rightarrow & T^*\otimes J_{r-1}(F_0) &\rightarrow ...
\end{array}    \]
Comparing to the Poincar\'{e} sequence, we get $dim(F_r)= n!/(r+2)!(n-r-2)!$ and it remains to find the geometric object providing ${\cal{D}}$. For this, we may introduce the dual density $\omega=({\omega}^{ij})$ of $\alpha =({\alpha}_{k_1, ...k_{n-2}})=\chi\wedge (d\chi)^{p-1}$ in a symbolic way by introducing $\epsilon \in {\wedge}^nT$ with $\mid {\epsilon}^{i_1,...,i_n}\mid=1$ and set ${\omega}^{ij}={\epsilon}^{ijk_1,...,k_{n-2}}{\alpha}_{k_1,...,k_{n-2}}$. Using jet notations, we have the {\it nonlinear system of finite Lie equations} in {\it Lie form}:\\
\[   {\omega}^{kl}(y)\frac{\partial x^i}{\partial y^k}\frac{\partial x^j}{\partial y^l} (\frac{\partial (y^1,...,y^n)}{\partial (x^1, ...,x^n)})^
{ - \frac{1}{p+1}} = {\omega}^{ij}(x)   \]
We obtain therefore by linearization the involutive system $R_1$ in of general infinitesimal Lie equations 
in {\it Medolaghi form}:  \\
\[   - {\omega}^{rj}(x){\partial}_r{\xi}^i - {\omega}^{ir}(x){\partial}_r{\xi}^j - \frac{1}{p+1}{\omega}^{ij}(x){\partial}_r{\xi}^r + {\xi}^r{\partial}_r{\omega}^{ij}(x)=0  \]
The Vessiot structure equations involve only one constant and become:ÊÊ\\
\[ {\chi}_i({\partial}_j{\chi}_k - {\partial}_k{\chi}_j) + {\chi}_j({\partial}_k{\chi}_i - {\partial}_i{\chi}_k) + {\chi}_k({\partial}_i{\chi}_j - {\partial}_j{\chi}_i) = c \,\,\,{\sigma}_{ijk}  \] 
where the $3$-form density $\sigma =({\sigma}_{ijk})$ is the dual of $(\omega)^{p-1}$. Finally, as $\chi$ is proportional to the dual density of 
$(\omega)^p$, it may be expressed rationally in terms of $\omega$. Linearizing as we did in Section 2 and with the Riemann tensor, we obtain 
${\cal{D}}_1$ with $dim(F_1)= n(n-1)(n-2)/6$ in a coherent way.  \\
\hspace*{12cm}    Q.E.D.   \\

\newpage

We use the previous results in order to revisit the Hamilton-Jacobi equation and prove the need to use differential algebra for studying the nonlinear systems involved. This is a difficult problem indeed, in the sense that no classical approach by means of exterior calculus can be adapted as formal integrability or involution become crucial tools that cannot be avoided. Let $z=f(t,x)$ be a solution of the non-linear PD equation $z_t+H(t,x,z,z_x)=0$ written with jet notations for the single unknown $z$. When dealing with applications, $t$ will be {\it time}, $x$ will be {\it space}, $z$ will be the {\it action} and, as usual, we shall set $p=z_x$ for the {\it momentum}. It is important to notice that, in this general setting, $H(t,x,z,p)$ {\it cannot be called Hamiltonian as it involves} $z$ (See [32] and in particular [33], p 506 for more details):    \\

\noindent
{\bf DEFINITION 6.2}: A {\it complete integral} $z=f(t,x;a,b)$ is a family of  solutions depending on two constant parameters $(a,b)$ in such a way that the Jacobian condition $\partial (z,p)/\partial (a,b)\neq 0$ whenever $p={\partial}_xf(t,x;a,b)$. Using the implicit function theorem, we may set\\

\noindent
{\bf THEOREM 6.3}: The search for a complete integral of the PD equation:  \\
 \[z_t+H(t,x,z,z_x)=0 \] 
is equivalent to the search for a {\it single} solution of the automorphic system ${\cal{A}}_1$ with $n=4,m=3$, obtained by eliminating $\rho(t,x,z,p)$ in the Pfaffian system:  \\
\[    dz-pdx+H(t,x,z,p) dt = \rho (dZ-PdX )  \]  
The corresponding Lie pseudogroup is the pseudogroup $\Gamma$ of {\it contact transformations} of $(X,Z,P)$ that reproduces the contact $1$-form $dZ-PdX$ up to a function factor.  \\

\noindent
{\it Proof}: If $z=f(t,x; a,b)$ is a complete integral, we have:  \\
\[   dz-pdx+H(t,x,z,p)dt= \frac{\partial f}{\partial a}da +\frac{\partial f}{\partial b} db   \]
Using the implicit function theorem and the Jacobian condition, we may set:  \\
\[  a =X(t,x,z,p), \, b=Z(t,x,z,p)  \Rightarrow \rho(t,x,z,p)=\frac{\partial f}{\partial b}, \,
 P(t,x,z,p)=\frac{\partial f}{\partial a}/\frac{\partial f}{\partial b}   \]
The converse is left to the reader.  \\

For another solution denoted wit a "bar", we have:  \\
\[ dz-pdx+H(t,x,z,p) dt = \bar{\rho} (d\bar{Z}-\bar{P} d\bar{X} )\,\,\Rightarrow \,\,d\bar{Z}-\bar{P} d\bar{X} = \frac{\rho}{\bar{\rho}}(dZ-PdX) \]
Closing this system, we obtain at once:  \\
\[    d\bar{X}\wedge d\bar{Z}\wedge d\bar{P}= (\frac{\rho}{\bar{\rho}})^2 dX\wedge dZ \wedge dP    \]
Closing again, we discover that $\rho/\bar{\rho}$ is in fact a function of $(X,Z,P)$, a result bringing the Lie pseudogroup of contact transformations and showing that no restriction must be imposed to $H(t,x,z,p)$.  \\
\hspace*{12cm}     Q.E.D.  \\

It is quite more dificult to exhibit the equations of the above automorphic sytem and the corresponding equations of the Lie pseudogroup $\Gamma$ in Lie form or even as involutive systems of PD equations. From what has been said, we obtain {\it at least}:   \\
\[  \frac{\frac{\partial \bar{Z}}{\partial X} - \bar{P}\frac{\partial \bar{X}}{\partial X}}{ \frac{\partial \bar{Z}}{\partial Z} - \bar{P}\frac{\partial \bar{X}}{\partial Z}}= - P , 
    \frac{\frac{\partial \bar{Z}}{\partial P} - \bar{P}\frac{\partial \bar{X}}{\partial P}}{ \frac{\partial \bar{Z}}{\partial Z} - \bar{P}\frac{\partial \bar{X}}{\partial Z}}= 0     \Rightarrow  \frac{\partial \bar{Z}}{\partial P} - \bar{P}\frac{\partial \bar{X}}{\partial P}=0  \]
 for defining ${\cal{R}}_1$, that is to say:  \\
 \[    \frac{ \partial \bar{Z}}{\partial X} - \bar{P} \frac{\partial \bar{X}}{\partial X} + P  (\frac{\partial \bar{Z}}{\partial Z} - \bar{P}\frac{\partial \bar{X}}{\partial Z})=0 , \hspace{1cm} \frac{\partial \bar{Z}}{\partial P} - \bar{P}\frac{\partial \bar{X}}{\partial P}=0   \]
Using now letters $(x,z,p)$ instead of the capital letters $(X,Z,P)$ and $(\xi, \eta, \zeta)$ for the corresponding vertical bundles, we obtain by linearization the system of first order infinitesimal Lie equations:  \\
\[  \frac{\partial \xi}{\partial x}- p \frac{ \partial \eta}{\partial  x} - \zeta + p ( \frac{\partial \xi}{\partial z}- p\frac{ \partial \eta}{\partial  z})=0, \,\, 
\frac{\partial \xi}{ \partial p }- p \frac{\partial \eta}{\partial p}=0  \]
This system is not involutive as it is not even formally integrable. Using crossed derivatives in $x/p$, we obtain the {\it only new first order} equation: \\
\[    \frac{\partial \eta}{\partial x} - \frac{\partial \xi}{\partial z}+ \frac{\partial \zeta }{\partial p} + 2p \frac{\partial \eta}{\partial z}=0  \]
and the resulting system ${\cal{R}}^{(1)}_1$ is involutive with two equations of class $x$ solved with respect to $(\frac{\partial \xi}{\partial x}, \frac{\partial \eta}{\partial x})$ and one equation of class $p$ solved with respecto $\frac{\partial \xi}{\partial p}$, that is $dim_Y({\cal{R}}^{(1)}_1)= (3+ 3 \times 3) - 3=9$. Accordingly, the non-linear system of Lie equations {\it must} become involutive by adding {\it only one equation in Lie form}, namely:   \\
\[   \frac{  \frac{\partial (\bar{Z},\bar{X}, \bar{P})}{\partial (Z,X,P)}}{ ( \frac{\partial \bar{Z}}{\partial Z} - \bar{P} \frac{\partial \bar{X}}{\partial X})^2} = 1   \]
and its linearization just provides:  \\
\[  \frac{\partial \eta}{\partial x} + \frac{\partial \xi}{\partial z}+ \frac{\partial \zeta }{\partial p}= 2  (\frac{\partial \xi}{\partial z} - p \frac{\partial \eta}{\partial z})  \]
that is exactly the previous equation. \\                            
Coming back to the original system and notations, we may suppose $\frac{\partial Z}{\partial z} -P\frac{\partial X}{\partial z}\neq 0$ and introduce the $7=3+4$ equations: \\
\[ \frac{\partial Z}{\partial x} - P\frac{\partial X}{\partial x} + p ( \frac{\partial Z}{\partial z} - P\frac{\partial X}{\partial z})=0, 
   \frac{\partial Z}{\partial t} - P\frac{\partial X}{\partial t} - H ( \frac{\partial Z}{\partial z} - P\frac{\partial X}{\partial z})=0,
   \frac{\partial Z}{\partial p} -  P \frac{\partial X}{\partial p}=0  \]
\[   \frac{\partial (Z,X,P)}{\partial (z,x,p)} - (\frac{\partial Z}{\partial z} - P\frac{\partial X}{\partial z})^2=0,
     \frac{\partial (Z,X,P)}{\partial (z,p,t)} - \frac{\partial H}{\partial p} (\frac{\partial Z}{\partial z} - P\frac{\partial X}{\partial z})^2=0, ...  \]  
                                
Starting now, the next results {\it canot} be obtained by exterior calculus and are therefore not known. Indeed, developping the $ 3 \times 3$ Jacobian determinant, the fourth equation provided can be written as:  \\
\[  \frac{\partial Z}{\partial x}. \frac{\partial (X,P)}{\partial (x,p)} - \frac{\partial Z}{\partial x} . \frac{\partial ( (X,P)}{\partial (z,p)} + \frac{\partial Z}{\partial p}. \frac{\partial (X,P)}{\partial (z,x)} - (\frac{\partial Z}{\partial z} - P\frac{\partial X}{\partial z})^2=0  \]
Using the previous equations in order to eliminate $\frac{\partial Z}{\partial x}$ and $\frac{\partial Z}{\partial p}$, we obtain:  \\
\[  \frac{\partial Z}{\partial x}. \frac{\partial (X,P)}{\partial (x,p)} + p(\frac{\partial Z}{\partial z}-P \frac{\partial X}{\partial z}) . \frac{\partial ( (X,P)}{\partial (z,p)}  - P\frac{\partial X}{\partial x}. \frac{\partial (X,P)}{\partial (z,p)}       + P \frac{\partial X}{\partial p}. \frac{\partial (X,P)}{\partial (z,x)}  =\] 
\[    (\frac{\partial Z}{\partial z} - P \frac{\partial X}{\partial z})( \frac{\partial (X,P)}{\partial (x,p)} + p \frac{\partial (X,P)}{\partial (z,p)})   = (\frac{\partial Z}{\partial z} - P\frac{\partial X}{\partial z})^2  \]
and thus:  \\
\[ \frac{\partial (X,P)}{\partial (x,p)} + p \frac{\partial (X,P)}{\partial (z,p)})  -  (\frac{\partial Z}{\partial z} - P\frac{\partial X}{\partial z}) =0   \]
which is nothing else than the first order equation that can be obtained from the first and third among the previous $7$ equations by using crossed derivatives in $x/p$. It follows that ${\cal{A}}^{(1)}_1$ may be defined by $6$ equations {\it only} and we have thus 
$dim_X ({\cal{A}}^{(1)}_1)= (3 + 4 \times 3)- 6=9$. This result proves that the involutive system ${\cal{A}}^{(1)}_1$ is an automorphic system for the involutive Lie groupoid ${\cal{R}}^{(1)}_1$. \\

\newpage

We finally show the link which is existing with differential algebra and the differential Galois theory because the Lie pseudogroup of contact transformations is an algebraic Lie pseudogroup. For this, using jet notations, let us consider the chain of strict inclusions of differential fields:  \\
\[ K= \mathbb{Q}< \frac{Z_x-PX_x}{Z_z-PX_z}, \,  \frac{Z_t-PX_t}{Z_z-PX_z}, \, \frac{Z_p-PX_p}{Z_z-PX_z}> , \,\, L=\mathbb{Q}< X,Z,P> \,\,
\Rightarrow \,\,  \mathbb{Q} \subset K \subset L \]
Using the chain rule for derivatives, we let the reader prove as an exercise that each fraction is a differential invariant for the Lie pseudogroup 
$\Gamma$ of contact transformations. Accordigly, $L/K$ is a {\it differential automorphic extension} in the sense that the corresponding infinite dimensional model differential variety is a {\it principal homogeneous space} (PHS) for $\Gamma$. It is not so evident that:  \\
\[    \frac{\partial(X,Z,P)}{\partial(x,z,p)} / (\frac{\partial Z}{\partial z}-P\frac{\partial X}{\partial z})^2 \in K  \]
because it is also a differential invariant of $\Gamma$. The intermediate differential field $K \subset K' \subset L$ with $K'=K<Z_z-PX_z> $ is the differential field of invariants of the Lie subpseudogroup ${\Gamma}'\subset \Gamma$ of {\it strict} or {\it unimodular} contact transformations preserving the contact form $dZ-PdX$ and thus the volume $3$-form $dZ\wedge dX \wedge dP$. We let the reader adapt the previous results to this particular case.  \\

\newpage

\noindent
{\bf 7) CONCLUSION}  \\

Whenever $R_q\subseteq J_q(E)$ is an involutive system of order $q$ on $E$, we may define the {\it Janet bundles} $F_r$ for $r=0,1,...,n$ by the short exact sequences:  \\
\[0 \rightarrow {\wedge}^rT^*\otimes R_q+\delta ({\wedge}^{r-1}T^*\otimes S_{q+1}T^*\otimes E)\rightarrow {\wedge}^rT^*\otimes J_q(E) \rightarrow F_r \rightarrow 0  \]
We may pick up a section of $F_r$, lift it up to a section of ${\wedge}^rT^*\otimes J_q(E)$ that we may lift up to a section of ${\wedge}^rT^*\otimes J_{q+1}(E)$ and apply $D$ in order to get a section of ${\wedge}^{r+1}T^*\otimes J_q(E) $ that we may project onto a section of $F_{r+1}$ in order to construct an operator ${\cal{D}}_{r+1}:F_r\rightarrow F_{r+1}$ generating the CC of ${\cal{D}}_r$ in the canonical {\it linear Janet sequence} ([34], p 145):  \\
\[  0 \longrightarrow  \Theta \longrightarrow E \stackrel{\cal{D}}{\longrightarrow} F_0 \stackrel{{\cal{D}}_1}{\longrightarrow}F_1 \stackrel{{\cal{D}}_2}{\longrightarrow} ... \stackrel{{\cal{D}}_n}{\longrightarrow} F_n \longrightarrow 0   \]
If we have two involutive systems $R_q \subset {\hat{R}}_q \subset J_q(E)$, {\it the Janet sequence for} $R_q$ {\it projects onto the Janet sequence for} ${\hat{R}}_q$ and we may define inductively {\it canonical epimorphisms} $F_r \rightarrow {\hat{F}}_r \rightarrow 0$ for 
$r=0, 1,...,n$ by comparing the previous sequences for $R_q$ and ${\hat{R}}_q$.  \\ 
A similar procedure can also be obtained if we define the Spencer bundles $C_r$ for $r=0,1,...,n$ by the short exact sequences:  \\
\[ 0 \rightarrow  \delta ({\wedge}^{r-1}T^*\otimes g_{q+1} )  \rightarrow  {\wedge}^rT^*\otimes R_q  \rightarrow  C_r  \rightarrow 0 \]
We may pick up a section of $C_r$, lift it to a section of ${\wedge}^rT^*\otimes R_q$, lift it up to a section of ${\wedge}^rT^*\otimes R_{q+1}$ and apply $D$ in order to construct a section of ${\wedge}^{r+1}\otimes R_q$ that we may project to $C_{r+1}$ in order to construct an operator $D_{r+1}:C_r \rightarrow C_{r+1}$ generating the CC of $D_r$ in the canonical {\it linear Spencer sequence} which is {\it another completely different resolution} of the set $\Theta$ of (formal) solutions of $R_q$:  \\
\[    0 \longrightarrow \Theta \stackrel{j_q}{\longrightarrow} C_0 \stackrel{D_1}{\longrightarrow} C_1 \stackrel{D_2}{\longrightarrow} C_2 \stackrel{D_3}{\longrightarrow} ... \stackrel{D_n}{\longrightarrow} C_n\longrightarrow 0  \]
However, if we have two systems as above, {\it the Spencer sequence for} $R_q$ {\it is now contained into the Spencer sequence for} 
${\hat{R}}_q$ and we may construct inductively {\it canonical monomorphisms} $0\rightarrow C_r \rightarrow {\hat{C}}_r$ for $r=0,1,...,n$ by comparing the previous sequences for $R_q$ and ${\hat{R}}_q$.   \\
When dealing with applications, we have set $E=T$ and considered systems of finite type Lie equations determined by Lie groups of transformations and $ad({\cal{D}}_r)$ generates the CC of $ad({\cal{D}}_{r+1})$ while $ad(D_r)$ generates the CC of $ad(D_{r+1})$. 
We have obtained in particular $C_r={\wedge}^rT^*\otimes R_q \subset {\wedge}^rT^*\otimes {\hat{R}}_q ={\hat{C}}_r$ when comparing the classical and conformal Killing systems, but {\it these bundles have never been used in physics}. Therefore, instead of the classical Killing system $R_2\subset J_2(T)$ defined by $\Omega \equiv {\cal{L}}(\xi)\omega=0$ {\it and} $\Gamma\equiv {\cal{L}}(\xi)\gamma=0$ or the conformal Killing system ${\hat{R}}_2\subset J_2(T)$ defined by $\Omega\equiv {\cal{L}}(\xi)\omega=A(x)\omega$ and ${\Gamma} \equiv {\cal{L}}(\xi)\gamma= ({\delta}^k_iA_j(x) +{\delta} ^k_j A_i(x) -{\omega}_{ij}{\omega}^{ks}A_s(x)) \in S_2T^*\otimes T$, we may introduce the {\it intermediate differential system} ${\tilde{R}}_2 \subset J_2(T)$ defined by ${\cal{L}}(\xi)\omega=A\omega$ with $A=cst$ and $\Gamma \equiv {\cal{L}}(\xi)\gamma=0 $, for the 
{\it Weyl group} obtained by adding the only dilatation with infinitesimal generator $x^i{\partial}_i$ to the Poincar\'e group. We have $R_1\subset {\tilde{R}}_1={\hat{R}}_1$ but the strict inclusions $R_2 \subset {\tilde{R}}_2 \subset {\hat{R}}_2$ and we discover {\it exactly} the group scheme used through this paper, both with the need to {\it shift by one step to the left} the physical interpretation of the various differential sequences used. Indeed, as ${\hat{g}}_2\simeq T^*$, the first Spencer operator ${\hat{R}}_2\stackrel{D_1}{\longrightarrow} T^*\otimes {\hat{R}}_2$ is induced by the usual Spencer operator ${\hat{R}}_3 \stackrel{D}{\longrightarrow} T^*\otimes {\hat{R}}_2:(0,0,{\xi}^r_{rj},{\xi}^r_{rij}=0) \rightarrow (0,{\partial}_i0-{\xi}^r_{ri}, {\partial}_i{\xi}^r_{rj}- 0)$ and thus projects by cokernel onto the induced operator $T^* \rightarrow T^*\otimes T^*$. Composing with $\delta$, it projects therefore onto $T^*\stackrel{d}{\rightarrow} {\wedge}^2T^*:A \rightarrow dA=F$ as in EM and so on by using the fact that $D_1$ 
{\it and} $d$ {\it are both involutive} or the composite epimorphisms ${\hat{C}}_r \rightarrow {\hat{C}}_r/{\tilde{C}}_r\simeq {\wedge}^rT^*\otimes ({\hat{R}}_2/{\tilde{R}}_2) \simeq {\wedge}^rT^*\otimes {\hat{g}}_2\simeq {\wedge}^rT^*\otimes T^*\stackrel{\delta}{\longrightarrow}{\wedge}^{r+1}T^*$. The main result we have obtained is thus to be able to increase the order and dimension of the underlying jet bundles and groups, proving therefore that any $1$-form with value in the second order jets ${\hat{g}}_2$ ({\it elations}) of the conformal Killing system (conformal group) can be decomposed uniquely into the direct sum $(R,F)$ where $R$ is a section of the {\it Ricci bundle} $S_2T^*$ and the EM field $F$ is a section of ${\wedge}^2T^*$ as in [41,42](Compare to [55]).  \\   
{\it The mathematical structures of electromagnetism and gravitation only depend on second order jets}.\\

\newpage

\noindent
{\bf REFERENCES}  \\

\noindent
[1] Adler, F.W.: \"{U}ber die Mach-Lippmannsche Analogie zum zweiten Hauptsatz, Anna. Phys. Chemie, 22, 578-594 (1907).  \\
\noindent
[2] Airy, G.B.:  On the Strains in the Interior of Beams, Phil. Trans. Roy. Soc.London, 153, 1863, 49-80 (1863).  \\  
\noindent
[3] Arnold, V.: M\'{e}thodes Math\'{e}matiques de la M\'{e}canique Classique, Appendice 2 (G\'{e}od\'{e}siques des m\'{e}triques invariantes \`{a} gauche sur des groupes de Lie et hydrodynamique des fluides parfaits), MIR, Moscow (1974,1976). \\
\noindent
[4] Assem, I.: Alg\`ebres et Modules, Masson, Paris (1997).  \\
\noindent
[5] Beltrami, E.: Osservazioni sulla Nota Precedente, Atti Reale Accad. Naz. Lincei Rend., 5, 141-142 (1892).  \\
\noindent
[6] Birkhoff, G.: Hydrodynamics, Princeton University Press (1954).  \\
\noindent
[7] Bjork, J.E. (1993) Analytic D-Modules and Applications, Kluwer (1993).  \\ 
\noindent
[8] Bourbaki, N.: Alg\`{e}bre, Ch. 10, Alg\`{e}bre Homologique, Masson, Paris (1980). \\
\noindent
[9] de Broglie, L.: Thermodynamique de la Particule isol\'{e}e, Gauthiers-Villars, Pris 1964).  \\
\noindent
[10] Choquet-Bruhat, Y.: Introduction to General Relativity, Black Holes and Cosmology, Oxford University Press (2015).  \\
\noindent
[11] Chyzak, F.,Quadrat, A., Robertz, D.:  Effective algorithms for parametrizing linear control systems over Ore algebras,
Appl. Algebra Engrg. Comm. Comput., 16, 319-376, 2005. \\
\noindent
[12] Chyzak, F., Quadrat, A., Robertz, D.: {\sc OreModules}: A symbolic package for the study of multidimensional linear systems,
Springer, Lecture Notes in Control and Inform. Sci., 352, 233-264, 2007.\\
http://wwwb.math.rwth-aachen.de/OreModules  \\
\noindent
[13] Cosserat, E., \& Cosserat, F.: Th\'{e}orie des Corps D\'{e}formables, Hermann, Paris, 1909.\\
\noindent
[14] Eisenhart, L.P.: Riemannian Geometry, Princeton University Press, Princeton (1926).  \\
\noindent
[15] Foster, J., Nightingale, J.D.: A Short Course in General relativity, Longman (1979).  \\
\noindent
[16] Gr\"{o}bner, W.: \"{U}ber die Algebraischen Eigenschaften der Integrale von Linearen Differentialgleichungen mit Konstanten Koeffizienten, Monatsh. der Math., 47, 247-284 (1939).\\
\noindent
[17] Hu,S.-T.: Introduction to Homological Algebra, Holden-Day (1968).  \\
\noindent
[18] Hughston, L.P., Tod, K.P.: An Introduction to General Relativity, London Math. Soc. Students Texts 5, Cambridge University Press 
(1990). \\
\noindent
[19] Janet, M.: Sur les Syst\`{e}mes aux D\'{e}riv\'{e}es Partielles, Journal de Math., 8, 65-151 (1920). \\
\noindent 
[20] Kashiwara, M.: Algebraic Study of Systems of Partial Differential Equations, M\'{e}moires de la Soci\'{e}t\'{e} 
Math\'{e}matique de France, 63 (1995) (Transl. from Japanese of his 1970 MasterÕs Thesis).  \\
\noindent
[21] Kolchin, E.R.: Differential Algebra and Algebraic groups, Academic Press, New York (1973).  \\
\noindent
[22] Kumpera, A., \& Spencer, D.C.: Lie Equations, Ann. Math. Studies 73, Princeton University Press, Princeton (1972).\\
\noindent
[23] Kunz, E.: Introduction to Commutative Algebra and Algebraic Geometry, BirkhaŸser (1985).  \\
\noindent
[24] Lippmann, G.: Extension du Principe de S. Carnot \`{a} la Th\'{e}orie des P\'{e}nom\`{e}nes \'{e}lectriques, C. R. Acad/ Sc. Paris, 82, 1425-1428 (1876).  \\
\noindent
[25] Lippmann, G.: \"{U}ber die Analogie zwischen Absoluter Temperatur un Elektrischem Potential, Ann. Phys. Chem., 23, 
994-996 (1907).  \\
\noindent
[26]  Macaulay, F.S.: The Algebraic Theory of Modular Systems, Cambridge (1916).  \\
\noindent
[27] Mach, E.: Die Geschichte und die Wurzel des Satzes von der Erhaltung der Arbeit, p 54, Prag: Calve (1872).  \\
\noindent
[28]ÊMach, E.: Prinzipien der W\"{a}rmelehre, 2, Aufl., p 330, Leipzig: J.A. Barth (1900).  \\
\noindent
[29] Maxwell, J.C.: On Reciprocal Figures, Frames and Diagrams of Forces, Trans. Roy. Soc. Ediinburgh, 26, 1-40 (1870).  \\
\noindent
[30] Morera, G.: Soluzione Generale della Equazioni Indefinite dellÕEquilibrio di un Corpo Continuo, Atti. Reale. Accad. dei Lincei, 1, 137-141+233(1892).  \\
\noindent
[31] Nordstr\"{o}m, G.: Einstein's Theory of Gravitation and Herglotz's Mechanics of Continua, Proc. Kon. Ned. Akad. Wet., 19, 884-891 (1917). \\
\noindent
[32] Northcott, D.G.: An Introduction to Homological Algebra, Cambridge university Press (1966).  \\
\noindent
[33] Northcott, D.G.: Lessons on Rings Modules and Multiplicities, Cambridge University Press (1968).  \\
\noindent
[34] Oberst, U.: Multidimensional Constant Linear Systems, Acta Appl. Math., 20, 1-175 (1990).  \\ 
\noindent
[35] Oberst, U.: The Computation of Purity Filtrations over Commutative Noetherian Rings of Operators and their Applications to Behaviours, Multidim. Syst. Sign. Process. (MSSP) 26, 389-404 (2013).  \\
http://dx.doi.org/10.1007/s11045-013-0253-4   \\
\noindent
[36] Ougarov, V.: Th\'{e}orie de la Relativit\'{e} Restreinte, MIR, Moscow, 1969, (french translation, 1979).\\
\noindent
[37] Poincar\'{e}, H.: Sur une Forme Nouvelle des Equations de la M\'{e}canique, C. R. Acad\'{e}mie des Sciences Paris, 132 (7) (1901) 369-371.  \\
\noindent
[38] Pommaret, J.-F.: Systems of Partial Differential Equations and Lie Pseudogroups, Gordon and Breach, New York, 1978; Russian translation: MIR, Moscow, 1983.\\
\noindent
[39] Pommaret, J.-F.: Differential Galois Theory, Gordon and Breach, New York, 1983.\\
\noindent
[40] Pommaret, J.-F.: Lie Pseudogroups and Mechanics, Gordon and Breach, New York, 1988.\\
\noindent
[41] Pommaret, J.-F.: Partial Differential Equations and Group Theory, Kluwer, 1994.\\
http://dx.doi.org/10.1007/978-94-017-2539-2    \\
\noindent
[42] Pommaret, J.-F.: Fran\c{c}ois Cosserat and the Secret of the Mathematical Theory of Elasticity, Annales des Ponts et Chauss\'ees, 82, 59-66 (1997) (Translation by D.H. Delphenich).  \\
\noindent
[43] Pommaret, J.-F.: Group Interpretation of Coupling Phenomena, Acta Mechanica, 149 (2001) 23-39.\\
http://dx.doi.org/10.1007/BF01261661  \\
\noindent
[44] Pommaret, J.-F.: Partial Differential Control Theory, Kluwer, Dordrecht, 2001.\\
\noindent
[45] POMMARET, J.-F.: Algebraic Analysis of Control Systems Defined by Partial Differential Equations, in "Advanced Topics in Control Systems Theory", Springer, Lecture Notes in Control and Information Sciences 311 (2005) Chapter 5, pp. 155-223.\\
\noindent
[46] Pommaret, J.-F.: Arnold's Hydrodynamics Revisited, AJSE-mathŽmatiques, 1, 1, 2009, pp. 157-174.  \\
\noindent
[47] Pommaret, J.-F.: Parametrization of Cosserat Equations, Acta Mechanica, 215 (2010) 43-55.\\
http://dx.doi.org/10.1007/s00707-010-0292-y  \\
\noindent
[48] Pommaret, J.-F.: Macaulay Inverse Systems revisited, Journal of Symbolic Computation, 46, 1049-1069 (2011). \\
\noindent
[49] Pommaret, J.-F.: Spencer Operator and Applications: From Continuum Mechanics to Mathematical Physics, in "Continuum Mechanics-Progress in Fundamentals and Engineering Applications", Dr. Yong Gan (Ed.), ISBN: 978-953-51-0447--6, InTech, 2012, Available from: \\
http://dx.doi.org/10.5772/35607   \\
\noindent
[50] Pommaret, J.-F.: The Mathematical Foundations of General Relativity Revisited, Journal of Modern Physics, 4 (2013) 223-239. \\
 http://dx.doi.org/10.4236/jmp.2013.48A022   \\
  \noindent
[51] Pommaret, J.-F.: The Mathematical Foundations of Gauge Theory Revisited, Journal of Modern Physics, 5 (2014) 157-170.  \\
http://dx.doi.org/10.4236/jmp.2014.55026  \\
 \noindent
[52] Pommaret, J.-F.: Relative Parametrization of Linear Multidimensional Systems, Multidim. Syst. Sign. Process., 26, 405-437 2015).  \\
DOI 10.1007/s11045-013-0265-0   \\
\noindent
[53] Pommaret,J.-F.:From Thermodynamics to Gauge Theory: the Virial Theorem Revisited, pp. 1-46 in "Gauge Theories and Differential geometry,", NOVA Science Publisher (2015).  \\
\noindent
[54] Pommaret, J.-F.: Airy, Beltrami, Maxwell, Einstein and Lanczos Potentials revisited, Journal of Modern Physics, 7, 699-728 (2016). \\
\noindent
http://dx.doi.org/10.4236/jmp.2016.77068   \\
\noindent
[55] Pommaret, J.-F.: Deformation Theory of Algebraic and Geometric Structures, Lambert Academic Publisher (LAP), Saarbrucken, Germany (2016). A short summary can be found in "Topics in Invariant Theory ", S\'{e}minaire P. Dubreil/M.-P. Malliavin, Springer 
Lecture Notes in Mathematics, 1478, 244-254 (1990).\\
http://arxiv.org/abs/1207.1964  \\
\noindent
[56] Pommaret, J.-F. and Quadrat, A.: Localization and Parametrization of Linear Multidimensional Control Systems, Systems \& Control Letters, 37, 247-260 (1999).  \\
\noindent
[57] Pommaret, J.-F., Quadrat, A.: Algebraic Analysis of Linear Multidimensional Control Systems, IMA Journal of Mathematical Control and Informations, 16, 275-297 (1999). \\
\noindent 
[58] Quadrat, A., Robertz, D.: Parametrizing all solutions of uncontrollable multidimensional linear systems, Proceedings of the 16th IFAC World Congress, Prague, July 4-8, 2005.  \\  
\noindent
[59] Quadrat, A.: An Introduction to Constructive Algebraic Analysis and its Applications, 
Les cours du CIRM, Journees Nationales de Calcul Formel, 1(2), 281-471 (2010).\\
\noindent
[60] Quadrat, A., Robertz, R.: A Constructive Study of the Module Structure of Rings of Partial Differential Operators, Acta Applicandae Mathematicae, 133, 187-234 (2014). \\
http://hal-supelec.archives-ouvertes.fr/hal-00925533   \\
\noindent
[61] Rotman, J.J.: An Introduction to Homological Algebra, Pure and Applied Mathematics, Academic Press (1979).  \\
\noindent
[62] Schneiders, J.-P.: An Introduction to D-Modules, Bull. Soc. Roy. Sci. Li\`{e}ge, 63, 223-295 (1994).  \\
\noindent
[63] Spencer, D.C.: Overdetermined Systems of Partial Differential Equations, Bull. Am. Math. Soc., 75 (1965) 1-114.\\
\noindent
[64] Teodorescu, P.P.: Dynamics of Linear Elastic Bodies,  Abacus Press, Tunbridge, Wells (1975) 
(Editura Academiei, Bucuresti, Romania).\\
\noindent
[65] Vessiot, E.: Sur la Th\'{e}orie des Groupes Infinis, Ann. Ec. Norm. Sup., 20, 411-451 (1903) (Can be obtained from 
http://numdam.org).  \\
\noindent
[66] Weyl, H.: Space, Time, Matter, Springer, 1918, 1958; Dover, 1952. \\
\noindent
[67] Zerz, E.: Topics in Multidimensional Linear Systems Theory, Lecture Notes in Control and Information Sciences (LNCIS) 256, 
Springer (2000).  \\
\noindent
[68] Zou, Z., Huang, P., Zang ,Y., Li, G.: Some Researches on Gauge Theories of Gravitation, Scientia Sinica, XXII, 6, 628-636 (1979).\\

\end{document}